\documentclass[12pt]{article}
\pdfoutput=1
\usepackage[utf8]{inputenc}
\usepackage[DIV13]{typearea}
\usepackage{indentfirst}
\usepackage{amsmath,amsfonts,mathrsfs,amssymb}
\usepackage{slashed,cancel,units}
\usepackage{array,longtable,multirow,multicol,colortbl}
\usepackage{lscape,fancyhdr}
\usepackage{graphicx,feynmp}
\usepackage[usenames,dvipsnames]{xcolor}
\usepackage{cite}
\usepackage[margin=0pt]{caption}
\RequirePackage[colorlinks=true,urlcolor=blue,anchorcolor=blue,citecolor=blue,filecolor=blue,
               linkcolor=black,menucolor=blue,linktocpage=true,pdfproducer=medialab]{hyperref}
\newcommand{\email}[1]{\href{mailto:#1}{\tt #1}}

\numberwithin{equation}{section}

\newcommand{\LLag}{\mathscr{L}}

\def\cF{{\cal F}}
\def\cO{{\cal O}}
\def\cQ{{\cal Q}}
\def\cP{{\cal P}}
\def\cR{{\cal R}}
\def\cS{{\cal S}}
\def\cY{{\cal Y}}
\def\Tr{{\rm Tr}}
\renewcommand{\a}{\alpha}
\renewcommand{\b}{\beta}
\renewcommand{\d}{\delta}
\newcommand{\g}{\gamma}
\newcommand{\e}{\varepsilon}
\newcommand{\s}{\sigma}
\newcommand{\st}{s_\theta}
\newcommand{\ct}{c_\theta}

\newcommand{\sdt}{s_{2\theta}}
\newcommand{\cdt}{c_{2\theta}}
\newcommand{\ssu}{\sigma^{\mu\nu}}

\newcommand{\de}{\partial}
\newcommand{\hc}{\text{h.c.}}
\newcommand{\unity}{1}

\newcommand{\tr}{\Tr}
\newcommand{\mean}[1]{\langle#1\rangle}
\DeclareMathOperator{\diag}{diag}
\newcommand{\U}{\mathbf{U}}
\newcommand{\T}{\mathbf{T}}
\newcommand{\V}{\mathbf{V}}
\newcommand{\D}{\mathbf{D}}
\newcommand{\WWd}{W_{\mu\nu}}
\newcommand{\WWu}{W^{\mu\nu}}
\newcommand{\BBd}{B_{\mu\nu}}
\newcommand{\BBu}{B^{\mu\nu}}
\newcommand{\GGd}{\mathcal{G}_{\mu\nu}}

\newcommand{\QBL}{\bar{Q}_L\,}
\newcommand{\QBR}{\bar{Q}_R\,}
\newcommand{\QL}{\,Q_L\,}
\newcommand{\QR}{\,Q_R\,}
\newcommand{\LBL}{\bar{L}_L\,}
\newcommand{\LBR}{\bar{L}_R\,}
\newcommand{\LL}{\,L_L\,}
\newcommand{\LR}{\,L_R\,}
\newcommand{\col}{\cellcolor{SkyBlue!20}}
\def\beq{\begin{equation}}
\def\eeq{\end{equation}}
\def\nn{\nonumber}
\newcommand{\tfqName}{\mathcal{N}^\mathcal{Q}}	\newcommand{\tfqC}{n^\mathcal{Q}}		\newcommand{\tfqA}{(na)^\mathcal{Q}}         
\newcommand{\tfqAA}{(naa')^\mathcal{Q}}
\newcommand{\tflName}{\mathcal{N}^\ell}		\newcommand{\tflC}{n^\ell}                      \newcommand{\tflA}{(na)^\ell}                
\newcommand{\tflAA}{(naa')^\ell}
\newcommand{\ffqName}{R^\mathcal{Q}}		\newcommand{\ffqC}{r^\mathcal{Q}}              
\newcommand{\fflName}{R^\ell}			\newcommand{\fflC}{r^\ell}                      
\newcommand{\ffqlName}{R^{\mathcal{Q}\ell}}	\newcommand{\ffqlC}{r^{\mathcal{Q}\ell}}        
\renewcommand{\P}{\mathcal{P}}

\newcommand{\tfq}{\refstepcounter{twofermionQ}\tfqName_{\arabic{twofermionQ}}}
\newcounter{twofermionQ}
\newcommand{\Otfq}[1]{\tfqName_{\ref{#1}}} 
\newcommand{\Ctfq}[1]{\tfqC_{\ref{#1}}}    
\newcommand{\Atfq}[1]{\tfqA_{\ref{#1}}}    
\newcommand{\AAtfq}[1]{\tfqAA_{\ref{#1}}}  
\newcommand{\tfl}{\refstepcounter{twofermionL}\tflName_{\arabic{twofermionL}}}
\newcounter{twofermionL}
\newcommand{\Otfl}[1]{\tflName_{\ref{#1}}} 
\newcommand{\Ctfl}[1]{\tflC_{\ref{#1}}}    
\newcommand{\Atfl}[1]{\tflA_{\ref{#1}}}    
\newcommand{\AAtfl}[1]{\tflAA_{\ref{#1}}}  
\newcommand{\ffq}{\refstepcounter{fourfermionQ}\ffqName_{\arabic{fourfermionQ}}}
\newcounter{fourfermionQ}
\newcommand{\Offq}[1]{\ffqName_{\ref{#1}}} 
\newcommand{\Cffq}[1]{\ffqC_{\ref{#1}}}    
\newcommand{\ffl}{\refstepcounter{fourfermionL}\fflName_{\arabic{fourfermionL}}}
\newcounter{fourfermionL}
\newcommand{\Offl}[1]{\fflName_{\ref{#1}}} 
\newcommand{\Cffl}[1]{\fflC_{\ref{#1}}}    
\newcommand{\ffql}{\refstepcounter{fourfermionQL}\ffqlName_{\arabic{fourfermionQL}}}
\newcounter{fourfermionQL}
\newcommand{\Offql}[1]{\ffqlName_{\ref{#1}}} 
\newcommand{\Cffql}[1]{\ffqlC_{\ref{#1}}}    

\newcommand{\Sas}{\renewcommand{\arraystretch}{2}}
\newcommand{\Sasp}{\renewcommand{\arraystretch}{1.7}}

\newcommand{\blue}[1]{\color{blue} #1 \color{black}}
\begin{document}
\begin{titlepage}
\vspace*{-1cm}
\phantom{hep-ph/***} 
{\flushleft
{\blue{FTUAM-16-13}}
\hfill{\blue{IFT-UAM/CSIC-16-034}}\\
{\blue{YITP-SB-16-14}}
\hfill\\}
\vskip 1cm
\begin{center}
\mathversion{bold}
{\LARGE\bf The Complete HEFT Lagrangian}\\[4mm]
{\LARGE\bf  after the LHC Run I}\\
\mathversion{normal}
\vskip .3cm
\end{center}
\vskip 0.5  cm
\begin{center}
{\large I.~Brivio}~$^{a)}$,
{\large J.~Gonzalez--Fraile}~$^{b)}$,
{\large M.~C.~Gonzalez--Garcia}~$^{c),d),e)}$,
{\large L.~Merlo}~$^{a)}$\\
\vskip .7cm
{\footnotesize
$^{a)}$~
Departamento de F\'isica Te\'orica and Instituto de F\'isica Te\'orica, IFT-UAM/CSIC,\\
Universidad Aut\'onoma de Madrid, Cantoblanco, 28049, Madrid, Spain\\
\vskip .1cm
$^{b)}$~
Institut f\"ur Theoretische Physik, Universit\"at Heidelberg, Germany\\
\vskip .1cm
$^{c)}$~
C.N.~Yang Institute for Theoretical Physics and Department of Physics and Astronomy, SUNY at Stony Brook, Stony Brook, NY 11794-3840, USA\\
\vskip .1cm
$^{d)}$~
Departament d'Estructura i Constituents de la Mat\`eria and ICC-UB, Universitat de Barcelona, 647 Diagonal, E-08028 Barcelona, Spain\\
\vskip .1cm
$^{e)}$~
Instituci\'o Catalana de Recerca i Estudis Avan\c{c}ats (ICREA)

\vskip .3cm
\begin{minipage}[l]{.9\textwidth}
\begin{center} 
\textit{E-mail:} 
\email{ilaria.brivio@uam.es},
\email{fraile@thphys.uni-heidelberg.de},
\email{concha@insti.physics.sunysb.edu},
\email{luca.merlo@uam.es}
\end{center}
\end{minipage}
}
\end{center}
\vskip 0.5cm
\begin{abstract}
The complete effective chiral Lagrangian for a dynamical Higgs is
presented and constrained by means of a global analysis including electroweak precision 
data together with Higgs and triple gauge boson coupling
data from the LHC Run~I. The operators' basis up to next-to-leading order in the
expansion consists of~148 (188~considering right-handed
neutrinos) flavour universal terms and it is presented here making explicit the 
custodial nature of the operators. This effective Lagrangian provides the most 
general description of the physical Higgs couplings once the electroweak symmetry
is assumed, and it allows for deviations from the $SU(2)_L$ doublet
nature of the Standard Model Higgs. The comparison with the effective
linear Lagrangian constructed with an exact $SU(2)_L$ doublet Higgs
and considering operators with at most canonical dimension six is
presented. A promising strategy to disentangle the two descriptions consists in
analysing (i) anomalous signals present only in the chiral Lagrangian and not expected
in the linear one, that are potentially relevant for
LHC searches, and (ii) decorrelation effects between observables
that are predicted to be correlated in the linear case and not in the 
chiral one. The global analysis presented here, that includes several kinematic distributions,
is crucial for reducing the allowed parameter space and for controlling the correlations between
parameters. This improves previous studies aimed at investigating the Higgs Nature and the 
origin of the electroweak symmetry breaking.
\end{abstract}
\end{titlepage}
\setcounter{footnote}{0}

\pdfbookmark[1]{Table of Contents}{tableofcontents}
\tableofcontents

\newpage

\section{Introduction}
\label{Sect:Intro}

The discovery of a resonance at
LHC~\cite{Aad:2012tfa,Chatrchyan:2012xdj} compatible with the Standard
Model (SM) scalar boson (``Higgs'' for
short)~\cite{Englert:1964et,Higgs:1964ia,Higgs:1964pj} opened a new
era in particle physics. Now, the on going LHC measurements of the Higgs
properties are a crucial step to understand the nature of the Higgs boson and of the
Electroweak (EW) symmetry breaking (EWSB).

Without entering into details of specific scenarios, the formalism of
Effective Field Theories (EFT) represents an optimal tool for studying
the phenomenology of the Higgs sector. In particular, an appropriate description of
scenarios in which the Higgs belongs to an elementary $SU(2)$ doublet
is provided by the Standard Model EFT (SMEFT). This consists of
operators constructed with the SM spectrum, invariant under the
Lorentz and SM gauge symmetries and respecting an expansion in
canonical mass dimensions $d$. Assuming lepton and baryon number
conservation, the first corrections to the SM are provided by
operators of dimension
six~\cite{Buchmuller:1985jz,Grzadkowski:2010es}, suppressed by two
powers of the cut-off scale $\Lambda$. Weakly coupled theories are the typical underlying
scenarios that can be matched to the SMEFT (also referred to
as ``linear'' Lagrangian) at low energy.

Scenarios where the Higgs does not belong to an elementary exact
$SU(2)_L$ doublet are still allowed within the current experimental
accuracy. This is the case, for example, of Composite Higgs
models~\cite{Kaplan:1983fs,Kaplan:1983sm,Banks:1984gj,Agashe:2004rs,Gripaios:2009pe}
or dilaton constructions~\cite{Halyo:1991pc,Goldberger:2008zz}. It is
then fundamental and necessary to identify observables that allow to
disentangle these different possibilities. 
When the Higgs is not required to belong to an exact EW doublet,
instead, a useful tool is the so-called Higgs EFT (HEFT) (also dubbed
``chiral'' Lagrangian). The main difference between SMEFT and HEFT
resides in the fact that, in the latter formalism, the physical Higgs
$h$ and the ensemble of the three EW Goldstone bosons $\vec{\pi}$ are
treated as independent objects, rather than being collectively
described by the Higgs doublet. In particular, the physical Higgs $h$
is assigned to a singlet representation of the SM gauge groups. The
Goldstone bosons' sector has been studied intensely in the
past~\cite{Appelquist:1980vg,Longhitano:1980iz,Longhitano:1980tm,Feruglio:1992wf}
in the context of Higgs-less EWSB scenarios. These works were the
first to describe the GBs by means of a dimensionless unitary matrix
transforming as a bi-doublet of the global symmetry $SU(2)_L\times
SU(2)_R$,
\beq
\U(x)\equiv e^{i\sigma_a \pi^a(x)/f_\pi}\,,\qquad\qquad
\U(x)\to L \U(x) R^\dag\,,
\eeq
being $f_\pi$ the scale associated to the SM GBs, and $L,\,R$ the
$SU(2)_{L,R}$ transformations. After EWSB, the invariance under the group
$SU(2)_L\times SU(2)_R$ is broken down to the diagonal $SU(2)_C$,
commonly called custodial symmetry, and explicitly broken by the
gauging of the hypercharge $U(1)_Y$ and by the fermion mass
splittings.  It is customary to introduce two objects, the vector and
scalar chiral fields, that transform in the adjoint of $SU(2)_L$.
They are defined, respectively, as
\beq\label{def_V_T}
\V_\mu\equiv (\D_\mu\U)\U^\dag\,,\qquad\qquad
\T\equiv\U\sigma_3\U^\dag\,,
\eeq
where the covariant derivative is given by
\beq
\D_\mu\U(x)\equiv \partial_\mu\U(x)+igW_\mu(x)\U(x)
-\dfrac{ig'}{2}B_\mu(x)\U(x)\sigma_3\,.
\eeq
Unlike $\V_\mu$, $\T$ is not invariant under $SU(2)_C$ and can
therefore be considered a custodial symmetry breaking spurion. The bosonic Higgs-less
EW chiral Lagrangian can then be constructed with $\V_\mu$, $\T$ and
the gauge-boson field strengths as building blocks, and the tower of
invariant operators shall be organised according to a chiral
(derivative) expansion~\cite{Weinberg:1978kz}.

In the last decade, the EW chiral Lagrangian has been extended with
the introduction of a light physical Higgs
$h$~\cite{Feruglio:1992wf,Grinstein:2007iv,Contino:2010mh,Alonso:2012px,Alonso:2012pz,
Brivio:2013pma,Gavela:2014vra,Buchalla:2013rka,Yepes:2015zoa,Yepes:2015qwa},
treated as an isosinglet of the SM gauge symmetries. The dependence on
the $h$ field is customarily encoded in generic functions $\cF(h)$,
that are used as building blocks for the construction of the effective
operators. These functions are made adimensional by implicitly
weighting the insertions of the Higgs field with an opportune
suppression scale $f_h$, so that one may rewrite the dependence as
$\cF(h/f_h)$.  It is worth underlining that the dependence on the
structure $(1+h/v)$, where $v$ is the EW vacuum expectation value
(vev), that characterises the SMEFT Lagrangian is lost in the HEFT and
substituted by a generic $h/f_h$ expansion.

The typical underlying scenarios that can be described at low-energy
in terms of the matrix $\U(x)$, the Higgs functions $\cF(h)$ and the
rest of the SM fields, are those of Composite Higgs 
models~\cite{Kaplan:1983fs,Kaplan:1983sm,Banks:1984gj,Agashe:2004rs,Gripaios:2009pe,Feruglio:2016zvt}. These
assume the existence of some strong (``ultracolour'') interaction at a
high energy, and initially invariant under some global symmetry group
$\mathcal{G}$. At the scale $\Lambda_s$, the formation of ultracolour
condensates breaks spontaneously this invariance, leaving a residual
symmetry $\mathcal{H}$ that can embed the EW group. This triggers the
appearance of a certain number of Goldstone bosons, among which three
can be identified with the would-be GBs of the EW group and a fourth one
with the Higgs.  In such scenarios, all the SM scalars are naturally
associated to the same scale $f_\pi=f_h\equiv f$, with
$\Lambda_s\leq 4\pi f$.  Spontaneous EWSB is triggered by some
explicit breaking of the $\mathcal{H}$ symmetry (provided either by
external symmetries~\cite{Kaplan:1983fs} or by gauging the SM symmetry together with
fermion interactions~\cite{Agashe:2004rs}) and takes place in a second stage. At this
level, the Higgs field acquires a vev $\mean{h}$ that does not need to
coincide with the EW scale $v$, defined by the EW gauge-boson mass:
the three quantities $v$, $f$ and $\mean{h}$ are instead related by a
model-dependent function. The splitting between $v$ and $f$
constitutes the well-known fine-tuning of composite Higgs models.
It is usually expressed in terms of the parameter
\beq
\xi\equiv \dfrac{v^2}{f^2}\,,
\label{xiDefinition}
\eeq
that substantially quantifies the degree of non-linearity of the Higgs
dynamics. The low-energy projection of composite Higgs models can be
described by the HEFT Lagrangian~\cite{Alonso:2014wta,Hierro:2015nna}
and the matching conditions allow to write the low-energy effective
operator coefficients in terms of the high-energy parameters, and the
generic functions $\cF(h)$ as trigonometric functions of $h/f$.
The HEFT Lagrangian can also be used to describe the
SMEFT~\cite{Alonso:2012px,Alonso:2012pz,Brivio:2013pma,
Alonso:2014wta,Hierro:2015nna,Brivio:2014pfa,Gavela:2014vra},
after identifying the operator coefficients of the effective
Lagrangians and writing all the $\cF(h)$ functions in terms of
$(1+h/v)$. Dilaton constructions~\cite{Halyo:1991pc,Goldberger:2008zz} 
or even more exotic models, where the Higgs is an EW singlet, can 
also be described by the HEFT Lagrangian.

Without assuming any specific underlying scenario or comparing with
SMEFT, the $v/f_h$ and $v/f_\pi$ parameters are not physical and can be
reabsorbed in the operators coefficients and in the coefficients of
the $\cF(h)$ functions. This is tantamount to substituting $f_\pi$ and $f_h$ by
$v$, which ensures canonical kinetic terms for the GBs and fixes the correct
order of magnitude for the gauge bosons masses, without fine-tunings. This notation will be
employed in the following, unless otherwise specified.

The disparities between the SMEFT and the HEFT originate from the
different nature of the building blocks used in the construction of
the effective operators. The independence between the GB field $\U(x)$
and the physical $h$, together with the fact that $h$ does not
transform under the SM gauge symmetries, leads to a different ordering
of the chiral effective operators compared to the linear ones.  As a
result, at any given order in the expansion the number of chiral
independent operators is much larger than in the SMEFT case.  The
corresponding phenomenology, focussing on the bosonic part of the
Lagrangian, has been studied in Refs.~\cite{Brivio:2013pma,
Gavela:2014vra}, where signatures that may allow one to discriminate
between an elementary and a dynamical Higgs have also been
identified. These signatures include sets of couplings that are
predicted to be correlated in an elementary Higgs scenario but are
generically decorrelated in the dynamical case, as well as effects
that are expected to be suppressed in the linear realisation but may
appear at the lowest order in the chiral expansion. These signatures are also 
typical in Dark Matter studies when the Higgs is not taken to 
be an exact $SU(2)_L$ doublet~\cite{Brivio:2015kia}. Complementary
signatures that can distinguish between SMEFT and HEFT also 
include the scattering of the longitudinal components of the 
gauge bosons~\cite{Murayama:2014yja,Delgado:2013hxa,Delgado:2014jda}.

The complete non-redundant HEFT Lagrangian including both bosonic and fermionic
operators has been constructed in this work and is presented in
Sect.~\ref{Sect:CompleteHEFT}, making explicit the
custodial nature of the operators.
The HEFT basis is formed by 148 independent flavour universal operators altogether, 
whose extension to generic flavour contractions is straightforward.
The Lagrangian does not account for the presence of right-handed neutrinos, whose
inclusion in the spectrum would imply the addition 40 extra operators to the basis, listed
in Appendix~\ref{APP:NR}.
Section~\ref{Sect:CompleteHEFT} also contains a comparison between the HEFT Lagrangian and the
SMEFT one, while a phenomenological analysis of the HEFT basis is presented in Sect.~\ref{sec.pheno}.
The study considers all the available collider data, which includes electroweak precision measurements
and Higgs and triple gauge-boson vertex (TGV) data from the LHC Run~I. To our knowledge,
this is the first time that such analysis has been done for
the complete HEFT description.
Finally, Sect.~\ref{Sect:EFTValidity}, contains a discussion of
the impact of higher order operators: a set of invariants that may become relevant at the increased
energies foreseen for the LHC and future colliders is also pointed out.
The conclusions are presented in Sect.~\ref{Sect:Conclusions},
while some more technical details are deferred to the appendices, together with the Feynman Rules 
for the CP-even subset of HEFT operators.

%
%
\section{The complete HEFT Lagrangian}
\label{Sect:CompleteHEFT}
In this section we review the construction of the HEFT Lagrangian, in
a notation similar to that of
Refs.~\cite{Alonso:2012jc,Alonso:2012px,Alonso:2012pz,
Brivio:2013pma,Gavela:2014vra,Brivio:2014pfa}. The
bosonic building blocks are the gauge
field strengths $\BBd$, $\WWd$, $\mathcal{G}_{\mu\nu}$, the vector and
scalar chiral fields $\V_\mu$ and $\T$ defined in Eq.~\eqref{def_V_T}
and the functions $\cF(h)$ introduced in the previous section.  The SM
fermions are conveniently grouped into doublets of the global
$SU(2)_{L,R}$ symmetries:
\begin{equation}
\begin{aligned}
 Q_L &= \binom{U_L}{D_L}\,,\qquad & 
 Q_R&=\binom{U_R}{D_R}\,,\qquad &
 L_L &= \binom{\nu_L}{E_L}\,,\qquad & 
 L_R&=\binom{0}{E_R}\,.
 \end{aligned}
\end{equation}
This choice allows one to have a more compact notation for the fermionic
operators. The $SU(2)_R$ doublet structure can easily be broken with 
the insertion of the custodial symmetry breaking spurion $\T$. 
Notice that the $\LR$ doublet only includes right-handed charged leptons.
The inclusion of right-handed neutrinos
requires an extension of the fermionic basis presented in Sec.~\ref{sec.Lfer} with the addition of the 
 operators listed in App.~\ref{APP:NR}.

The HEFT Lagrangian can be written as a sum of two terms,
\beq
\LLag_\text{HEFT}\equiv \LLag_0 + \Delta \LLag\,,
\eeq
where the first term contains the leading order (LO) operators and the
second one accounts for new interactions and for deviations from the
LO.

The LO Lagrangian includes the kinetic terms for all the particles in
the spectrum, the Yukawa couplings and the scalar
potential\footnote{Comments on the construction of the LO Lagrangian in Eq.~(\ref{Lag0}) are given in App.~\ref{APP:LOLag}.}:
\begin{equation}
\begin{split}
\LLag_0=& -\dfrac{1}{4} \GGd^\alpha \mathcal{G}^{\alpha\,\mu\nu}
-\dfrac{1}{4} \WWd^a W^{a\,\mu\nu}-\dfrac{1}{4} \BBd\BBu+\\
&+\dfrac{1}{2}\de_\mu h \de^\mu h-\dfrac{v^2}{4}\Tr(\V_\mu \V^\mu)\cF_C(h)-V(h)
+\\
&+i\bar{Q}_L\slashed{D}Q_L+i\bar{Q}_R\slashed{D}Q_R+i\bar{L}_L\slashed{D}L_L
+i\bar{L}_R\slashed{D}L_R+\\
&-\dfrac{v}{\sqrt2}\left(\bar{Q}_L\U \cY_Q(h) Q_R+\hc\right)
-\dfrac{v}{\sqrt2}\left(\bar{L}_L\U \cY_L(h) L_R+\hc\right)+\\
&-\dfrac{g_s^2}{16\pi^2}\lambda_s\,\GGd^\alpha\, 
\tilde{\mathcal{G}}^{\alpha\,\mu\nu}\,,
\end{split}
\label{Lag0}
\end{equation}
where $\tilde{\mathcal{G}^{\mu\nu}}\equiv 
\frac12\epsilon^{\mu\nu\rho\sigma} \mathcal{G}_{\rho\sigma}$. The
first line describes the kinetic terms of the gauge bosons; the second
line contains the Higgs and Goldstone bosons' kinetic term, the scalar potential, and the mass terms for the EW gauge bosons; the
third line presents the kinetic terms for all the fermions, while the
fourth line accounts for the Yukawa interactions. Finally, the last
line contains the theta term of QCD. The function $\cF_C(h)$ appearing
in the kinetic term for the GBs can be expanded as
\begin{equation}\label{FC}
 \cF_C(h) = 1 + 2 a_C \frac{h}{v}+b_C \frac{h^2}{v^2}+\dots
\end{equation} 
where the dots account for higher powers of $(h/v)$. For the
the phenomenological analysis it is convenient to single out the BSM
part of the coefficients $a_C,\, b_C$, using the notation
\begin{equation}
 a_C = 1 + \Delta a_C\,,\qquad 
 b_C = 1 + \Delta b_C\,,
\end{equation} 
where $\Delta a_C,\, \Delta b_C$ will be assumed to be of the same
order as the coefficients accompanying the operators appearing in $\Delta \LLag$.  The functions
$\cY_{Q,L}(h)$ appearing in the Yukawa couplings have an analogous structure to $\cF_C(h)$:
\begin{equation}
\begin{aligned}
\cY_{Q}(h)\equiv& \diag\left(\sum_n Y_{U}^{(n)}\dfrac{h^n}{v^n},
\sum_n Y_{D}^{(n)}\dfrac{h^n}{v^n}\right)\,,\qquad
\cY_{L}(h)\equiv& \diag\left(0,\sum_n Y_{\ell}^{(n)}\dfrac{h^n}{v^n}\right)\,.
\end{aligned}
\end{equation}
The $n=0$ terms yield fermion masses, while the higher orders describe
the interaction with $n$ insertions of the Higgs field $h$, accounting
in general for non-aligned contributions.

The kinetic terms of the fermions and of the physical Higgs are not
accompanied by any $\cF(h)$ since, as shown in
App.~\ref{APP:LOLag}, it is always possible to reabsorb their
contributions inside the generic functions $\cF_C(h)$ and
$\cY_{Q,L}(h)$. This can be done either via a field redefinition or,
alternatively, applying the Equations of Motion (EOMs) (the two procedures
are \emph{not} equivalent in general, but lead to the same result at
first order in the deviations from the LO). Moreover, the kinetic
terms of the gauge bosons in the first line of Eq.~(\ref{Lag0}) 
do not come associated with any $\cF(h)$,
assuming that the transverse components of the gauge fields, described
by the gauge field strength, do not couple strongly to the Higgs
sector. These couplings can be neglected at the LO and be considered, instead, at the next-to-leading order (NLO).

$\Delta \LLag$ contains higher order operators with respect to those appearing in $\LLag_0$. 
The precise ordering of these operators depends on the choice 
of a specific power counting rule. The HEFT can be seen
as a fusion of two theories, the chiral perturbation approach associated 
to the SM GBs -- i.e. the longitudinal components of the gauge bosons -- 
and the traditional linear description that applies to the transverse components of
the gauge bosons and to fermions. The physical $h$ should also
undergo the chiral perturbation description as it enters in the Lagrangian via the
adimensional functions $\cF(h)$: the latter can be interpreted as playing the same role as the
adimensional GB matrix field $\U(x)$. Indeed, in concrete composite Higgs
models, the pseudo-GB nature of the Higgs forces the $\cF(h)$ functions to take trigonometric structures~\cite{Alonso:2014wta}. 
Being the HEFT a merging between linear and
chiral descriptions, the counting rules which apply singularly to each
of the expansions hold simultaneously for the HEFT~\cite{Gavela:2016bzc}. 
As a result, the LO Lagrangian in Eq.~(\ref{Lag0}) itself does not strictly respect 
the chiral expansion:
$\LLag_0$ contains both operators
with two derivatives and the gauge-boson kinetic terms, which has four derivatives; at the
same time, some two-derivative operators have been excluded from the LO. On the other
hand, $\LLag_0$ does not even follow an expansion in canonical dimensions, as for instance the
Yukawa interactions and the gauge-boson mass term present an infinite series of $h$ legs,
contrary to all the other terms in the LO Lagrangian.

The renormalisability conditions are also different in the two descriptions. In the linear
expansion  an $n$-loop diagram containing one single $d = 6$ vertex generates divergent
contributions that can be reabsorbed by other $d = 6$ operators and do not require the
introduction of any higher-dimensional operator. On the contrary, in the chiral case, 1-loop
diagrams with $n$ insertions of a two-derivative coupling, usually listed in the LO Lagrangian,
produce divergences that require the introduction of operators with four-derivatives, which
generically constitute the NLO Lagrangian.

Finally, the HEFT presents an additional aspect that makes it hard
to identify a proper counting rule: the presence of multiple
scales. Besides the cut-off of the theory $\Lambda$, one should
consider the presence of the GB scale $f_\pi$ and of the $h$-scale $f_h$.
Although it may happen that the last two coincide with $f_\pi =f_h=f$ and that they are related to the first one by the constraint
$\Lambda\leq4\pi f$ (which is the case in composite Higgs models),
the three scales are in principle independent and associated to different physical quantities.
On top of this, one should not forget the fine-tuning
associated to the EW scale $v$ and parametrised by $\xi$ defined in
Eq.~(\ref{xiDefinition}). 
In practice, the counting
rule associated to the HEFT depends on more than one expansion 
parameters and may vary depending on the typical energy scale 
of the observables considered in the phenomenological analysis.

In conclusion, rather than basing the choice of the NLO Lagrangian
operators on a sophisticated counting rule whose applicability 
is not valid in full generality, here the selection is performed with the following
strategy. An NLO operator should satisfy at least one of the criteria below:
\begin{itemize}
\item It is necessary for reabsorbing 1-loop 
divergences arising from the renormalisation of $\LLag_0$.
\item It presents the same suppression as the operators in 
the first class and receives finite 1-loop contributions: for instance,
all the four-fermion operators are included in the NLO, 
in spite of the fact that only a subset 
of these is required to reabsorb 1-loop divergences.
\item It has been left out from the LO Lagrangian 
due to phenomenological reasons.
\end{itemize}

The suppression factor of each operator is determined using the NDA master formula, first proposed in Ref.~\cite{Manohar:1983md} and later modified in Refs.~\cite{Cohen:1997rt} and \cite{Gavela:2016bzc}. 
Following the notation of Ref.~\cite{Gavela:2016bzc}:
\beq
\frac{\Lambda^4}{16 \pi^2 } \left[\frac{\partial}{\Lambda}\right]^{N_p}  
\left[\frac{ 4 \pi\,  \phi}{ \Lambda} \right]^{N_\phi}
 \left[\frac{ 4 \pi\,  A}{ \Lambda } \right]^{N_A}  
\left[\frac{ 4 \pi \,  \psi}{\Lambda^{3/2}}\right]^{N_\psi} 
\left[ \frac{g}{4 \pi }  \right]^{N_g}
\left[\frac{y}{4 \pi } \right]^{N_y}\,,
\label{MasterFormula}
\eeq
where $\phi$ represents either the SM GBs or $h$, $\psi$ a generic
fermion, $A$ a generic gauge field, $g$ the gauge couplings and $y$
the Yukawa couplings. All the operators appearing in the LO Lagrangian
in Eq.~(\ref{Lag0}) are normalised according to this formula, apart
from the operators providing gauge-boson masses,
$(v^2/4)\tr(\V_\mu\V^\mu)\cF_C(h)$, and fermions' masses
$(v\sqrt2)\bar\psi_L\U\cY_\psi(h)\psi_R$, which are multiplied by
powers of the EW scale $v$ and not by $\Lambda$ or $f$ as expected. This is
due to the well-known fine-tuning, typical of theories where the EWSB 
sector is non-linearly realised. Notice that with these conventions all
the kinetic terms are canonically normalised, differently from what
follows using the original version of the NDA master formula from
Ref.~\cite{Manohar:1983md}.

The master formula also ensures that the operators belonging to the NLO
Lagrangian are typically suppressed with respect to those of $\LLag_0$
by powers of $(4\pi)^{(n\leq2)}$, reflecting the renormalisation of the
chiral sector, and/or by powers of $\Lambda^{(n\leq2)}$, associated to possible
new physics contributions. Different cases will be
discussed when necessary.

\subsection{The NLO Lagrangian}
The second part of the HEFT Lagrangian, $\Delta\LLag$, contains in
general all the invariant operators appearing beyond the leading
order. They include corrections to the interactions contained in
$\LLag_0$ as well as completely new couplings.  This Lagrangian can be
generically written as a sum of two parts
\begin{equation}
 \Delta\LLag = \Delta\LLag_\text{bos}+ \Delta\LLag_\text{fer}\,,
\end{equation}
where $\Delta\LLag_\text{bos}$ contains all the purely-bosonic operators,
while $\Delta\LLag_\text{fer}$ accounts for the interactions that involve
fermions.

In this work, $\Delta\LLag$ will be restricted to the NLO, defined
according to the rules presented in the previous section.  
An alternative construction of a NLO Lagrangian was derived in Ref.~\cite{Buchalla:2013rka}.
We present a set of invariants that forms a complete, non-redundant basis
at this order in the effective expansion, which has been constructed
identifying first a complete basis for each of the two sectors
individually (bosonic and fermionic) and subsequently employing the
EOMs to remove redundant terms.

Given the large number of invariants, the operators are classified as
follows: the bosonic basis is split into CP conserving and CP violating
subsets (the field $h$ is assumed to be a CP-even scalar): 
\beq
\Delta\LLag_\text{bos} =\Delta \LLag_\text{bos}^{CP} + \Delta\LLag_\text{bos}^{\cancel{CP}}\,,
\eeq
while in the fermionic sector the distinction is between fermionic single- 
and double-current structures:
 \beq
\Delta \LLag_\text{fer} =\Delta \LLag_{2F} +\Delta \LLag_{4F}\,.
\end{equation} 
The operators are named differently according to the category to which
they belong and each of them includes a function $\cF_i(h)$
conventionally parametrised as
\begin{equation}
 \cF_i(h) =1 + 2a_i\frac{h}{v}+b_i\frac{h^2}{v^2}+\dots
\end{equation} 
Moreover, each effective operator is multiplied with a real
coefficient, indicated with a lowercase letter
($c,\,\tilde{c},\,n,\,r$) associated to each class.  The following
table defines the notation and summarises the number of independent
invariants for each set, in the absence of right-handed neutrinos and
after the application of the EOMs.
\begin{center}
\renewcommand{\arraystretch}{1.5}
 \begin{tabular}{>{$}l<{$}l>{$ }c<{ $}c}
  \hline
  \mathbf{\LLag}& \text{\bf Sub-category}& \text{\bf Notation}& {\bf \# operators} \\\hline
  \\[-5mm]
  \Delta\LLag_\text{bos}^{CP}& 		&c_j\;\cP_j &	      26\\
  \Delta\LLag_\text{bos}^{\cancel{CP}}&	&\tilde{c}_j\;\cS_j & 16\\[2mm]\hline
  \\[-5mm]
  \Delta\LLag_{2F}&  Quark current&	\tfqC_j\;\tfqName_j&  36\\
  & 		Lepton current&		\tflC_j\;\tflName_j&  14\\[2mm]
 \Delta\LLag_{4F}& Four quarks&		\ffqC_j\;\ffqName_{j}& 26\\
  & 	Four leptons&			\fflC_j\;\fflName_{j}& 7\\
  & 	Two quarks and two leptons& 	\ffqlC_j\;\ffqlName_{j}&23\\\hline
  \text{Tot}& & & 148\\\hline
\end{tabular}
\end{center}
Forty additional operators should be considered if right-handed
neutrinos are added to the spectrum: 17 in $\LLag_{2F}$, 8
four-leptons interactions and 15 mixed two-quark-two-lepton terms.

The complete list of NLO operators is provided in the following:
Sects.~\ref{sec.L_bos} and~\ref{sec.Lfer} are, respectively, dedicated
to the bosonic and fermionic sectors.  Further details of the
construction of the invariants and of how the EOMs have
been employed to remove redundant terms can be found in
Appendices~\ref{APP:ReductionFermionBasis} and~\ref{APP:EOM}.  The Feynman
rules of the complete CP conserving basis are reported in Appendix~\ref{APP:FR}, in
unitary gauge and for vertices with up to four legs.

\boldmath
\subsection{NLO basis: bosonic sector \texorpdfstring{$\Delta\LLag_\text{bos}$}{Lbos}}
\unboldmath
\label{sec.L_bos}
At NLO in the chiral expansion, the Lagrangian $\Delta\LLag_\text{bos}$
contains purely bosonic operators. Complete bases for the CP even and CP odd sectors have
been already constructed in Refs.~\cite{Alonso:2012px,Brivio:2013pma}
and~\cite{Gavela:2014vra} respectively. In this work only a subset of
those ensembles are retained as, once the fermionic sector is
introduced, some of the terms become redundant and can be removed
using the EOMs (see App.~\ref{APP:EOM}). Nonetheless,
the original numeration of the operators has been kept, in order to
simplify the comparison with the literature. Finally, the explicit
formal dependence on $h$ in the generic functions $\cF_i(h)$ is
dropped in the following for brevity.

\boldmath
\subsubsection{CP even bosonic basis \texorpdfstring{$\Delta\LLag_\text{bos}^{CP}$}{Lbos-CP}}
\unboldmath

The CP even NLO Lagrangian reads 
\begin{equation}
\Delta\LLag_\text{bos}^{CP} = \sum_{j} c_j\cP_j(h)\,,
\end{equation} 
with
\begin{equation}
j=\{T,B,W,G,DH,1-6,8,11-14,17,18,20-24,26, WWW, GGG\}
\end{equation} 
where all the operators contain four derivatives, with the exception of
\beq
\cP_T(h) = \frac{v^2}{4} \tr(\T\V_\mu)\tr(\T\V^\mu) \cF_{T}\,,
\eeq
and
\beq
\begin{aligned}
\cP_{WWW}(h) &=\dfrac{4\pi\varepsilon_{abc}}{\Lambda^2}W_\mu^{a\nu}W_\nu^{b\rho} W_{\rho}^{c\mu}\cF_{WWW}\,,\\
\cP_{GGG}(h) &=\dfrac{4\pi f_{\alpha\beta\gamma}}{\Lambda^2}G_\mu^{\alpha\nu}G_\nu^{\beta\rho} G_{\rho}^{\gamma\mu}\cF_{GGG}\,,
\end{aligned}
\eeq
where $f_{\a\b\g}$ denotes the structure constants of $SU(3)$.

The two-derivative operator $\cP_T(h)$ is very similar 
to $v^2\Tr(\V_\mu \V^\mu)\cF_C$ and, therefore, it could have been included 
in $\LLag_0$ {\it a priori}. However, it is customary to move it to
$\Delta\LLag$ because  the bounds existing on its coefficient are
quite strong: $c_T\lesssim 10^{-2}$. In fact, this operator violates
the custodial symmetry and contributes to the $T$ parameter, which is
constrained to a high accuracy by electroweak precision data
(EWPD). In order to avoid irrelevant contributions to the EOMs, this operator has been moved to the NLO,
which is justifiable assuming an approximately preserved custodial symmetry.\footnote{Although 
the $T$ parameter only constrains the $h$-independent coupling of $\P_T(h)$, 
the whole operator has been moved to the NLO Lagrangian. This follows the basic 
assumption that for a given operator the $h^{n>0}$ coefficients are of the same
order as the $h^0$ coefficient. Indeed, if an operator is suppressed due to a 
symmetry principle, this applies to any of the $h^{n\geq0}$
couplings.}  
The two operators $\cP_{WWW}(h)$ and $\cP_{GGG}(h)$ 
are not required to absorb divergences due to the 1-loop 
renormalisation. However, they can be listed among the NLO operators: 
containing only the transverse components of the gauge bosons, they 
follow the linear description; then they come suppressed by $\Lambda^2$, 
on the same foot as the four-fermion operators.
It will be shown in the following that they have a non-trivial impact
at the phenomenological level.

The remaining 23 operators in $\Delta\LLag_\text{bos}^{CP}$, in the numeration 
of Ref.~\cite{Brivio:2013pma}, are the following:
\begin{center}
\renewcommand{\arraystretch}{1.8}
\begin{tabular}{>{$}l<{$}@{\hspace*{0.5cm}}>{$}l<{$}}
\cP_{B}(h) =-\dfrac{1}{4}\BBd \BBu \cF_B&
\cP_{W}(h) =-\dfrac{1}{4}\WWd^a W^{a\mu\nu} \cF_W \\
\cP_G(h) = -\dfrac{1}{4}G_{\mu\nu}^a G^{a\mu\nu}\cF_G& 
\cP_{D H}(h) = \left(\partial_\mu\cF_{D H}(h)\partial^\mu\cF'_{D H}(h)\right)^2\\ 
\cP_{1}(h) = \BBd \tr(\T \WWu) \cF_{1}&
\cP_{2}(h) = \dfrac{i}{4\pi}\BBd \tr(\T[\V^\mu,\V^\nu]) \cF_{2} \\
\cP_{3}(h)= \dfrac{i}{4\pi}\tr(\WWd [\V^\mu,\V^\nu]) \cF_{3}&
\cP_{4}(h) =\dfrac{i}{4\pi} \BBd \tr(\T\V^\mu) \de^\nu\cF_{4} \\
\cP_{5}(h)=\dfrac{i}{4\pi} \tr(\WWd\V^\mu) \de^\nu\cF_{5}&
\cP_{6}(h) =\dfrac{1}{(4\pi)^2} (\tr(\V_\mu\V^\mu))^2 \cF_{6} \\
\cP_{8}(h) =\dfrac{1}{(4\pi)^2}  \tr(\V_\mu\V_\nu) \de^\mu\cF_{8}\de^\nu\cF_{8}' &
\cP_{11}(h) =\dfrac{1}{(4\pi)^2}  (\tr(\V_\mu\V_\nu))^2 \cF_{11} \\
\cP_{12}(h) = (\tr(\T\WWd))^2 \cF_{12} &
\cP_{13}(h) = \dfrac{i}{4\pi} \tr(\T\WWd)\tr(\T[\V^\mu,\V^\nu])\cF_{13}  \\
\cP_{14}(h)  =\dfrac{\e^{\mu\nu\rho\lambda}}{4\pi}\tr(\T\V_\mu)\tr(\V_\nu W_{\rho\lambda}) \cF_{14} &
\cP_{17}(h) =\dfrac{i}{4\pi} \tr(\T \WWd) \tr(\T\V^\mu) \de^\nu\cF_{17} \\
\cP_{18}(h) = \dfrac{1}{(4\pi)^2}\tr(\T[\V_\mu,\V_\nu])\tr(\T\V^\mu)\de^\nu\cF_{18} &
\cP_{20}(h) = \dfrac{1}{(4\pi)^2}\tr(\V_\mu\V^\mu) \de_\nu\cF_{20}\de^\nu\cF_{20}' \\
\cP_{21}(h) = \dfrac{1}{(4\pi)^2}(\tr(\T\V_\mu))^2 \de_\nu\cF_{21}\de^\nu\cF_{21}'&
\cP_{22}(h) = \dfrac{1}{(4\pi)^2}\tr(\T\V_\mu)\tr(\T\V_\nu)\de^\mu\cF_{22}\de^\nu\cF_{22}' \\
\cP_{23}(h) =\dfrac{1}{(4\pi)^2} \tr(\V_\mu\V^\mu) (\tr(\T\V_\nu))^2 \cF_{23}&
\cP_{24}(h) = \dfrac{1}{(4\pi)^2}\tr(\V_\mu\V_\nu)\tr(\T\V^\mu)\tr(\T\V^\nu) \cF_{24} \\
\cP_{26}(h) =\dfrac{1}{(4\pi)^2} (\tr(\T\V_\mu)\tr(\T\V_\nu))^2 \cF_{26} \;. &
\end{tabular} 
\end{center}
As anticipated in the previous section, while the kinetic terms for the gauge bosons
are listed at the LO, the interactions obtained after introducing the dependence on $h$
are reported in the list of NLO operators, under the assumption that
the coupling of the transverse components of the gauge fields with the
Higgs sector is a subleading effect. 

It is also worth commenting on the operators $\cP_1(h)$ and $\cP_{12}(h)$:
these two structures, including the terms without $h$ insertions, are customarily listed among 
the NLO terms despite their similarity with the gauge-boson kinetic terms. 
This is justified, {\it a posteriori}, by the fact that they 
contribute to the $S$ and $U$ parameters respectively (see Sect.~\ref{SubSectConstraintsEWPD}),
which are strongly constrained. In this sense, their treatment is analogous to that of $\cP_T(h)$. 

The operators $\cP_C(h)$ and $\cP_H(h)$
of Ref.~\cite{Brivio:2013pma} have not been included in this list, as
their effects can be reabsorbed in redefinitions of the arbitrary
functions $\cF_C(h)$ and $\cY_{Q,L}(h)$ appearing in $\LLag_0$ in
Eq.~\eqref{Lag0} (see App.~\ref{APP:LOLag}). Moreover, compared to Ref.~\cite{Brivio:2013pma},
a different normalisation for the operators has been chosen: the $4\pi$ suppression factors
determined by the NDA master formula in Eq.~\eqref{MasterFormula} have
been made explicit (see Ref.~\cite{Gavela:2016bzc} for details of the
advantages of the NDA normalisation), while the dependence on the
coupling constants has been removed, in order to emphasise the
generality of the EFT approach.  It is customary, indeed, to include
in the definition of the HEFT operators the numerical factors arising
from the 1-loop renormalisation procedure 
(see Refs.~\cite{Biekoetter:2014jwa,Contino:2016jqw} for a general 
discussion in the SMEFT case): for instance, the operator
$\cP_1(h)$ is often defined proportionally to
$gg'/(4\pi)^2$~\cite{Longhitano:1980tm,Feruglio:1992wf,Alonso:2012px,Brivio:2013pma}.
However, in principle the coefficients $c_i$ account not only for
renormalisation effects, but also for possible external contributions,
originating by sources that do not need to share the same
dependence on the gauge couplings. This normalisation choice is common
in many EFTs, such as Fermi's theory, the EFT for mesons processes and the SMEFT.

\boldmath
\subsubsection{CP odd bosonic basis 
\texorpdfstring{$\Delta\LLag_\text{bos}^{\cancel{CP}}$}{Lbos-CPV}}
\unboldmath

In the CP-odd sector the bosonic Lagrangian contains 16 operators: 
according to Ref.~\cite{Gavela:2014vra},
\begin{equation}
\Delta\LLag_\text{bos}^{\cancel{CP}} = \sum_{j} \tilde{c}_j \cS_j, \qquad 
j= \{2D,\tilde{B},\tilde{W},\tilde{G},1-9,15,\widetilde{W}WW,\widetilde{G}GG\}\,,
\end{equation} 
where, as for $\Delta\LLag_\text{bos}^{CP}$, all the operators have 
four derivatives, with the exception of  
\beq
\cS_{2D}(h)\equiv i\,\frac{v^2}{4}\,\text{Tr}\left(\T\,\V_\mu\right)\,
\partial^\mu\cF_{2D}
\eeq
and
\beq
\begin{aligned}
\cS_{\widetilde{W}WW}(h) &=\dfrac{4\pi\varepsilon_{abc}}{\Lambda^2}\widetilde{W}_\mu^{a\nu}W_\nu^{b\rho} W_{\rho}^{c\mu}\cF_{\widetilde{W}WW}\,,\\
\cS_{\widetilde{G}GG}(h) &=\dfrac{4\pi f_{\alpha\beta\gamma}}{\Lambda^2}\widetilde{G}_\mu^{\alpha\nu}G_\nu^{\beta\rho} G_{\rho}^{\gamma\mu}\cF_{\widetilde{G}GG}\,.
\end{aligned}
\eeq
The rest of operators entering $\Delta\LLag_\text{bos}^{\cancel{CP}} $ are
\begin{center}
\begin{tabular}{>{$}l<{$}@{\hspace*{1cm}}>{$}l<{$}}
\cS_{\widetilde{B}}(h)\equiv -B^{\mu\nu}\, \widetilde{B}_{\mu\nu}\,\cF_{\widetilde{B}}&
\cS_{\widetilde{W}}(h)\equiv -\text{Tr}\left(W^{\mu\nu}\widetilde{W}_{\mu\nu}\right)\cF_{\widetilde{W}}\\
\cS_{\widetilde{G}}(h)\equiv -G^{a\mu\nu}\, \widetilde{G}^a_{\mu\nu}\,\cF_{\widetilde{G}} &
\cS_1(h)\equiv \widetilde{B}^{\mu\nu}\text{Tr}\left(\T W_{\mu\nu}\right)\,\cF_1\\
\cS_{2}(h)\equiv \dfrac{i}{4\pi}\,\widetilde{B}^{\mu\nu}\,\text{Tr}\left(\T\,\V_\mu\right)\,\partial_{\nu}\cF_{2} &
\cS_{3}(h)\equiv \dfrac{i}{4\pi}\,\text{Tr}\left(\widetilde{W}^{\mu\nu}\,\V_\mu\right)\,\partial_{\nu}\cF_{3}\\
\cS_{4}(h)\equiv \dfrac{1}{4\pi}\text{Tr}\left(W^{\mu\nu}\V_\mu\right)\text{Tr}\left(\T\,\V_\nu\right)\cF_{4} &
\cS_{5}(h)\equiv \dfrac{i}{(4\pi)^2}\text{Tr}\left(\V^\mu\,\V^\nu\right)\text{Tr}\left(\T\,\V_\mu\right)\partial_{\nu}\cF_{5}\\
\cS_{6}(h)\equiv \dfrac{i}{(4\pi)^2}\text{Tr}\left(\V^\mu\,\V_\mu\right)\text{Tr}\left(\T\,\V^\nu\right)\partial_{\nu}\cF_{6} &
\cS_{7}(h)\equiv \dfrac{1}{4\pi}\text{Tr}\left(\T\,\left[W^{\mu\nu},\V_\mu\right]\right)\,\partial_{\nu}\cF_{7}\\
\cS_{8}(h)\equiv \text{Tr}\left(\T\,\widetilde{W}^{\mu\nu}\right)\text{Tr}\left(\T\,W_{\mu\nu}\right)\cF_{8} &
\cS_{9}(h)\equiv \dfrac{i}{4\pi}\text{Tr}\left(\T\,\widetilde{W}^{\mu\nu}\right)\text{Tr}\left(\T\,\V_\mu\right)\partial_{\nu}\cF_{9}\\
\cS_{15}(h)\equiv \dfrac{i}{(4\pi)^2}\text{Tr}\left(\T\,\V^\mu\right)\,\left(\text{Tr}\left(\T\,\V^\nu\right)\right)^2\,\partial_{\mu}\cF_{15} \;.
\end{tabular}
\end{center}
As for the CP-even part of the bosonic basis, the explicit dependence
on the gauge couplings is not part of the definition of the operators,
while the $4\pi$ factors are reported according to
Eq.~(\ref{MasterFormula}).

The operator $\cS_{2D}(h)$ deserves a special remark. Being a two-derivative
operator, it would be naturally listed at the LO. However, restricting for simplicity
the discussion to the unitary gauge, $\cS_{2D}(h)$ introduces a
mixing between the gauge boson $Z$ and the physical $h$, that can be rotated away via a proper
redefinition of the Goldstone bosons' matrix, as detailed in
Ref.~\cite{Georgi:1986df,Gavela:2014vra}:
\begin{equation} \U\to \tilde{\U}\exp\left[-ia_{2D}\tilde{c}_{2D}
\frac{h}{v}\,\s_3\right]\,.
\end{equation} 
At leading order in the effective coefficients, the effects of this operator are 
eventually recast into CP-odd contributions to the Yukawa couplings
with arbitrary number of $h$ legs and to the vertices $Zh^n$, $n\geq2$. Furthermore, $\cS_{2D}(h)$ induces,
at 1-loop, corrections to the Higgs gauge-boson 
couplings that are bounded by the strong experimental limits on fermionic EDMs, 
as discussed in Ref.~\cite{Gavela:2014vra}. For this reason, it is considered as a NLO operator, similarly to $\cP_T(h)$.

Finally, the two operators $\cP_{\widetilde{W}WW}(h)$ and $\cP_{\widetilde{G}GG}(h)$ 
are the CP-odd counterparts of $\cP_{WWW}(h)$ and $\cP_{GGG}(h)$; comments similar to  
those given for the latter apply here too.

\boldmath
\subsection{NLO basis: fermionic sector 
\texorpdfstring{$\Delta\LLag_\text{fer}$}{Lfer}}
\unboldmath
\label{sec.Lfer}
The fermionic Lagrangian at NLO is constituted by single-current
operators with up to two derivatives and by four-fermion operators. 
Flavour indices are left implicit, unless necessary for the discussion.
This section presents a set of independent terms that completes the
NLO basis in the bosonic sector $\Delta\LLag_\text{bos}$: some redundant
structures have been removed using the EOMs, as
detailed in App.~\ref{APP:EOM}. Only baryon and lepton number conserving 
operators are considered (see Ref.~\cite{MSStoappear} for the baryon 
and lepton number violating basis). Moreover, as already 
stated in the previous sections,
right-handed neutrinos are not considered in the present
description. Their inclusion in the spectrum would require an
extension of the basis presented in this section, with the addition of
the operators in App.~\ref{APP:NR}.

The numbering of the functions $\cF_i(h)$ is dropped in the following
for brevity.  The Pauli matrices that act on the $SU(2)_L$ components
are denoted by $\s^i$, while the Gell-Mann matrices that contract
colour indices are indicated by $\lambda^A$. Whenever they are not
specified, the colour (uppercase) and isospin (lowercase) contractions
are understood to be diagonal. Flavour contractions are also assumed
to be diagonal. The tensor structure $\ssu$ entering the dipole
operators is defined as $\ssu=\frac{i}{2}[\g^\mu,\g^\nu]$.  Finally,
the mark \cancel{CP} on the left of an operator indicates that it is
intrinsically CP-odd.

\boldmath
\subsubsection{Single fermionic current 
\texorpdfstring{$\Delta\LLag_{2F}$}{L2F}}
\unboldmath
The operators with a single fermionic current and up to two
derivatives (including those in $\V_\mu$) are contained in the
Lagrangian
\begin{equation}
\begin{aligned}
   \Delta\LLag_{2F} =& \sum_{j=1}^{8} \tfqC_j\tfqName_j+ \sum_{j=9}^{28}
\frac{1}{\Lambda}\left(\tfqC_j+i
\tilde{n}^\mathcal{Q}_j\right)\tfqName_j +\sum_{j=29}^{36}
\frac{4\pi}{\Lambda}\left(\tfqC_j+i
\tilde{n}^\mathcal{Q}_j\right)\tfqName_j\\ 
&+ \sum_{j=1}^{2}
\tflC_j\tflName_j + \sum_{j=3}^{11} \frac{1}{\Lambda}\left(\tflC_j+ i
\tilde{n}^\ell_j\right) \tflName_j+ \sum_{j=12}^{14}
\frac{4\pi}{\Lambda}\left(\tflC_j+ i \tilde{n}^\ell_j\right) \tflName_j+\hc\,,
\end{aligned}
\end{equation} 
where we recall that the coefficients $\tfqC_j,\,\tflC_j,\,
\tilde{n}^\mathcal{Q}_j,\, \tilde{n}^\ell_j$ are real and smaller than
unity. \\ The terms with two derivatives have overall canonical mass
dimension 5 and are therefore suppressed by $\Lambda^{-1}$. Moreover,
they necessarily require chirality-flipping (scalar or tensor) Lorentz
structures. These structures do not have definite CP character, as the
scalar ($\bar{\psi}\psi$) and pseudo-scalar ($\bar{\psi}i\g_5\psi$)
contractions have opposite parity. As a consequence, each $SU(2)$
structure yields two contributions with opposite CP properties, that
have been parameterised by two independent real coefficients: for the
quark bilinears, the terms $\tfqC_j(\tfqName_j + \hc)$ with the
$\tfqName_j$'s defined below are CP even, while the combinations
${\tilde{n}^\mathcal{Q}_j(i\tfqName_j+\hc)}$ are CP odd. A similar
notation has been adopted for the lepton bilinears.

\subsubsection*{Quark Current Operators}

All the non-redundant terms that can be constructed coupling one
derivative or one chiral vector field $\V_\mu$ to a fermionic bilinear
necessarily have a vector-axial Lorentz structure, that preserves
chirality. For the quarks case, they are:
\begin{center}
\renewcommand{\arraystretch}{1.5}
 \begin{tabular}{>{\footnotesize}r>{$}l<{$}@{\hspace*{1cm}}>{\footnotesize}r>{$}l<{$}}
&\tfq(h)\equiv i\QBL \g_\mu \V^\mu \QL \cF			\label{Qll_v}&
&\tfq(h)\equiv i\QBR \g_\mu \U^\dag\V^\mu\U \QR \cF		\label{Qrr_v}\\
\cancel{CP}&\tfq(h)\equiv\QBL \g_\mu [\V^\mu,\T]\QL \cF		\label{Qll_cvt}&
\cancel{CP}&\tfq(h)\equiv \QBR \g_\mu \U^\dag[\V^\mu,\T]\U\QR \cF\label{Qrr_cvt}\\
&\tfq(h)\equiv i\QBL \g_\mu \{\V^\mu,\T\}\QL \cF		\label{Qll_avt}&
&\tfq(h)\equiv i\QBR \g_\mu \U^\dag\{\V^\mu,\T\}\U\QR \cF	\label{Qrr_avt}\\
&\tfq(h)\equiv i\QBL \g_\mu \T\V^\mu\T\QL \cF			\label{Qll_tvt}&
&\tfq(h)\equiv i\QBR \g_\mu \U^\dag\T\V^\mu\T\U\QR \cF\,. 	\label{Qrr_tvt}
\end{tabular}
\end{center}
Invariants with a derivative acting on a fermion field or on a
$\cF(h)$ function are redundant upon application of the EOMs
and integration by parts, and they have therefore been removed from the final basis.

Operators with two derivatives require a fermionic current with an
even number (zero or two) of gamma matrices: therefore only
chirality-flipping Lorentz structures are allowed.  All the operators
with a scalar structure are required as counter-terms in the 1-loop
renormalisation of $\LLag_0$:
{\renewcommand{\arraystretch}{1.5}
 \begin{longtable*}{>{$}l<{$}@{\hspace*{1cm}}>{$}l<{$}}
\tfq(h)\equiv  \QBL \U\QR{\de_\mu\cF\de^\mu\cF'} \label{Qlr_1_deFdeF}&
\tfq(h)\equiv  \QBL\T\U\QR{\de_\mu\cF\de^\mu\cF'}\label{Qlr_t_deFdeF}\\
\tfq(h)\equiv \QBL \V_\mu \U\QR{\de^\mu\cF}      \label{Qlr_v_deF}&
\tfq(h)\equiv \QBL\{\V_\mu,\T\}\U\QR{\de^\mu\cF} \label{Qlr_cvt_deF}\\
\tfq(h)\equiv \QBL[\V_\mu,\T]\U\QR{\de^\mu\cF}	 \label{Qlr_avt_deF}&
\tfq(h)\equiv \QBL\T\V_\mu\T\U\QR{\de^\mu\cF}    \label{Qlr_tvt_deF}\\
\tfq(h)\equiv \QBL \V_\mu\V^\mu \U\QR\cF	\label{Qlr_vv}&
\tfq(h)\equiv \QBL \V_\mu\V^\mu\T \U\QR\cF	\label{Qlr_vvt}\\
\tfq(h)\equiv \QBL \T\V_\mu\T\V^\mu \U\QR\cF	\label{Qlr_tvtv}&
\tfq(h)\equiv \QBL \T\V_\mu\T\V^\mu\T \U\QR\cF  \label{Qlr_tvtvt}\\
\tfq(h)\equiv \QBL \V_\mu\T\V^\mu \U\QR\cF	\label{Qlr_vtv}&
\tfq(h)\equiv \QBL \V_\mu\T\V^\mu\T \U\QR\cF\,.	\label{Qlr_vtvt}
\end{longtable*}}
Operators with tensor structure are also included in the NLO basis,
although they are not needed to reabsorb the 1-loop divergences of
$\LLag_0$, as the loop diagrams that generate them in the EFT are
finite. Nonetheless, these interactions may result from the
(tree-level) exchange of a heavy BSM resonance and therefore they 
may be as relevant as those in the previous lists:
\begin{center}
\renewcommand{\arraystretch}{1.5}
 \begin{tabular}{>{$}l<{$}@{\hspace*{1cm}}>{$}l<{$}}
\tfq(h)\equiv \QBL\ssu\V_\mu\U\QR{\de_\nu\cF}		\label{Qlr_s_v_deF}&
\tfq(h)\equiv \QBL\ssu[\V_\mu,\T]\U\QR{\de_\nu\cF}	\label{Qlr_s_cvt_deF}\\
\tfq(h)\equiv \QBL\ssu\{\V_\mu,\T\}\U\QR{\de_\nu\cF}	\label{Qlr_s_avt_deF}&
\tfq(h)\equiv \QBL\ssu\T\V_\mu\T\U\QR{\de_\nu\cF}	\label{Qlr_s_tvt_deF}\\
\tfq(h)\equiv \QBL\ssu\V_\mu\T\V_\nu\U\QR\cF		\label{Qlr_s_vtv}&
\tfq(h)\equiv \QBL \ssu\V_\mu\T\V_\nu\T\U\QR\cF		\label{Qlr_s_tvtv}\\
\tfq(h)\equiv \QBL \ssu[\V_\mu,\V_\nu]\U\QR\cF		\label{Qlr_s_vv}&
\tfq(h)\equiv \QBL \ssu[\V_\mu,\V_\nu]\T\U\QR\cF	\label{Qlr_s_vvt} \\
\tfq(h)\equiv ig'\,\QBL\ssu\U\QR \BBd\cF	\label{Qlr_s_b}&
\tfq(h)\equiv ig'\,\QBL\ssu\T\U\QR \BBd\cF	\label{Qlr_s_tb}\\
\tfq(h)\equiv ig_s\,\QBL\ssu\GGd\U\QR\cF 	\label{Qlr_s_g}&
\tfq(h)\equiv ig_s\,\QBL\ssu\GGd\T\U\QR\cF 	\label{Qlr_s_tg}\\
\tfq(h)\equiv ig\,\QBL\ssu\WWd\U\QR\cF		\label{Qlr_s_w}&
\tfq(h)\equiv ig\,\QBL\ssu\{\WWd,\T\}\U\QR\cF	\label{Qlr_s_atw}\\
\tfq(h)\equiv ig\,\QBL\ssu[\WWd,\T]\U\QR\cF	\label{Qlr_s_ctw}&
\tfq(h)\equiv ig\,\QBL\ssu\T\WWd\T\U\QR\cF\,.	\label{Qlr_s_twt}
\end{tabular}
\end{center}

\subsubsection*{Leptonic Current Operators}
Leptonic bilinears can be constructed along the same lines as the
quark ones. The absence of right-handed neutrinos, however, reduces
notably the number of independent invariants. Making use of 
Eq.~(\ref{EOMForLeptonicDominantOperators}), only two independent 
operators can be constructed with the insertion of a
single derivative or $\V_\mu$:
\begin{center}
\renewcommand{\arraystretch}{1.5}
 \begin{tabular}{>{\footnotesize}r>{$}l<{$}@{\hspace*{1cm}}>
{\footnotesize}r>{$}l<{$}}
\cancel{CP}&\tfl(h)\equiv\LBL \g_\mu [\V^\mu,\T]\LL \cF\label{Lll_cvt}&
&\tfl(h)\equiv i\LBR \g_\mu \U^\dag\{\V^\mu,\T\}\U\LR \cF\,.\label{Lrr_avt}\\
\end{tabular}
\end{center}
Notice that, if flavour effects are also taken into consideration, three other structures should be considered:
\beq
i\bar{L}_{Li}\g_\mu \V^\mu L_{Lj}\cF\,,\qquad\qquad
i\bar{L}_{Li}\g_\mu\{\T,\V^\mu\}L_{Lj}\cF\,,\qquad\qquad
i\bar{L}_{Li}\g_\mu \T\V^\mu\T L_{Lj}\cF\,.
\eeq
only for the case with $i\neq j$. Indeed, as shown in 
Eq.~(\ref{EOMForLeptonicDominantOperators}), the flavour-diagonal contractions 
do not represent independent terms as they are related via EOMs to 
bosonic operators that have been retained in the basis.

With two derivatives, two $\V_\mu$ or a combination of them, the following structures can be constructed:
{\renewcommand{\arraystretch}{1.5}
 \begin{longtable*}{>{$}l<{$}@{\hspace*{1cm}}>{$}l<{$}}
\tfl(h)\equiv  \LBL \U\LR{\de_\mu\cF\de^\mu\cF'}	\label{Llr_1_deFdeF}&
\tfl(h)\equiv \LBL\{\V_\mu,\T\}\U\LR{\de^\mu\cF}	 \label{Llr_cvt_deF}\\
\tfl(h)\equiv \LBL[\V_\mu,\T]\U\LR{\de^\mu\cF}		\label{Llr_avt_deF}&
\tfl(h)\equiv \LBL \V_\mu\V^\mu \U\LR\cF		\label{Llr_vv}\\
\tfl(h)\equiv \LBL \T\V_\mu\T\V^\mu \U\LR\cF		\label{Llr_tvtv}&
\tfl(h)\equiv \LBL\ssu[\V_\mu,\T]\U\LR{\de_\nu\cF}	\label{Llr_s_cvt_deF}\\
\tfl(h)\equiv \LBL\ssu\{\V_\mu,\T\}\U\LR{\de_\nu\cF}	\label{Llr_s_avt}&
\tfl(h)\equiv \LBL\ssu\V_\mu\T\V_\nu\U\LR\cF		\label{Llr_s_vtv}\\
\tfl(h)\equiv \LBL \ssu[\V_\mu,\V_\nu]\U\LR\cF		\label{Llr_s_vv}&
\tfl(h)\equiv ig'\,\LBL\ssu\U\LR \BBd\cF		\label{Llr_s_b}\\
\tfl(h)\equiv ig\,\LBL\ssu\WWd\U\LR\cF			\label{Llr_s_w}&
\tfl(h)\equiv ig\,\LBL\ssu[\WWd,\T]\U\LR\cF\,.		\label{Llr_s_cwt}
\end{longtable*}}
where, as explained above, all these operators are required as
counter-terms in the 1-loop renormalisation of $\LLag_0$ with the
exception of those with tensor structure, that correspond to finite
contributions. It is also worth recalling that all the
chirality-flipping structures listed here are CP even in the
combination ($\tflName_j+\hc$) but independent CP violating terms of
the form ($i\tflName_j+\hc$) should also be considered.

\boldmath
\subsubsection{Four-fermion operators \texorpdfstring{$\Delta\LLag_{4F}$}{L4F}}
\unboldmath
Four-fermion operators can be classified into four-quark,
four-lepton and two-quark-two-lepton sets. The overall Lagrangian reads
\begin{equation}
\begin{aligned}
 \Delta\LLag_{4F} = \frac{(4\pi)^2}{\Lambda^2}\Bigg[&
 \sum_{j=1}^{8} \left(\ffqC_j+i\tilde{r}^\mathcal{Q}_j\right)\ffqName_{j}+
 \sum_{j=9}^{26} \ffqC_j\ffqName_{j}+
 \left(\fflC_1+i\tilde{r}^\ell_1\right)\fflName_1+ \sum_{j=2}^7 
\fflC_j\fflName_j+\\
  &+ \sum_{j=1}^{6} \left(\ffqlC_j+i\tilde{r}^{\mathcal{Q}\ell}_j\right)
\ffqlName_j
  +\sum_{j=7}^{23} \ffqlC_j\ffqlName_j +\hc\Bigg]\,.
\end{aligned}
\end{equation}
Details of the construction and reduction of this subset of operators
can be found in App.~\ref{APP_4F}.  As for the bilinears case, the
chirality-flipping contractions
$(\bar{\psi}_L\psi_R)(\bar{\psi}_L\psi_R)$ listed here are CP even in
the combination ($R^f_j+\hc$) but independent CP violating terms of
the form ($iR^f_j+\hc$) should also be considered.

\subsubsection*{Pure Quark Operators}
The only four-quark operators required to remove divergences
originating at one-loop are the following:
\begin{center}
\renewcommand{\arraystretch}{1.5}
 \begin{tabular}{>{$}l<{$}@{\hspace*{0.5cm}}>{$}l<{$}}
\ffq(h)\equiv(\QBL\U\QR)(\QBL\U\QR)\cF		\label{Qlrlr_11}&	
\ffq(h)\equiv(\QBL\s^i\U\QR)(\QBL\s^i\U\QR)\cF \label{Qlrlr_ss}\\
\ffq(h)\equiv(\QBL\U\QR)(\QBL\T\U\QR)\cF	\label{Qlrlr_t}&
\ffq(h)\equiv(\QBL\T\U\QR)(\QBL\T\U\QR)\cF	\label{Qlrlr_tt}\\
\ffq(h)\equiv(\QBL\lambda^A\U\QR)(\QBL\lambda^A\U\QR)\cF\label{Qlrlr_ll}&
\ffq(h)\equiv(\QBL\lambda^A\s^i\U\QR)(\QBL\lambda^A\s^i\U\QR)\cF\label{Qlrlr_lsls}\\
\ffq(h)\equiv(\QBL\lambda^A\U\QR)(\QBL\lambda^A\T\U\QR)\cF\label{Qlrlr_llt}&
\ffq(h)\equiv(\QBL\lambda^A\T\U\QR)(\QBL\lambda^A\T\U\QR)\cF\,.\label{Qlrlr_ltlt}
\end{tabular}
\end{center}
A large number of additional structures can be constructed, that are
listed below and included in the basis. Although they do not
correspond to counter-terms in the renormalisation of $\LLag_0$, they
are potentially generated by the exchange of BSM resonances:\\
\begin{center}
\renewcommand{\arraystretch}{1.5}
\hspace*{-8mm}
\begin{tabular}{>{$}l<{$}@{\hspace*{0.5cm}}>{$}l<{$}}
\ffq(h)\equiv(\QBL\g_\mu\QL)(\QBL\g^\mu\QL)\cF\label{Qllll_11}& 
\ffq(h)\equiv(\QBL\g_\mu\QL)(\QBL\g^\mu\T\QL)\cF\label{Qllll_1t}\\
\ffq(h)\equiv(\QBL\g_\mu\T\QL)(\QBL\g^\mu\T\QL)\cF\label{Qllll_tt}&
\ffq(h)\equiv(\QBL\g_\mu\s^j\QL)(\QBL\g^\mu\s^j\QL)\cF\label{Qllll_ss}\\
\ffq(h)\equiv(\QBR\g_\mu\QR)(\QBR\g^\mu\QR)\cF\label{Qrrrr_11}&                      
\ffq(h)\equiv(\QBR\g_\mu\QR)(\QBR\g^\mu\U^\dag\T\U\QR)\cF\label{Qrrrr_1t}\\
\ffq(h)\equiv(\QBR\g_\mu\U^\dag\T\U\QR)(\QBR\g^\mu\U^\dag\T\U\QR)\cF\label{Qrrrr_tt}&
\ffq(h)\equiv(\QBR\g_\mu\s^j\QR)(\QBR\g^\mu\U^\dag\s^j\U\QR)\cF\label{Qrrrr_ss}\\
\ffq(h)\equiv(\QBL\g_\mu\QL)(\QBR\g^\mu\QR)\cF\label{Qllrr_11}&
\ffq(h)\equiv(\QBL\g_\mu\QL)(\QBR\g^\mu\U^\dag\T\U\QR)\cF\label{Qllrr_1t}\\
\ffq(h)\equiv(\QBL\g^\mu\T\QL)(\QBR\g_\mu\QR)\cF\label{Qllrr_t1} &
\ffq(h)\equiv(\QBL\g_\mu\T\QL)(\QBR\g^\mu\U^\dag\T\U\QR)\cF\label{Qllrr_tt}\\
\ffq(h)\equiv(\QBL\g_\mu\s^i\QL)(\QBR\g^\mu\U^\dag\s^i\U\QR)\cF \label{Qllrr_ss}& 
\ffq(h)\equiv(\QBL\g_\mu\lambda^A\QL)(\QBR\g^\mu\lambda^A\QR)\cF\label{Qllrr_ll}\\
\ffq(h)\equiv(\QBL\g_\mu\lambda^A\QL)(\QBR\g^\mu\lambda^A\U^\dag\T\U\QR)\cF\label{Qllrr_llt}&
\ffq(h)\equiv(\QBL\g^\mu\lambda^A\T\QL)(\QBR\g_\mu\lambda^A\QR)\cF\label{Qllrr_ltl}\\
\ffq(h)\equiv(\QBL\g_\mu\lambda^A\T\QL)(\QBR\g^\mu\lambda^A\U^\dag\T\U\QR)\cF\label{Qllrr_ltlt}&       
\ffq(h)\equiv(\QBL\g_\mu\lambda^A\s^i\QL)(\QBR\g^\mu\lambda^A\U^\dag\s^i\U\QR)\cF \,.\label{Qllrr_lsls}
\end{tabular}
\end{center}

\subsubsection*{Pure Leptonic Operators}
The set of independent four-lepton operators is considerably smaller
than that with four quarks, due to the absence of right-handed
neutrinos and of colour charges. Only one operator is required as a
1-loop counter-term:
\begin{equation*}
\ffl(h)\equiv(\LBL\U\LR)(\LBL\U\LR)\cF\,.	\label{Llrlr_11}
\end{equation*}
Six additional structures, that are not required as counter-terms, complete the list of possible invariants:
\begin{center}
\renewcommand{\arraystretch}{1.5}
 \begin{tabular}{>{$}l<{$}@{\hspace*{0.5cm}}>{$}l<{$}}
\ffl(h)\equiv(\LBL\g_\mu\LL)(\LBL\g^\mu\LL)\cF		\label{Lllll_11}&
\ffl(h)\equiv(\LBR\g_\mu\LR)(\LBR\g^\mu\LR)\cF 		\label{Lrrrr_11}\\
\ffl(h)\equiv(\LBL\g_\mu\LL)(\LBL\g^\mu\T\LL)\cF	\label{Lllll_1t}&
\ffl(h)\equiv(\LBL\g_\mu\T\LL)(\LBL\g^\mu\T\LL)\cF	\label{Lllll_tt}\\
\ffl(h)\equiv(\LBL\g_\mu\LL)(\LBR\g^\mu\LR)\cF		\label{Lllrr_11} &
\ffl(h)\equiv(\LBL\g^\mu\T\LL)(\LBR\g_\mu\LR)\cF 	\label{Lllrr_t1}\,.
\end{tabular}
\end{center}

\subsubsection*{Mixed Quark-Lepton Operators}
Finally, barring any $B$ or $L$ violation effects, mixed four-fermion
operators can only contain two quarks and two leptons in either of the
current structures $\bar{L}L\bar{Q}Q$ and $\bar{L}Q\bar{Q}L$.

Among the constructed invariants, the following are required to 
reabsorb 1-loop divergences:
\begin{center}
\renewcommand{\arraystretch}{1.5}
 \begin{tabular}{>{$}l<{$}@{\hspace*{1cm}}>{$}l<{$}}
\ffql(h)\equiv(\LBL\U\LR)(\QBL\U\QR)\cF			\label{QLlrlr_11}&
\ffql(h)\equiv(\LBL\U\QR)(\QBL\U\LR)\cF	 		\label{QLlrlrmix_11}\\
\ffql(h)\equiv(\LBL\U\LR)(\QBL\T\U\QR)\cF		\label{QLlrlr_t1}&
\ffql(h)\equiv(\LBL\T\U\QR)(\QBL\U\LR)\cF	 	\label{QLlrlrmix_1t}\\
\ffql(h)\equiv(\LBL\s^i\U\LR)(\QBL\s^i\U\QR)\cF    	 \label{QLlrlr_ss}  & 
\ffql(h)\equiv(\LBL\s^i\U\QR)(\QBL\s^i\U\LR)\cF\,,		\label{QLlrlrmix_ss}
 \end{tabular}
 \end{center}
while the remaining correspond to finite diagrams and are included 
for completeness:
\begin{center}
\renewcommand{\arraystretch}{1.5}
\begin{tabular}{>{$}l<{$}@{\hspace*{1cm}}>{$}l<{$}}
\ffql(h)\equiv(\LBL\g_\mu\LL)(\QBL\g^\mu\QL)\cF\label{QLl_11}&
\ffql(h)\equiv(\LBR\g_\mu\LR)(\QBR\g^\mu\QR)\cF\label{QLr_11}\\
\ffql(h)\equiv(\LBL\g_\mu\LL)(\QBL\g^\mu\T\QL)\cF\label{QLl_1t}	&
\ffql(h)\equiv(\LBR\g_\mu\LR)(\QBR\g^\mu\U^\dag\T\U\QR)\cF\label{QLr_1t}\\
\ffql(h)\equiv(\LBL\g_\mu\T\LL)(\QBL\g^\mu\QL)\cF\label{QLl_t1}&
\ffql(h)\equiv(\LBL\g_\mu\T\LL)(\QBL\g^\mu\T\QL)\cF\label{QLl_tt}\\
\ffql(h)\equiv(\LBL\g_\mu\s^i\LL)(\QBL\g^\mu\s^i\QL)\cF\label{QLl_ss}	&
\ffql(h)\equiv(\LBL\g_\mu\LL)(\QBR\g^\mu\QR)\cF\label{QLllrr_11}\\
\ffql(h)\equiv(\QBL\g_\mu\QL)(\LBR\g^\mu\LR)\cF\label{QLrrll_11}&
\ffql(h)\equiv(\LBL\g^\mu\T\LL)(\QBR\g_\mu\QR)\cF\label{QLllrr_t1}\\
\ffql(h)\equiv(\QBL\g_\mu\T\QL)(\LBR\g^\mu\LR)\cF\label{QLrrll_t1}&
\ffql(h)\equiv(\LBL\g_\mu\LL)(\QBR\g^\mu\U^\dag\T\U\QR)\cF\label{QLllrr_1t}\\
\ffql(h)\equiv(\LBL\g^\mu\T\LL)(\QBR\g_\mu\U^\dag\T\U\QR)\cF\label{QLllrr_tt}&
\ffql(h)\equiv(\LBL\g^\mu\s^j\LL)(\QBR\g_\mu\U^\dag\s^j\U\QR)\cF\label{QLllrr_ss}\\
\ffql(h)\equiv(\QBL\g_\mu\LL)(\LBR\g^\mu\QR)\cF\label{QLrrllmix_11}&
\ffql(h)\equiv(\QBL\g_\mu\T\LL)(\LBR\g^\mu\QR)\cF\label{QLrrllmix_t1}\\
\ffql(h)\equiv(\QBL\g^\mu\s^j\LL)(\LBR\g_\mu\U^\dag\s^j\U\QR)\cF\,.\label{QLrrllmix_ss}&
\end{tabular}
\end{center}

\subsection{Comparison with the SMEFT basis}

The comparison with the SMEFT is crucial for the identification of signals 
able to shed some light on the Higgs nature. 

For the bosonic sector, the relation between
the HEFT and its linear counterpart has already 
been identified in Ref.~\cite{Brivio:2013pma}, adopting the so-called 
HISZ basis~\cite{Hagiwara:1993ck,Hagiwara:1996kf}, which is also used in Refs.~\cite{Corbett:2012dm,Corbett:2012ja,Corbett:2013pja}.
Those results still hold here, up to the fact that some operators have been traded for fermionic ones:
the correspondence is summarised in Table~\ref{tab:Linear_siblings}, where the relation to the basis of Ref.~\cite{Grzadkowski:2010es} is also reported. The fermionic sector of the HEFT has also been matched with the linear bases of Refs.~\cite{Grzadkowski:2010es} and~\cite{Corbett:2012dm,Corbett:2012ja,Corbett:2013pja}, as indicated in Table~\ref{tab:Linear_siblings_fer}.

\begin{table}[t]\centering\footnotesize
\renewcommand{\arraystretch}{1.6}
\begin{tabular}{|*2{>{$}l<{$}>{$}l<{$}>{$}l<{$}|}}
\hline
\multicolumn{1}{|l}{Ref.~\cite{Grzadkowski:2010es}}&\text{Refs.~\cite{Corbett:2012ja,Corbett:2013pja}}&\text{HEFT}
&\multicolumn{1}{|l}{Ref.~\cite{Grzadkowski:2010es}}&\text{Refs.~\cite{Corbett:2012ja,Corbett:2013pja}}&\text{HEFT}\\
\hline
\cQ_\varphi & \cO_{\Phi,3}& \text{scalar pot.}&
\cQ_{\varphi\square}& \cO_{\Phi_2}& \cF_C+\cF_Y (\cP_H)\\
\cQ_{\varphi D}& \cO_{\Phi,1}& \cP_T&
\cQ_{\varphi G} &\cO_{GG}& \cP_G\\
\cQ_{\varphi W} &\cO_{WW}& \cP_W&
\cQ_{\varphi B} &\cO_{BB}& \cP_B\\
\cQ_{\varphi WB} &\cO_{BW}& \cP_1&
- &\cO_{B} & \cP_2+\cP_4\\
- &\cO_{W} & \cP_3+\cP_5&
&&\\
\cQ_{G}&``{\cQ_{G}}\text{''}& \cP_{GGG}&
\cQ_{W}&\cO_{WWW}& \cP_{WWW}\\
\cQ_{\varphi \widetilde{G}} &``{\cQ_{\varphi\widetilde{G}}}\text{''}& \cS_{\tilde{G}}&
\cQ_{\varphi \widetilde{B}} &``{\cQ_{\varphi\widetilde{B}}}\text{''} &\cS_{\tilde{B}}\\
\cQ_{\varphi \widetilde{W}} &``{\cQ_{\varphi\widetilde{W}}}\text{''}& \cS_{\tilde{W}}&
\cQ_{\varphi \widetilde{W}B} &``{\cQ_{\varphi\widetilde{W}B}}\text{''}& \cS_1\\
\cQ_{\widetilde G}&``{\cQ_{\widetilde G}}\text{''}&\cP_{\widetilde{G}GG}&
\cQ_{\widetilde W}&``{\cQ_{\widetilde W}}\text{''}&\cP_{\widetilde{W}WW}\\
\hline
\end{tabular}
\caption{\it Correspondence between the SMEFT operators from 
Refs.~\cite{Grzadkowski:2010es} and~\cite{Corbett:2012ja,Corbett:2013pja}, 
and the HEFT terms presented here for the bosonic sector. The - refers to the absence of an 
equivalent operator. The use of ``$\cQ_i$'' notation for the second column means 
that a particular operator does not explicitly appear in 
Refs.~\cite{Corbett:2012ja,Corbett:2013pja}, but that anyway enters the 
SMEFT basis and is defined as in Ref.~\cite{Grzadkowski:2010es}.
Numerical coefficients and signs in the combinations of the HEFT operators are not indicated.}
\label{tab:Linear_siblings}
\end{table}
 
It is worth pointing out a few points that should be kept into account when performing this comparison:
\begin{itemize}
\item In the HEFT, right-handed fermions are grouped in the $SU(2)_R$ doublets, $L_R$ and $Q_R$, and
the different components of each bilinear fermionic structure are disentangled inserting $\U^\dag\T\U=\s^3$ or $\U^\dag\s^j\U$. Each linear operator, written in the traditional notation, is then easily matched with 
a linear combination of HEFT invariants.
 
\item The adimensional scalar field $\T$ corresponds, in the linear context,
to a quadratic combination of Higgs doublets. As a consequence, the counterparts of fermionic invariants containing $\T$
are mostly linear operators of dimension $d>6$, that are therefore not present in the list of
Refs.~\cite{Grzadkowski:2010es},\cite{Corbett:2012ja,Corbett:2013pja}. 

The insertions of $\T$ into right-handed currents, mentioned in the previous point, represent an exception. In
fact, in these cases $\T$ appears in the combination ${\U^\dag \T\U=\s^3}$, that 
does not contain any field and in fact is not associated to dimensional objects in the linear language.

\item The adimensionality of $\T$ also leads to the presence of CP-odd operators in $\Delta\LLag$, whose corresponding structures in the SMEFT would appear only at $d>6$. An example is the operator $\Otfq{Qll_cvt}(h)$ that has been already studied in Ref.~\cite{Alonso:2012jc,Alonso:2012pz} for its impact on flavour physics.

\item The two-derivative object $\V_\mu \V^\mu$ is typically described, in the SMEFT,
by a quantity proportional to $D_\mu\Phi^\dag D^\mu\Phi$, which has canonical dimension 4.
Thus, fermionic bilinears containing this structure correspond to SMEFT operators with $d\geq 7$.
\end{itemize}

\begin{table}[h!]\centering\footnotesize
\renewcommand{\arraystretch}{1.6}
\begin{tabular}{|*2{>{$}l<{$}>{$}l<{$}>{$}l<{$}|}}
\hline
\multicolumn{1}{|l}{Ref.~\cite{Grzadkowski:2010es}}&\text{Refs.~\cite{Corbett:2012ja,Corbett:2013pja}}&\text{HEFT}
&\multicolumn{1}{|l}{Ref.~\cite{Grzadkowski:2010es}}&\text{Refs.~\cite{Corbett:2012ja,Corbett:2013pja}}&\text{HEFT}\\
\hline

\cQ_{\varphi u}&\cO_{u\Phi}& \cY_U(h)&
\cQ_{\varphi e}&\cO_{e\Phi}& \cY_E(h)\\
\cQ_{\varphi d}&\cO_{d\Phi}& \cY_D(h)& 
\cQ_{\varphi l,ii}^{(1)}&-& -\\
\cQ_{\varphi q}^{(1)}&\cO_{\Phi Q}^{(1)}& \Otfq{Qll_avt}&
\cQ_{\varphi l,ij}^{(1)}&\cO_{\Phi L,ij}^{(1)}&i\bar{L}_{L_i}\g_\mu\{\T,\V^\mu\}L_{L_j}\cF\\
\cQ_{\varphi q}^{(3)}&\cO_{\Phi Q}^{(3)}& \Otfq{Qll_v}& 
\cQ_{\varphi l,ii}^{(3)}&- &-\\
\cQ_{\varphi u}&\cO_{\Phi u}^{(1)}& \Otfq{Qrr_v}+\Otfq{Qrr_avt}+\Otfq{Qrr_tvt}&
\cQ_{\varphi l,ij}^{(3)}&\cO_{\Phi L,ij}^{(3)}&i\bar{L}_{L_i}\g_\mu \V^\mu L_{L_j}\cF\\
\cQ_{\varphi d}&\cO_{\Phi d}^{(1)}& \Otfq{Qrr_v}+\Otfq{Qrr_avt}+\Otfq{Qrr_tvt}& 
\cQ_{\varphi e}& \cO_{\Phi e}^{(1)}& \Otfl{Lrr_avt}\\
\cQ_{\varphi ud}&\cO_{\Phi ud}^{(1)}& \Otfq{Qrr_v}+\Otfq{Qrr_tvt}& & &\\
\cQ_{uG}& ``{\cQ_{uG}}\text{''} & \Otfq{Qlr_s_g}+\Otfq{Qlr_s_tg}& & &\\
\cQ_{dG}& ``{\cQ_{dG}}\text{''} & \Otfq{Qlr_s_g}+\Otfq{Qlr_s_tg}& & &\\
\cQ_{uW}& ``{\cQ_{uW}}\text{''} & \Otfq{Qlr_s_w}+\Otfq{Qlr_s_atw}+\Otfq{Qlr_s_ctw}& & &\\
\cQ_{dW}& ``{\cQ_{dW}}\text{''} & \Otfq{Qlr_s_w}+\Otfq{Qlr_s_atw}+\Otfq{Qlr_s_ctw}& 
\cQ_{eW}& ``{\cQ_{eW}}\text{''} & \Otfl{Llr_s_w}\\
\cQ_{uB}& ``{\cQ_{uB}}\text{''} & \Otfq{Qlr_s_b}+\Otfq{Qlr_s_tb}& & &\\
\cQ_{dB}& ``{\cQ_{dB}}\text{''} & \Otfq{Qlr_s_b}+\Otfq{Qlr_s_tb}&
\cQ_{eB}& ``{\cQ_{eB}}\text{''} & \Otfl{Llr_s_b}\\
\hline
\cQ_{qq}^{(1)}& ``{\cQ_{qq}^{(1)}}\text{''} & \Offq{Qllll_11}&
\cQ_{ll}& ``{\cQ_{ll}}\text{''} & \Offl{Lllll_11}\\
\cQ_{qq}^{(3)}& ``{\cQ_{qq}^{(3)}}\text{''} & \Offq{Qllll_ss}& 
\cQ_{lq}^{(1)}& ``{\cQ_{lq}^{(1)}}\text{''} & \Offql{QLl_11}\\
\cQ_{uu}& ``{\cQ_{uu}}\text{''} & \Offq{Qrrrr_11}+\Offq{Qrrrr_1t}+\Offq{Qrrrr_tt}& 
\cQ_{lq}^{(3)}& ``{\cQ_{lq}^{(3)}}\text{''} & \Offql{QLl_ss}\\
\cQ_{dd}& ``{\cQ_{dd}}\text{''} & \Offq{Qrrrr_11}+\Offq{Qrrrr_1t}+\Offq{Qrrrr_tt}&
\cQ_{ee}& ``{\cQ_{ee}}\text{''} & \Offl{Lrrrr_11}\\
\cQ_{ud}^{(1)}& ``{\cQ_{ud}^{(1)}}\text{''} & \Offq{Qrrrr_11}+\Offq{Qrrrr_tt}& 
\cQ_{eu}& ``{\cQ_{eu}}\text{''} & \Offql{QLr_11}+\Offql{QLr_1t}\\
\cQ_{ud}^{(8)}& ``{\cQ_{ud}^{(8)}}\text{''} & \Offq{Qrrrr_11}+\Offq{Qrrrr_ss} +\Offq{Qrrrr_tt}& 
\cQ_{ed}& ``{\cQ_{ed}}\text{''} & \Offql{QLr_11}+\Offql{QLr_1t}\\
\cQ_{qu}^{(1)}&``{\cQ_{qu}^{(1)}}\text{''} & \Offq{Qllrr_11}+\Offq{Qllrr_1t}& 
\cQ_{le}& ``{\cQ_{le}}\text{''} & \Offl{Lllrr_11}\\
\cQ_{qu}^{(8)}& ``{\cQ_{qu}^{(8)}}\text{''} & \Offq{Qllrr_ll}+\Offq{Qllrr_llt}& 
\cQ_{lu}& ``{\cQ_{lu}}\text{''} & \Offql{QLllrr_11}+\Offql{QLllrr_1t}\\
\cQ_{qd}^{(1)}& ``{\cQ_{qd}^{(1)}}\text{''} & \Offq{Qllrr_11}+\Offq{Qllrr_1t}& 
\cQ_{ld}& ``{\cQ_{ld}}\text{''} & \Offql{QLllrr_11}+\Offql{QLllrr_1t}\\
\cQ_{qd}^{(8)}& ``{\cQ_{qd}^{(8)}}\text{''} & \Offq{Qllrr_ll}+\Offq{Qllrr_llt}& 
\cQ_{qe}& ``{\cQ_{qe}}\text{''} & \Offql{QLrrll_11}\\
\cQ_{quqd}^{(1)}& ``{\cQ_{quqd}^{(1)}}\text{''} & \Offq{Qlrlr_11}+\Offq{Qlrlr_ss}&
\cQ_{ledq}& ``{\cQ_{lelq}}\text{''} & \Offql{QLrrllmix_11}+\Offql{QLrrllmix_t1}\\
\cQ_{quqd}^{(8)}& ``{\cQ_{quqd}^{(8)}}\text{''} & \Offq{Qlrlr_ll}+\Offq{Qlrlr_lsls}&
\cQ_{lequ}^{(1)}& ``{\cQ_{lequ}^{(1)}}\text{''} & \Offql{QLlrlrmix_11}+\Offql{QLlrlrmix_ss}\\
& &&
\cQ_{lequ}^{(3)}& ``{\cQ_{lequ}^{(3)}}\text{''} & \Offql{QLlrlr_11}+\Offql{QLlrlrmix_11}+\Offql{QLlrlr_t1}+\Offql{QLlrlr_ss}+\Offql{QLlrlrmix_ss}\\
\hline
\end{tabular}
\caption{\it Correspondence between the SMEFT operators from 
Refs.~\cite{Grzadkowski:2010es} and~\cite{Corbett:2012ja,Corbett:2013pja}, 
and the HEFT terms presented here for the fermionic sector. The - refers to the absence of an 
equivalent operator. The use of ``$\cQ_i$'' notation for the second column means 
that a particular operator does not explicitly appear in 
Refs.~\cite{Corbett:2012ja,Corbett:2013pja}, but that anyway enters the 
SMEFT basis and is defined as in Ref.~\cite{Grzadkowski:2010es}.
Flavour indices are omitted, unless 
explicitly indicated. Numerical coefficients and signs in the combinations 
of the HEFT operators are not indicated.}
\label{tab:Linear_siblings_fer}
\end{table}

Tables~\ref{tab:Linear_siblings} and~\ref{tab:Linear_siblings_fer} summarise the relations between operators of the HEFT,
defined in the previous section, and those of the SMEFT from Refs.~\cite{Grzadkowski:2010es} 
and~\cite{Corbett:2012ja,Corbett:2013pja}. The only difference between these two linear bases
(the first two columns in both tables) lies in the choice of two invariants:
in Refs.~\cite{Corbett:2012ja,Corbett:2013pja} the EOMs have been used for removing the fermionic terms corresponding to $\cQ_{\varphi l, ii}^{(1)}$ and $\cQ_{\varphi l, ii}^{(3)}$ in Ref.~\cite{Grzadkowski:2010es}, replacing them with the bosonic operators $\cO_B$ and $\cO_W$.
In the HEFT construction, the EOMs have been applied analogously to Refs.~\cite{Corbett:2012ja,Corbett:2013pja}, namely retaining $\cP_B$ and $\cP_W$, rather than two leptonic invariants (see Eq.~\eqref{EOMForLeptonicDominantOperators}).

All the HEFT operators that do not appear in this list have SMEFT 
counterparts (dubbed also ``linear siblings'') of dimension larger than six 
and therefore are not contained in the bases 
of Refs.~\cite{Grzadkowski:2010es} and~\cite{Corbett:2012ja,Corbett:2013pja}. 

\clearpage
\section{Phenomenology}\label{sec.pheno}

\subsection{Physical Parameters Definitions}
The phenomenological analysis is carried out in the
Z-scheme, defined by the following set of observables, 
that are taken as input parameters:
\beq
\begin{aligned}
\a_s&&&\text{world average~\cite{Agashe:2014kda},}\\
G_F&&&\text{extracted from the muon decay rate~\cite{Agashe:2014kda},}\\
\a_\text{em}&&&\text{extracted from Thomson scattering
~\cite{Agashe:2014kda},}\\
M_Z&&&\text{extracted from the $Z$ lineshape at LEP I
~\cite{Agashe:2014kda},}\\
M_h&&&\text{measured at LHC~\cite{Aad:2015zhl}.}
\end{aligned}
\label{inputs}
\eeq 
All the other quantities appearing in the Lagrangian will be
implicitly interpreted as corresponding to the combinations of 
experimental inputs as follows:
\beq
\begin{aligned}
 e^2 &= 4 \pi \a_\text{em}\,, \qquad
 &\sin^2\theta_W &= \dfrac{1}{2}\left
(1-\sqrt{1-\dfrac{4\pi\a_\text{em}}{\sqrt{2}G_F M_Z^2}}\right)\,,\\
 v^2 &= \dfrac{1}{\sqrt{2}G_F}\,,\qquad
 &\Big(g &= \dfrac{e}{\sin\theta_W}\,,\qquad g' = \left.
\dfrac{e}{\cos\theta_W}\,\Big)\right|
_{\theta_W,\,e \text{ as above}}\,.
\end{aligned}
\label{param}
\eeq
The trigonometric functions $\sin\theta_W$, $\cos\theta_W$ 
will be conveniently shortened to $\st$, $\ct$.\\
The kinetic terms are made canonical and diagonal with the following 
field redefinitions:
\begin{equation}\label{renorm_fields}
\begin{aligned}
 A_\mu &\to A_\mu \;\,\left[1+ \sdt c_1 + 2\st^2 c_{12}
-\frac{1}{2}(\ct^2 c_B+\st^2 c_W)\right] +\\
 &\qquad+ Z_\mu \,2\left[\cdt c_1+\sdt\left(c_{12}+\frac{c_B-c_W}{4}\right)
\right]+\mathcal{O}(c_i^2)\\
 Z_\mu &\to Z_\mu\;\,\left[1-\sdt c_1+2\ct^2c_{12}-\frac{1}{2}
\left(\ct^2c_W+ \st^2c_B\right)\right ]+\mathcal{O}(c_i^2)\\
 W^+_\mu &\to W^+_\mu\left[1-\frac{1}{2}c_W\right]+\mathcal{O}(c_i^2)\,.
\end{aligned}
\end{equation}
The contributions to the input parameters at first order 
in the effective coefficients read:
\begin{equation}\label{renorm_inputs}
\begin{aligned}
\frac{\d \a_\text{em}}{\a_\text{em}}&\simeq 2\sdt c_1 + 4\st^2 c_{12}-\ct^2 c_B-\st^2 c_W &
\frac{\d G_F}{G_F}&\simeq  -32\pi^2 \frac{v^2}{\Lambda^2}(\Cffl{Lllll_11}-\Cffl{Lllll_tt})\\
\frac{\d M_Z}{M_Z}&\simeq -c_T -\sdt c_1 +2\ct^2c_{12} -\frac{1}{2}(\ct^2c_W+\st^2 c_B)&
\frac{\d M_h}{M_h}&\simeq 0\,.
\end{aligned}
\end{equation}
The resulting shifts for the $W$ mass and fermion couplings to gauge bosons with respect to their corresponding SM expectations due to these finite renormalisation effects are summarised below:
\begin{description}
\item{\bf \boldmath$W$ mass:}
\begin{equation}\label{d_mW}
\frac{\Delta M_W}{M_W}= \frac{\ct^2}{\cdt}c_T+\frac{\sdt}{\cdt}c_1-2c_{12}
 +\frac{16\pi^2\st^2}{\cdt}\frac{v^2}{\Lambda^2}(\Cffl{Lllll_11}-\Cffl{Lllll_tt})\,.
\end{equation}
\item{\bf Fermionic couplings:}\\
It is convenient to adopt the following compact notation:
\begin{equation}
\label{delta_g_s}
  \begin{aligned}
   \Delta g_1 &=
c_T+16\pi^2\frac{v^2}{\Lambda^2}(\Cffl{Lllll_11}-\Cffl{Lllll_tt})\\
\Delta g_W &= 
\frac{\ct^2}{\cdt}c_T+t_{2\theta}c_1-2c_{12}+\frac{16\pi^2\ct^2}{\cdt}\frac{v^2}{\Lambda^2}(\Cffl{Lllll_11}-\Cffl{Lllll_tt})\\
\Delta g_2 &=
-\st^2\left(-\frac{\d\st^2}{\st^2}-\frac{\b}{t_\theta}\right)=\frac{\sdt^2}{2\cdt}\left(c_T+\frac{2c_1}{\sdt}+16\pi^2\frac{v^2}{\Lambda^2}(\Cffl{Lllll_11}-\Cffl{Lllll_tt})\right)\,,
  \end{aligned}
 \end{equation} 
where $\Delta g_1$ accounts for the renormalisation of $Z_\mu$, 
$g$ and $c_\theta$ in the combination $gZ_\mu /\ct$; $\Delta g_W$ for the renormalisation of $W_\mu$
and $g$ in the combination $gW_\mu$; $\Delta g_2$ for the renormalisation of $\st^2$
and for the contribution to the $Z$ couplings that comes from the
redefinition of the photon field: $A\to \a A +\b Z$ (see
Eq.~\eqref{renorm_fields}).  With this notation, the renormalisation
of  $Z$ couplings  to left-handed and right-handed fermions, 
$g_L^f=(T_3^f-s_\theta^2 Q^f)$  and $g_R^f= -s_\theta^2 Q^f$, and of
the  $W$ to left handed fermions can be written as
\begin{eqnarray}
\Delta g_{L,R}^f=g_{L,R}^f\Delta g_1+Q^f\Delta g_{2}
\qquad\qquad
\Delta g^{ff'}_{W}=\Delta g_W\,,
\end{eqnarray}
where $Q_f$ and $T_{3f}$ are, respectively. the electric and isospin
charges of the fermion $f$, and where the $W$ couplings to left-handed fermions  
is normalised to 1 in the SM.
\end{description}

The next sections are dedicated to the discussion of the constraints imposed on
the operator coefficients considering respectively electroweak precision data, Higgs results from the LHC and
the Tevatron, and measurements of the triple gauge-bosons couplings. For the sake of simplicity
we will assume fermion universality as well as the absence of new sources of flavour violation.

\subsection{Constraints from EWPD}
\label{SubSectConstraintsEWPD}
After accounting for finite renormalisation effects in the gauge bosons' 
wavefunctions and couplings as well as for direct contributions to
the vertices, 12 operators modify the
$Z$ and $W$ gauge boson couplings to fermions with the same Lorentz 
structure as the SM and the $W$ mass, which correspondingly 
lead to linear modifications of the EWPD. 

Five operators, $\cP_T(h)$, $\cP_1(h)$, $\cP_{12}(h)$, $\cR^\ell_2(h)$, $\cR^\ell_5(h)$
give tree level contributions to universal modifications 
of the couplings and of the $W$ mass, which can be recast in terms of the oblique
$S,T,U$ parameters \cite{Peskin:1990zt,Peskin:1991sw}
and of the shift in the Fermi constant $\Delta G_F$. In particular
\begin{equation}
\begin{array}{llll}
\alpha\,S=-8 \st\ct c_1\,, &\alpha\, T=2 c_T\, ,  & 
\alpha\, U=-16 \st^2 c_{12}\, , &
\frac{\delta G_F}{G_F}=-32\pi^2
\frac{v^2}{\Lambda^2}
\left(r^\ell_2-r^\ell_5\right)\, , 
\end{array}
\end{equation}
so, for example, the correction to the $W$ mass in Eq.~(\ref{d_mW}) reads
\begin{equation}
\frac{\Delta M_W}{M_W}=\frac{\ct^2}{2\cdt} \alpha\, T -\frac{1}{4\cdt} \alpha\, S
+\frac{1}{8\st^2}\alpha\, U -\frac{\st^2}{2\cdt} \frac{\delta G_F}{G_F}\; . 
\label{eq:dmw}
\end{equation}
The other seven operators, 
$\tfqName_1(h)$, $\tfqName_2(h)$, $\tfqName_5(h)$, $\tfqName_6(h)$, 
$\tfqName_7(h)$, $\tfqName_8(h)$, $\tflName_2(h)$,
give fermion dependent contributions to the $W$ and $Z$ couplings. Altogether
the shifts to the SM $Z$ couplings 
can be written as
\begin{equation}
\Delta g_{L,R}^f=g_{L,R}^f\Delta g_1+Q^f\Delta g_{2}+\Delta\tilde g^f_{L,R}\,,
\end{equation}
where the finite renormalisation shifts of the fermion couplings 
in Eq.~(\ref{delta_g_s}) can be rewritten as:
\begin{eqnarray}
\Delta g_1=\frac{1}{2}\left(\alpha\, T-\frac{\delta G_F}{G_F}\right) \,,
&&
\Delta g_2=\frac{\st^2}{\cdt}\left(\ct^2 \left(\alpha\, T-\frac{\delta G_F}{G_F}\right) -
\frac{1}{4\st^2}\alpha\, S\right)\,,
\label{eq:dg12}
\end{eqnarray}
while the fermion dependent modification of the couplings read\footnote{One 
could expect $\Delta \tilde g^{\nu,e}_{L}$ to have a similar contributions 
as $\Delta \tilde g^{u,d}_{L}$. This is not the case as the corresponding 
leptonic operators have been removed from the basis by using the EOMs, 
as discussed in Eq.~(\ref{EOMForLeptonicDominantOperators}). This choice
simplifies the renormalisation procedure as $\Delta \tilde g^{\nu,e}_{L}$ are 
vanishing.}
\begin{equation}
\begin{array}{lll}
\Delta \tilde g^u_{L}= n_1^\mathcal{Q}+2 n_5^\mathcal{Q}
+n_7^\mathcal{Q}\,, 
\qquad\qquad
&&
\Delta \tilde g^u_{R}= n_2^\mathcal{Q}+2 n_6^\mathcal{Q}
+n_8^\mathcal{Q}\,, 
\\
\Delta \tilde g^d_{L}= -n_1^\mathcal{Q}+2 n_5^\mathcal{Q}
-n_7^\mathcal{Q}\, ,
&&
\Delta \tilde g^d_{R}= -n_2^\mathcal{Q}+2 n_6^\mathcal{Q}
-n_8^\mathcal{Q}\,, \\
\Delta \tilde g^\nu_{L}=0\,,
&&
\Delta \tilde g^\nu_{R}= 0\,,\\
\Delta \tilde g^e_{L}=0\,,
&&
\Delta \tilde g^e_{R}=2 n^\ell_2 \,.
\end{array}
\end{equation}
The corresponding shifts to the $W$ couplings to left-handed  fermions  
(normalised to 1 in the SM) are
\begin{equation}
\Delta g_W^{ff'}=\Delta g_W+ \Delta\tilde g_W^{ff'}\,,
\end{equation}
with the universal shift due to the finite renormalisation 
defined in Eq.~(\ref{delta_g_s}) given by
\begin{equation}
\Delta g_W=\frac{\Delta M_W}{M_W}-\frac{1}{2}\frac{\delta G_F}{G_F}\,, 
\end{equation}
and the fermion dependent shifts induced by the fermionic operators by
\begin{eqnarray}
\Delta\tilde g_W^{ud}=
2 n_1^\mathcal{Q}-2 n_7^\mathcal{Q}\,, \qquad\qquad
& \Delta\tilde g_W^{e\nu}=0\,. 
\label{eq:dgw}
\end{eqnarray}

There are two main differences with respect to the corresponding contributions
to EWPD obtained assuming a linear realisation of the $SU(2)_L\times U(1)_Y$
gauge symmetry breaking with operators up to dimension six (see for
example Refs.~\cite{Pomarol:2013zra, Ciuchini:2014dea}). 
\begin{itemize}
\item First, in the SMEFT 
no contribution to the $U$ parameter is generated at
dimension six, while a contribution is generated in the HEFT at NLO, $\mathcal O(p^4)$.
\item  Second, in the linear description and
assuming universality, the fermion dependent shifts of the $W$
couplings to fermions are directly determined by those of the $Z$ as
there are only five independent dimension-six operators entering those
vertices with SM Lorentz structure 
(which can be chosen for example to
be $\mathcal O^{(3)}_{\phi q}$, $\mathcal O^{(1)}_{\phi q}$, $\mathcal
O_{\phi u}$, $\mathcal O_{\phi d}$, $\mathcal O_{\phi e}$ in the notation of
Ref.~\cite{Grzadkowski:2010es}). In the chiral description at order $p^4$
the fermion dependent contributions come in contrast from the seven
operators given above, of which six combinations contribute
independently to EWPD.
\end{itemize}

So altogether 10 combinations of the 12 operator coefficients can be 
determined by the analysis of  EWPD which have been chosen here to be
$c_T$, $c_1$, $c_{12}$, $(r^\ell_2-r^\ell_5)$, $n^\mathcal{Q}_1$, $(
n^\mathcal{Q}_2+n^\mathcal{Q}_8)$,  $n^\mathcal{Q}_5$, 
$n^\mathcal{Q}_6$, $n^\mathcal{Q}_7$ and $n^\ell_2$. 
In order to obtain the corresponding constraints on these 10 parameters
a fit including 16 experimental data points is performed. These are 
13 $Z$ observables: $\Gamma_Z$,
$\sigma_{h}^{0}$, $P_\tau^{\rm pol}$, 
$\sin^2\theta^\ell{\rm eff}$, $R^0_l$, $\mathcal{A}_{l}({\rm SLD})$, 
$A_{\rm  FB}^{0,l}$, $R^0_c$, $R^0_b$, $\mathcal{A}_{c}$,
$\mathcal{A}_{b}$, $A_{\rm  FB}^{0,c}$, and $A_{\rm  FB}^{0,b}$ from  
SLD/LEP-I~\cite{ALEPH:2005ab},  plus three $W$ observables: 
the average  of the $W$-boson mass, 
from~\cite{Group:2012gb}, the $W$ width, $\Gamma_W$, 
from LEP-II/Tevatron~\cite{ALEPH:2010aa}, and the leptonic
$W$ branching ratio, $Br_W^{e\nu}$, for which the average in 
Ref.~\cite{Agashe:2014kda} is taken.  
The correlations among the inputs can be 
found in Ref.~\cite{ALEPH:2005ab} and have been taken into 
consideration in the analysis.  As mentioned above, unlike in the fits to
dimension-6 SMEFT operators, the
independent experimental information on the $W$ couplings to fermions
have been included in the present study: this is done by considering in the fit
the leptonic $W$ branching ratio, as it is measured independently  
of the total $W$ width, which is determined from kinematic distributions. 
The corresponding predictions for the observables in the analysis 
in terms of the shifts of the SM couplings defined above 
are given by:
\begin{eqnarray}
&&\Delta \Gamma_Z=2 \Gamma_{Z,\rm SM} \left(
\frac{\displaystyle{\sum_f} (g_L^f\Delta g_L^f+g_R^f\Delta g_R^f)N_C^f}{
\displaystyle{\sum_f}(|g_L^f|^2+|g_R^f|^2)N_C^f}\right)\\
&&\Delta\sigma_h^0=
2\sigma_{h,\rm SM}^0\left(
\frac{(g_L^e\Delta g_L^e+g_R^e\Delta g_R^e)}{|g_L^e|^2+|g_R^e|^2}
+\frac{\displaystyle{\sum_q} (g_L^q\Delta g_L^q+g_R^q\Delta g_R^q)}
{\displaystyle{\sum_q}(|g_L^q|^2+|g_R^q|^2)}
-\frac{\Delta \Gamma_Z}
{\Gamma_{Z,\rm SM}}\right)\\
&&\Delta R_l^0\equiv
\Delta\left(\frac{\Gamma_Z^{\rm had}}{\Gamma_Z^{l}}\right)=
2 R_{l,\rm SM}^0
\left(\frac{\displaystyle{\sum_q} (g_L^q\Delta g_L^q+g_R^q\Delta g_R^q)}
{\displaystyle{\sum_q}(|g_L^q|^2+|g_R^q|^2)}
- \frac{(g_L^l\Delta g_L^l+g_R^l\Delta g_R^l)}{|g_L^l|^2+|g_R^l|^2}
\right)\\
&&\Delta R_q^0\equiv\Delta\left(\frac{\Gamma_Z^{q}}{\Gamma_Z^{\rm had}}\right)=
2R_{q,\rm SM}^0\left(
\frac{(g_L^q\Delta g_L^q+g_R^q\Delta g_R^q)}
{|g_L^q|^2+|g_R^q|^2}
- \frac{\displaystyle{\sum_{q'}} (g_L^{q'}\Delta g_L^{q'}+
g_R^{q'}\Delta g_R^{q'})}
{\displaystyle{\sum_{q'}}(|g_L^{q'}|^2+|g_R^{q'}|^2)}
\right)
\end{eqnarray}
\begin{eqnarray}
&&\Delta\sin^2\theta_{\rm eff}^l=\sin^2\theta_{\rm eff,SM}^l\;
\frac{g_L^l}{g_L^l-g_R^l}
\left(\frac{\Delta g_R^f}{g_R^f}-\frac{\Delta g_L^f}{g_L^f}\right)\\
&&\Delta\mathcal{A}_f=4 \mathcal A_{f,\rm SM}\;
\frac{g_L^fg_R^f}{|g_L^f|^4-|g_R^f|^4}
\left(g_R^f\Delta g_L^f-g_L^f\Delta g_R^f\right) \\
&&\Delta P_\tau^{\rm pol}=\Delta\mathcal{A}_l\\
&&\Delta A_{\rm  FB}^{0,f}=A_{\rm  FB,SM}^{0,f}\left(
\frac{\Delta\mathcal{A}_l}{\mathcal{A}_l}+
\frac{\Delta\mathcal{A}_f}{\mathcal{A}_f}\right) \\
&&\Delta \Gamma_W=\Gamma_{W,\rm SM}
\left(\frac{4}{3}\Delta g_W^{ud}+\frac{2}{3}\Delta g_W^{e\nu}+\Delta M_W\right)\\
&&\Delta Br_W^{e\nu}=Br_{W,\rm SM}^{e\nu}
\left(-\frac{4}{3}\Delta g_W^{ud}+\frac{4}{3}\Delta g_W^{e\nu}\right)\,.
\end{eqnarray}

\begin{figure}[h!]
\centering
\includegraphics[width=0.9\textwidth]{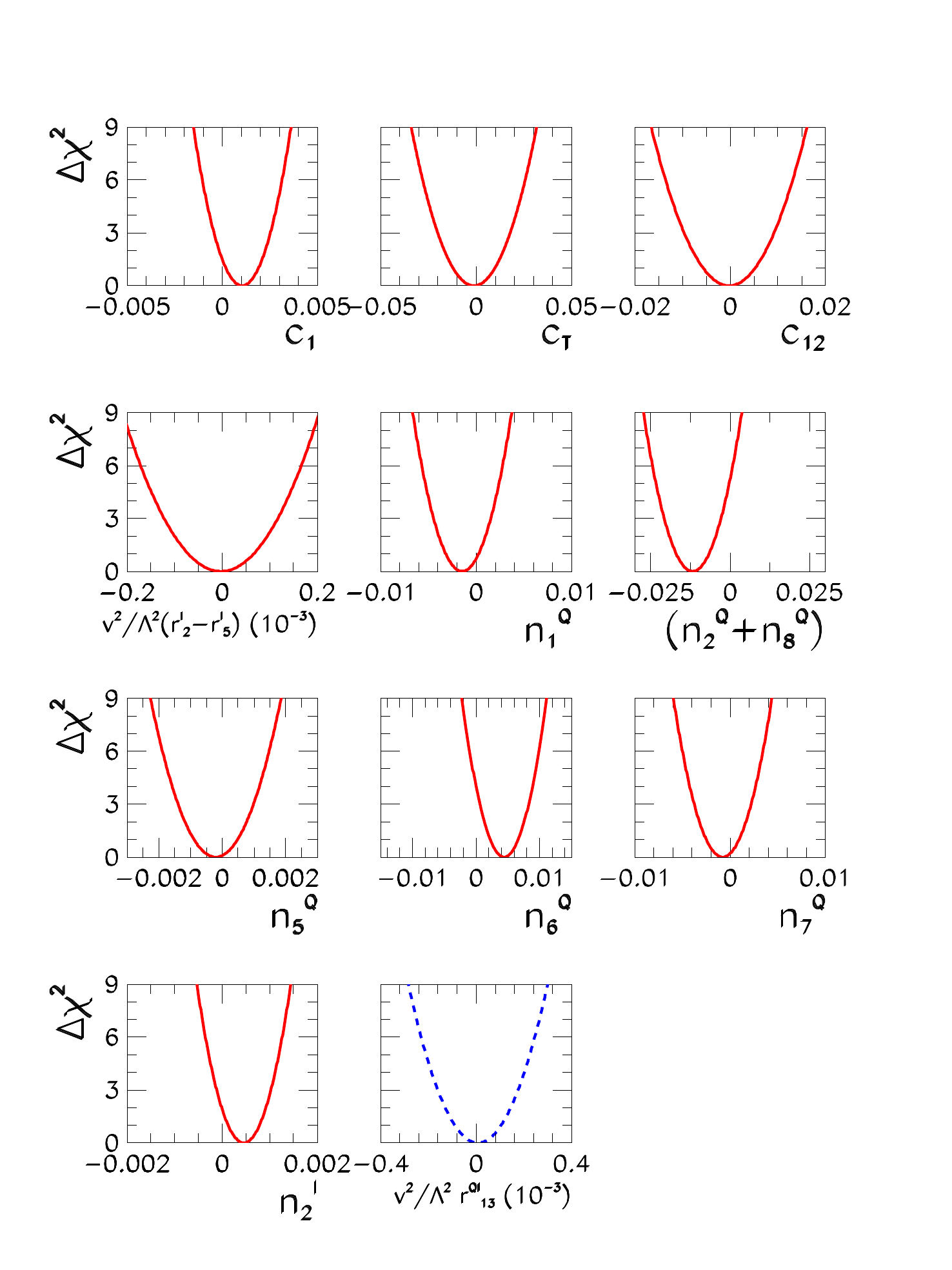}
\caption{\it Dependence of $\Delta\chi^2_{\rm EWPD+CKM}$ 
($=\Delta\chi^2_{\rm EWPD}$ for all, but last panel)   
on the 11 independent operator coefficients as labeled in 
the figure. In each panel $\Delta\chi^2_{\rm EWPD+CKM}$ is shown after 
marginalising over the other  undisplayed parameters.}
\label{fig:ewpd}
\end{figure}

When performing the fit within the context of the SM the result is 
$\chi^2_{\rm EWPD,SM}=18.3$, while when including the 10 new parameters
it gets reduced to $\chi^2_{\rm EWPD,min}=6$. 
The results of the analysis are shown in Fig.~\ref{fig:ewpd} which 
displays the  $\Delta\chi^2_{\rm EWPD}$  
dependence of the 10 independent operator coefficients. In each panel
$\Delta\chi^2_{\rm EWPD}$   is shown after marginalising over the other
nine coefficients.  The figure shows the corresponding 95\% allowed
ranges given in Table~\ref{tab:ewpd}: the only operator
coefficient not compatible with zero at $2\sigma$ is 
$n^\mathcal{Q}_2+n^\mathcal{Q}_8$, a result driven by the 2.7$\sigma$
discrepancy between the observed $A_{\rm  FB}^{0,b}$ and the SM expectation.

It is interesting to notice that the resulting constraints on
the coefficients contributing to $T$, $U$ and
$\delta G_F$ are considerably weaker than what one would obtain in 
the standard 3 parameter fits to
$S$, $T$, $U$. Quantitatively, the results of the 10 parameter analysis 
performed here give the following $1\sigma$ ranges for $S,T,U$ and $\delta G_F$:  
\begin{equation}
\begin{array}{llll}
S=-0.45\pm 0.37\,,  & T=-0.3\pm 2.8\,, & U=-0.1 \pm 2.5\,, & 
\frac{\delta G_F}{G_F}=(0.08\pm 2.2)\times 10^{-2}\,,
\end{array}
\label{STUGFtot}
\end{equation}
to be compared with the results of the standard 3 parameter fit for $S,T,U$ 
\cite{Ciuchini:2014dea},
\begin{eqnarray}
S=0.08\pm 0.1\,, &  T=-0.1\pm 0.12\,, & U=0.0 \pm 0.09\, \;.
\end{eqnarray}
While the range for $S$  is only about 4 times broader when including 
the effects of all  the additional operators, the bounds on $T$ and $U$ 
are weakened  by more than a factor 20.
The main reason is that when $\delta G_F$ is also included in the 
analysis cancellations can occur. 
In particular  as can be seen in Eq.~(\ref{eq:dmw})--(\ref{eq:dg12}) for
\begin{equation}
\alpha\,T=\frac{\delta G_F}{G_F}=-\frac{1}{4\st^2} \alpha U
\end{equation}
the contributions from $T$, $U$, and $\delta G_F$ cancel both in the $Z$ observables and in
$\Delta M_W$.Therefore, along this direction in the
parameter space, the bounds on these three quantities come from the
contribution of $\delta G_F$ to $\Gamma_W$ and $Br_W^{e\nu}$ in
Eq.~(\ref{eq:dgw}), but these observables are less precisely
determined.

It is important to notice that this ``weakening'' arises even 
if the $n_i^f$ coefficients, that is
all the fermion dependent contributions, but the four-fermionic ones,
are set to zero and only the 
four contributions $c_1$, $c_T$, $c_{12}$ and $r_2^\ell-r_5^\ell$ 
are retained. In this particular case, the result of the fit is
\begin{equation}
\begin{array}{llll}
S=-0.1\pm 0.1\,,  & T=0.43\pm 2.86\,, & U=-0.3 \pm 2.4\,, 
\,\,\,\frac{\delta G_F}{G_F}=(-0.26\pm 2.0) \,
\times 10^{-2},
\end{array}
\end{equation}
to be compared with Eq.~(\ref{STUGFtot}).
On the contrary, in the framework of linear dimension-6 operators, the condition
$U=0$ makes this cancellation not possible, so bounds on the corresponding
operator coefficients are generically stronger. 
In other words, when making the EWPD analysis in the context of HEFT at 
${\cal O}(p^4)$ the bounds on the operators contributing
to $T$ and $U$  are generically weaker by more than one order of magnitude. \\
 
\begin{table}[h!]\centering
\renewcommand{\arraystretch}{1.2}
\begin{tabular}{|c|c|}
\hline 
coupling & 95\% allowed range \\
\hline
$c_1$     &  $(-0.66 \,,\, 2.7)\times 10^{-3}$\\
$c_T$     &  $(-0.023\,,\, 0.021)$\\
$c_{12}$  &  $(-0.011\,,\,0.011)$	\\
$\frac{v^2}{\Lambda^2}
\left(r^\ell_2-r^\ell_5\right)$
& $(-1.4\,,\, 1.3)\times 10^{-4}$ \\	
$n^\mathcal{Q}_1$ &$(-4.9\,,\,2.0)\times 10^{-3}$\\ 
$n^\mathcal{Q}_2+n^\mathcal{Q}_8$& 
$(-22 \,,\,-1.5)\times 10^{-3}$ \\
$n^\mathcal{Q}_5$ &$(-1.6\,,\,	1.2)\times 10^{-3}$\\ 
$n^\mathcal{Q}_6$ & $(-0.025\,,\,	8.8)\times 10^{-3}$ \\
$n^\mathcal{Q}_7$ &$(-4.2 \,,\,	2.7)\times 10^{-3}$ \\
$n^\ell_2$ &  $(-0.2\,,\,1.1)\times 10^{-3}$ \\
$\frac{v^2}{\Lambda^2}
 r^{\mathcal{Q}\ell}_{13}$
&  $(-2.0\,,\,1.9)\times 10^{-4}$ \\
\hline
\end{tabular}
\caption{\it 95\% allowed ranges for the  combinations of operator
coefficients entering the EWPD analysis and  the CKM unitarity test.}
\label{tab:ewpd}
\end{table}

The fermionic operators can also lead to modifications of the 
semileptonic decay amplitudes used to determine the elements of the 
CKM matrix and to test its unitarity.
In particular,
$\tfqName_1(h)$, $\tfqName_7(h)$, $\cR^\ell_2(h)$, $\cR^\ell_5(h)$,
$\cR^{\mathcal{Q}\ell}_{13}(h)$ induce linear shifts to the corresponding 
amplitudes (normalised to $G_F$ as determined from $\mu$ decay)  
which can be parameterised as a shift
in the effective CKM matrix,
\begin{equation}
\Delta {V_{\rm CKM}}_{ij} = {V_{\rm CKM,SM}}_{ij}
\left(-32\pi^2 
\frac{v^2}{\Lambda^2}
r^{\mathcal{Q}\ell}_{13}+\Delta\tilde g_W^{ud} -\frac{\delta G_F}{G_F}\right)\,,
\label{eq:ckm}
\end{equation}
and which can lead to violations of unitarity of the CKM matrix which
are strongly constrained. In the case of SMEFT with operators up 
to dimension six,  
three operators enter this observable after equivalent application of the EOMs
~\cite{Pomarol:2013zra, Ciuchini:2014dea} 
 (which can be chosen for example to be $\mathcal O^{(3)}_{\phi q}$, 
$\mathcal O_{ll}$, and, $\mathcal O^{(3)}_{l q}$
Ref.~\cite{Grzadkowski:2010es}). From the global analysis in 
Ref.\cite{Agashe:2014kda}  
\begin{equation}
\sum_i|V_{ui}|^2-1=2 \left(-32\pi^2 
\frac{v^2}{\Lambda^2}
r^{\mathcal{Q}\ell}_{13}+\Delta\tilde g_W^{ud} -\frac{\delta G_F}{G_F}\right)
=(-1\pm 6)\times 10^{-4}\, .
\end{equation}
In combination with the analysis of the EWPD, this allows for constraining 
the coefficient of an 11th operator $\cR^{\mathcal{Q}\ell}_{13}(h)$.  
Adding this data point to the 16 of the EWPD allows one to construct
 $\chi^2_{\rm EWPD+CKM}$, which is now a function of 11 parameters
(with ${\chi^2_{\rm EWPD+CKM,SM}=18.4}$ and 
${\chi^2_{\rm EWPD+CKM,min}=6}$). 
The marginalised distributions verify 
${\Delta \chi^2_{\rm EWPD+CKM}(x)=\Delta \chi^2_{\rm EWPD}(x)}$
for the first 10 parameters, i.e. the inclusion of the CKM unitarity
constraint has no impact in the previous analysis as long as 
$r^{\mathcal{Q}\ell}_{13}$ is allowed to vary free in the fit. 
The new $\Delta \chi^2_{\rm EWPD+CKM}(r^{\mathcal{Q}\ell}_{13})$
is shown in the curve in the last panel in Fig.~\ref{fig:ewpd} and its
95\% CL range is listed in the last row in Table~\ref{tab:ewpd}.

\subsection{Effects in Higgs Physics}

This section is dedicated to the study of the current bounds stemming
from the Higgs searches at the LHC. 
Restricting the analysis to the subset of $C$ and $P$--even
operators\footnote{The extension of the analysis to $CP$--odd non
linear operators could be performed after the inclusion of $CP$
sensitive observables, see Ref.~\cite{Gavela:2014vra}.}, the focus is on
those terms that contribute to the trilinear Higgs
interactions with fermions and gauge bosons (deviations in the
Higgs triple vertex will only become observable in the future).
The deviations on Higgs quartic vertices ($HVf\bar{f^\prime}$) generated
by some of the single fermionic current operators have been omitted
  from this analysis. Those contributions to Higgs physics could also be studied at the
  LHC~\cite{Isidori:2013cla,Isidori:2013cga,Gonzalez-Alonso:2015bha} and, if analysed in combination
  with gauge-fermion data, they would potentially improve
  the comparison between linear and non-linear scenarios~\cite{Isidori:2013cga,Gonzalez-Alonso:2015bha}.
  Nevertheless the generalisation
  of the analysis with the inclusion of these effects is out of the scope of the present study.
The list of operators analysed includes then $\cP_T(h)$, $\cP_{B,G,W}(h)$ and
$\cP_{1,4,5,12,17}(h)$, in addition to the contributions from
$Y_U^{(1)}$, $Y_D^{(1)}$, $Y_\ell^{(1)}$ and to the deviations in the GBs
kinetic term parameterised by $\Delta a_C$. 
This set can be further reduced considering the strong
constraints imposed on $\cP_{T,1,12}(h)$ by the
global analysis of EWPD at the $Z$ pole: 
the impact of these operators on Higgs physics can be safely neglected,
given the accuracy at which these observables are currently measured.
Moreover, the current Higgs 
searches are only sensitive to $Hff$ vertices with $f=t,\,b,\,\tau$ 
(the addition of $\mu$ to the analysis will be straightforward
once the sensitivity to this coupling increases). Therefore, only a subset of 10 operators 
is relevant for the analysis of the available Higgs data.
Their contributions to the several Higgs trilinear interactions can be
illustrated with the usual HVV phenomenological Lagrangian in the unitary gauge:
\begin{eqnarray}
{\cal L} 
 & = & g_{Hgg} \; H G^a_{\mu\nu} G^{a\mu\nu} 
+  g_{H \gamma \gamma} \; H A_{\mu \nu} A^{\mu \nu} 
+ g^{(1)}_{H Z \gamma} \; A_{\mu \nu} Z^{\mu} \partial^{\nu} H
+  g^{(2)}_{H Z \gamma} \; H A_{\mu \nu} Z^{\mu \nu} \notag \\
& & + g^{(1)}_{H Z Z}  \; Z_{\mu \nu} Z^{\mu} \partial^{\nu} H 
+  g^{(2)}_{H Z Z}  \; H Z_{\mu \nu} Z^{\mu \nu} 
+  g^{(3)}_{H Z Z}  \; H Z_\mu Z^\mu \notag \\
& & + g^{(1)}_{H W W}  \; \left (W^+_{\mu \nu} W^{- \, \mu} \partial^{\nu} H 
                            +\text{h.c.} \right) 
+  g^{(2)}_{H W W}  \; H W^+_{\mu \nu} W^{- \, \mu \nu} 
+  g^{(3)}_{H W W}  \; H W^+_{\mu} W^{- \, \mu}\notag\\
& &+ \sum_{f=\tau,b,t} \left( g_f H \bar f_{L} f_{R} + \text{h.c.} \right)
\,.
\label{eq:higgsphenolag}
\end{eqnarray}
The 13 parameters in this Lagrangian can be re-written in terms of the following ten
coefficients\footnote{Notice the implicit redefinitions $a_i\equiv  c_ia_i$ for the bosonic operators.}:
\begin{equation}
 \Delta a_C,\; a_B,\; a_G,\; a_W,\; a_4,\; a_5,\; a_{17},\; Y^{(1)}_t,\; Y^{(1)}_b,\; Y^{(1)}_\tau\,,
 \label{eq:higgsparameters}
\end{equation}
and explicitly they read
\beq
\begin{gathered}
g_{Hgg}=-\frac{1}{2 v} a_G\,,\qquad
g^{(1)}_{H Z \gamma}= -\frac{g \st}{4\pi v \ct}\left(a_5+2\frac{c_\theta}{s_\theta}a_4+2a_{17}\right)\,,\qquad
 g^{(2)}_{H Z \gamma}= \frac{\st \ct}{v}\left(a_B-a_W\right)\,, \\
g^{(1)}_{H Z Z}=\frac{g}{4\pi v}\left(2\frac{\st}{\ct} a_4 - a_5 - 2 a_{17}\right)\,,\qquad \qquad
g^{(2)}_{H Z Z}= -\frac{1}{2v}\left(\st^2 a_B + \ct^2a_W\right)\,,\\
g^{(3)}_{H Z Z}= M_Z^2\left(\sqrt{2}G_F\right)^{1/2}\left(1+\Delta a_C\right)\,,\qquad\qquad
g_{H \gamma \gamma}= -\frac{1}{2v}\left(\st^2a_W+\ct^2 a_B\right)\,,\\
g^{(1)}_{H W W}= -\frac{g}{4\pi v} a_5\,,\qquad
g^{(2)}_{H W W}= \frac{1}{v} a_W\,,\qquad
g^{(3)}_{H W W}= 2 M_W^2\left(\sqrt{2}G_F\right)^{1/2}\left(1+\Delta a_C\right)\,,\\
g_f = -\frac{Y_f^{(1)}}{\sqrt{2}}\,.
\end{gathered}
\label{eq:higgsphenolag2}
\eeq
The anomalous Higgs interactions described 
by these 10 operators can be studied and constrained in a model
independent way by means of a global analysis of all the Higgs
experimental measurements that were performed at the LHC during the
Run~I. This includes not only event rate data in several
Higgs production and decay categories, but also some kinematic 
distributions, that have an interesting phenomenological impact, as shown 
in the context of SMEFT in Ref.~\cite{Corbett:2015ksa,Masso:2012eq,Banerjee:2013apa,Ellis:2014dva,Ellis:2014jta,Edezhath:2015lga}.
Indeed, they are important for allowing to obtain finite constraints in the large-dimensional 
parameter space spanned in the global analysis~\cite{Corbett:2015ksa}. 
Moreover, they make it possible to disentangle the non-SM Lorentz
structures from the SM-like shifts.  

The global analysis of all Run I Higgs,
data using the \textsc{SFitter} framework~\cite{Lafaye:2009vr,Klute:2012pu,
Plehn:2012iz,Klute:2013cx,Lopez-Val:2013yba}
for the SMEFT~\cite{Corbett:2012ja,Corbett:2013pja}, has been presented in Ref.~\cite{Corbett:2015ksa}: 
in that case, 
the 13 parameters of the phenomenological Lagrangian in 
Eq.~(\ref{eq:higgsphenolag}) received contributions from 9 linear operators. 
Here, that analysis is extended to account for the 10th  coefficient~$a_{17}$. 
All the details regarding the data set and the kinematic
distributions analysed, as well as the statistical treatment performed
in this log-likelihood analysis follow exactly the description
presented in Ref.~\cite{Corbett:2015ksa} and will not be repeated here.

\begin{table}[htb!]
\renewcommand{\arraystretch}{1.5}
\small
\begin{tabular}{|c|c|c|}
\cline{2-3} 
\multicolumn{1}{c|}{} &
\multicolumn{1}{c|}{Best fit} & 
\multicolumn{1}{c|}{95\% CL region} \\ [1mm]
\hline
\multicolumn{1}{|c|}{$a_G$} & \parbox[t][2.2cm][t]{2cm}{\centering -0.0125\\[2mm]
						      -0.0030 \\
						      0.0029\\[2mm]
						      0.0123}
 &  \parbox[t][2.2cm][t]{3cm}{\centering
  $(-0.018,-0.0080)$ \\[4mm]
  $(-0.0054,0.0058)$ \\[4mm]
  $(0.0091,0.017)$ }\\
 \hline
\multicolumn{1}{|c|}{$a_W$} & -0.017 & $(-0.11,0.088)$ \\\hline
\multicolumn{1}{|c|}{$a_B$} & 0.0052 & $(-0.025,0.041)$ \\\hline
\multicolumn{1}{|c|}{$a_4$} & 0.041 & $(-0.85,1.1)$ \\\hline
\multicolumn{1}{|c|}{$a_5$} & 0.13 & $(-0.81,0.60)$ \\\hline
\multicolumn{1}{|c|}{$\Delta a_C$} & -0.13 & $(-0.30,0.23)$ \\\hline
\multicolumn{1}{|c|}{$a_{17}$} & 0.055 & $(-0.52,0.65)$ \\\hline
$Y_t^{(1)}/Y_t^{\text{(0)}}$ & -1.11 & $(-1.7,-0.53)$\\
\multicolumn{1}{|c|}{} & 1.31 & $(0.56,1.7)$\\\hline
$Y_b^{(1)}/Y_b^{\text{(0)}}$ & -0.70 & $(-1.7,-0.39)$\\
\multicolumn{1}{|c|}{} & 0.66 & $(0.35,1,7)$\\\hline
$Y_\tau^{(1)}/Y_\tau^{\text{(0)}}$ & -0.94 & $(-1.37,-0.63)$\\
\multicolumn{1}{|c|}{} & 0.82 & $(0.66,1.47)$\\
\hline
\hline
$c_2$ & 0.041 & $(-0.24,0.27)$ \\\hline
$c_3$ & 0.15 & $(-0.093,0.39)$\\\hline
$c_{WWW}$ & 0.006 & $(-0.013,0.018)$ \\\hline
\end{tabular}
\includegraphics[width=0.44\textwidth,bb=0 235 235 600 ]{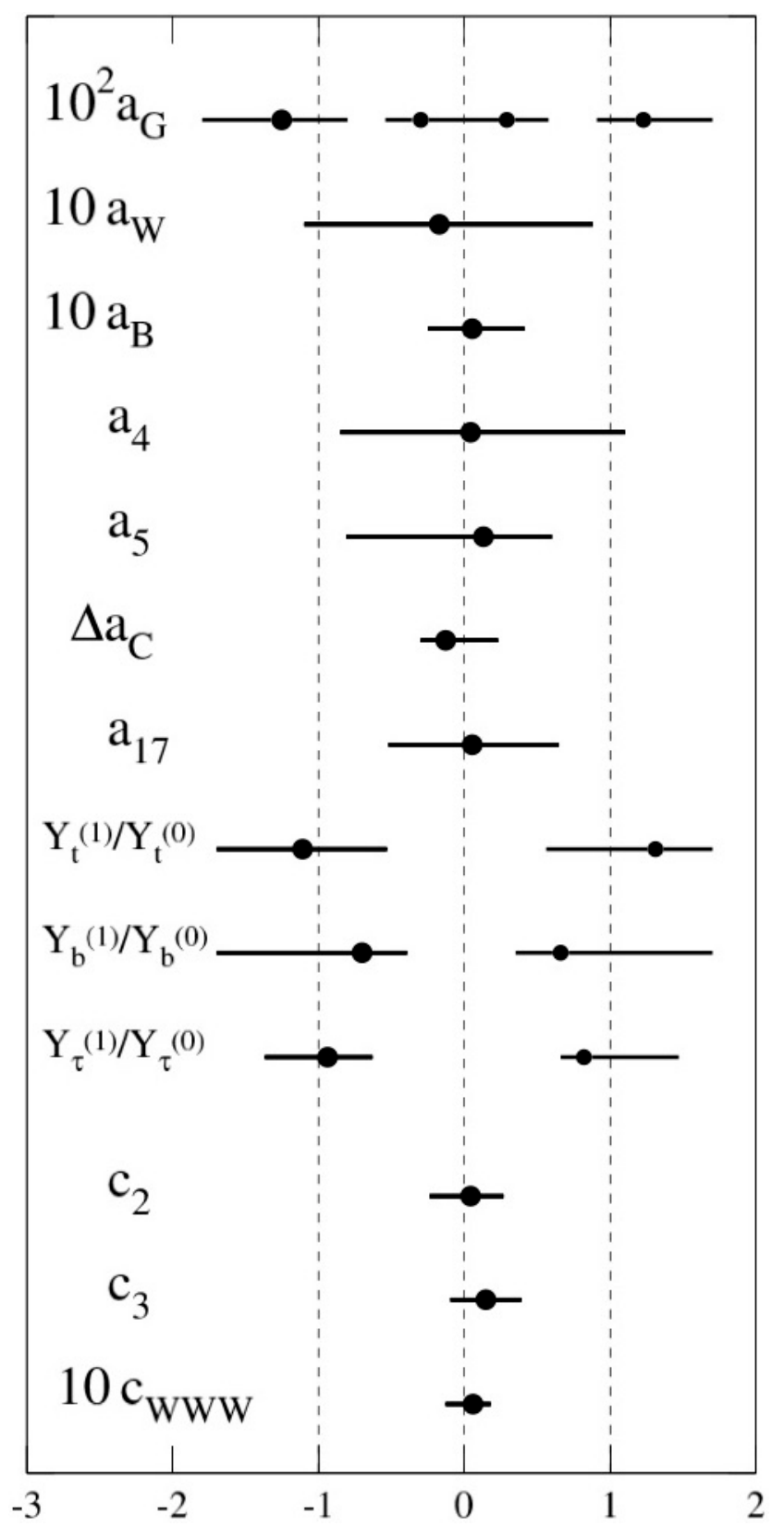}
\caption{\it Best fit and 95\% C.L. allowed ranges of the coefficients of
  the operators contributing to Higgs data ($a_G$, $a_W$, $a_B$,
  $a_4$, $a_5$, $a_{17}$, $\Delta a_C$, $Y_t^{(1)}$, $Y_b^{(1)}$ and
  $Y_\tau^{(1)}$) and to TGV analyses ($c_2$, $c_3$ and $c_{WWW}$).
  $Y_t^{(1)}$, $Y_b^{(1)}$ and $Y_\tau^{(1)}$ are normalised 
  to the SM expectation.}
\label{tab:higgstgvresults}
\end{table}

The results of the global analysis on the parameters in
Eq.~(\ref{eq:higgsparameters}) using the available Higgs data,
including all the kinematic distributions described in
Ref.~\cite{Corbett:2015ksa}, are reported in
Table~\ref{tab:higgstgvresults}. On the right figure we
graphically display the corresponding values where error bars refer to
the 95\% C.L. allowed ranges, obtained profiling for each
coefficient on the other 9 parameters that are included in the global
analysis. The off-shell $m_{4\ell}$ distributions, which have been
implemented in Ref.~\cite{Corbett:2015ksa}, are not included here, as
their impact in the present analysis is subdominant with respect to
the rest of kinematic distributions considered.

\begin{figure}[t]
  \centering
  \hspace*{-5mm}
  \includegraphics[width=0.52\textwidth]{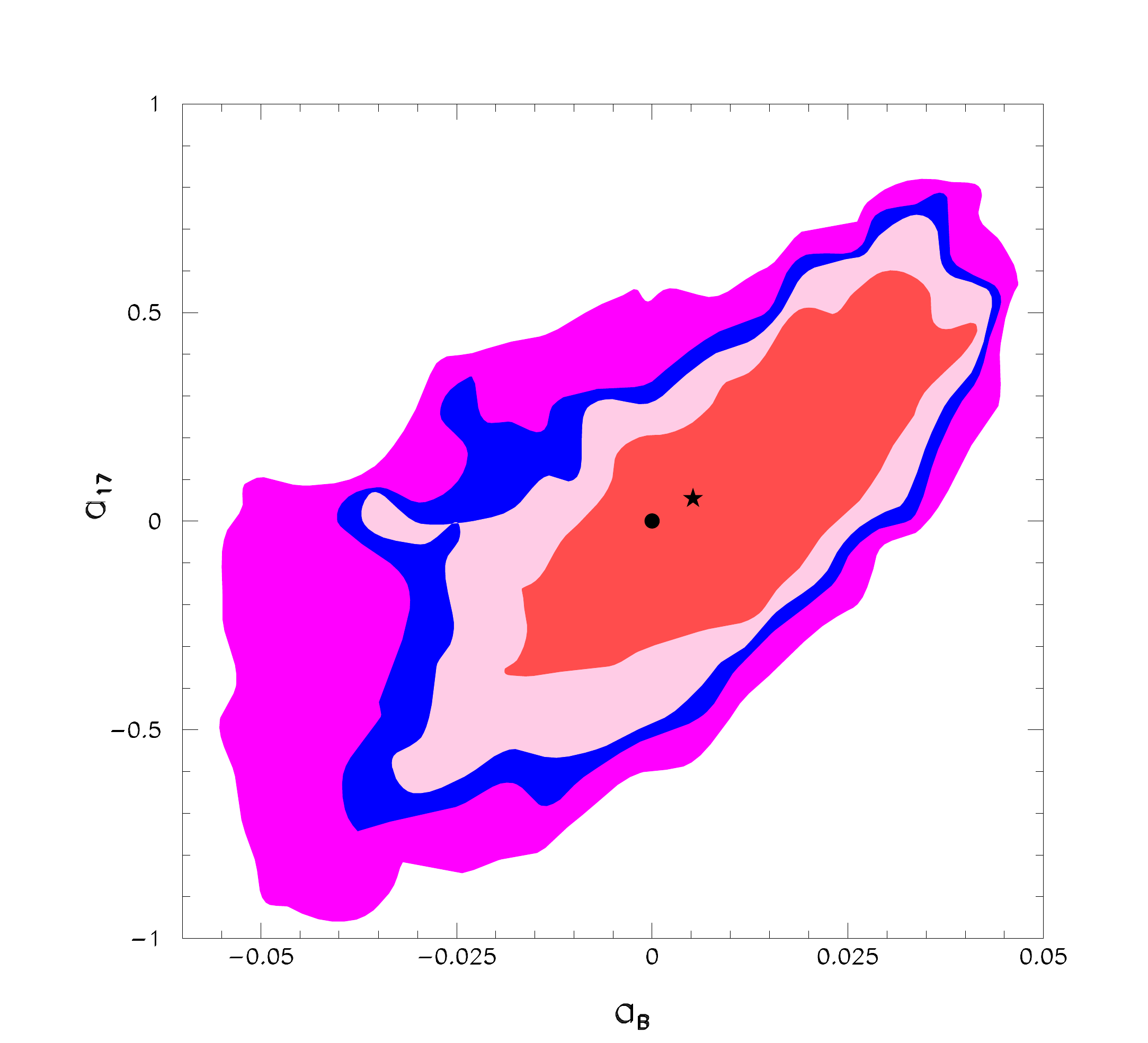}\hspace*{-5mm}
  \includegraphics[width=0.52\textwidth]{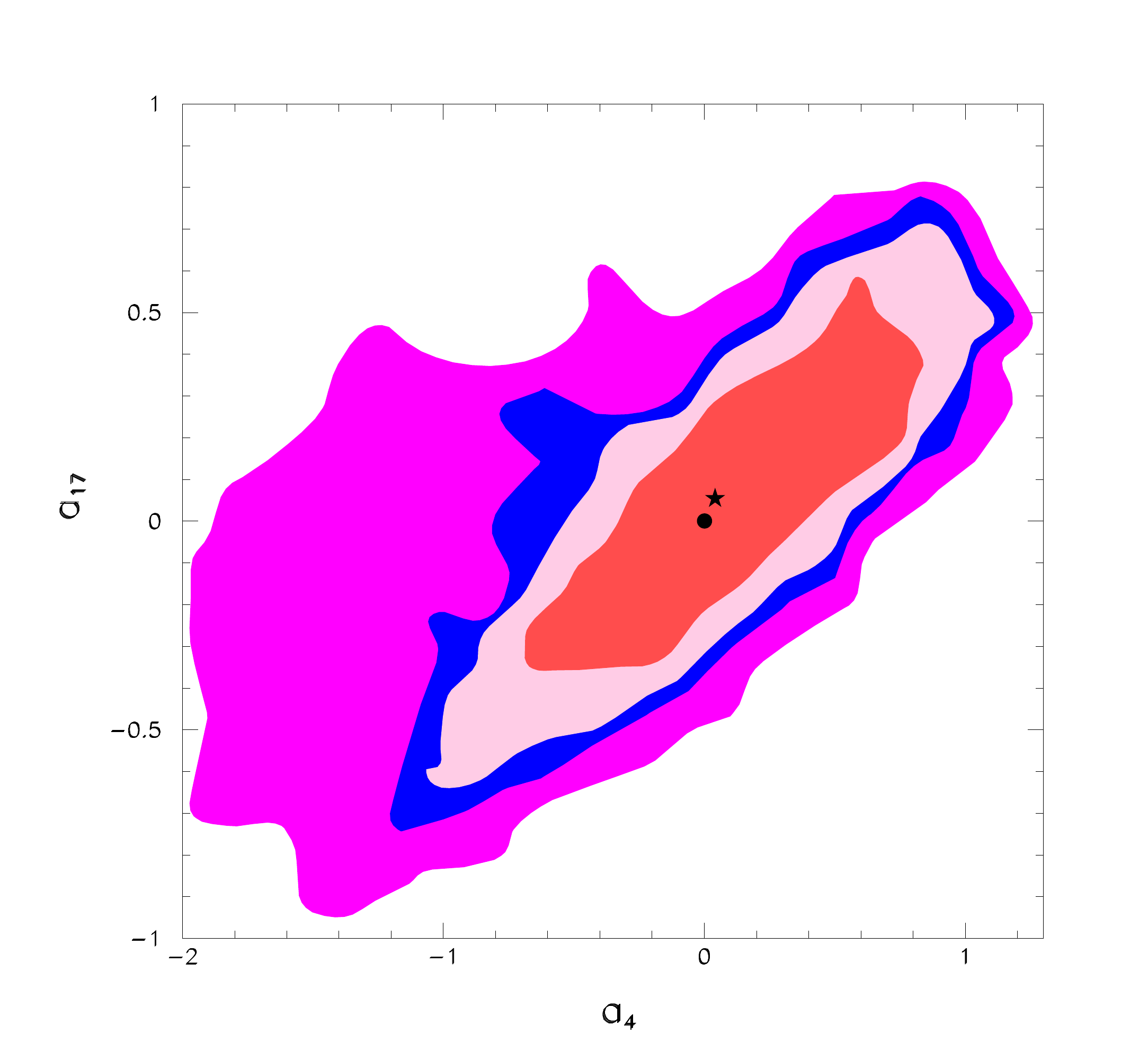}
  \caption{\it Results of the global analysis of LHC Higgs run I data,
  including kinematic distributions, for $\{a_B,\,a_4,\,a_{17}\}$,
  profiling on the undisplayed parameters. The colours refers to the
  different C.L. regions: from the inner to outer, 68\%, 90\%, 95\%,
  99\% C.L..}
\label{fig:higgs}
\end{figure}

The addition of the extra parameter $a_{17}$ has enlarged the
allowed range for all the rest of coefficients contributing to the
bosonic Higgs trilinear interactions ($a_4$, $a_5$, $a_W$, $a_B$ and
$\Delta a_C$) in comparison with the results in Ref.~\cite{Corbett:2015ksa,Corbett:2015mqf}
(after taking into account the different normalisations used between
the two analyses).  This was expected given the larger dimensionality
of the parameter space analysed in here.  The new contributions from
$\cP_{17}(h)$ are consequently strongly correlated to 
some of the other operators, as illustrated in Figure~\ref{fig:higgs}, where 
the 2-dimensional planes $a_B$ vs. $a_{17}$ and $a_{4}$
vs. $a_{17}$ are shown, after profiling on the rest of undisplayed
coefficients for each of the panels.

In the present analysis the addition of kinematic distributions 
is crucial both for closing the allowed regions on all the
considered parameters, and for controlling the correlations among
the anomalous couplings~\cite{Corbett:2015ksa}. To the best of our
knowledge, the results derived here present the most complete set of
Higgs based constraints on the set of
operators of the HEFT Lagrangian. They highlight, in addition, the potential of the EFT
expansion to describe and study the Higgs interactions at the LHC.

\subsection{Triple gauge boson couplings and Higgs interplay}
The study of triple gauge boson vertices is complementary to the analysis of Higgs physics,
and it is fundamental for obtaining a more complete description of the EWSB sector. Focusing again on the $C$ and
$P$ even operators and after including the strong constraints from EWPD,
only four operators, $\cP_2(h)$, $\cP_3(h)$, $\cP_{13}(h)$ and $\cP_{WWW}(h)$, enter
this analysis\footnote{An additional operator,  $\cP_{14}(h)$, generates a $CP$ conserving but $C$ and $P$
violating coupling, whose effects and numerical analysis 
have been discussed in Ref.~\cite{Brivio:2013pma,Eboli:2010qd} and also hold here.}.
They can give observable deviations from the SM predictions for the
triple gauge boson vertices $WWZ$ and $WW\gamma$.  These  anomalous
contributions can be parameterised in terms of the usual phenomenological 
TGV Lagrangian presented in Ref.~\cite{Hagiwara:1986vm}:
\begin{equation}
{\cal L}_{WWV} = - \,i g_{WWV} \Bigg\{ 
g_1^V \Big( W^+_{\mu\nu} W^{- \, \mu} V^{\nu} - 
W^+_{\mu} V_{\nu} W^{- \, \mu\nu} \Big) 
   \,+\, \kappa_V W_\mu^+ W_\nu^- V^{\mu\nu}
+\frac{\lambda_V}{m_W^2} \; W_{\mu \nu}^+ W^{- \nu \rho} V_\rho^\mu
\Bigg\}\,, \label{eq:classical} 
\end{equation}
with deviations from the SM predictions $g_1^Z=\kappa_Z=\kappa_\gamma=1$,
$\lambda_\gamma=\lambda_Z=0$ 
\begin{eqnarray}
 && \Delta g_1^Z =g_1^Z-1\equiv\frac{g}{4\pi c_\theta^2} c_3\,, 
\nonumber \\
&&  \Delta \kappa_Z= \kappa_Z-1\equiv\frac{g}{4\pi} \left(c_3 +2 c_{13} - 2t_\theta c_2\right)\,, 
\\
&&  \Delta \kappa_\gamma  =\kappa_\gamma-1\equiv\frac{g}{4\pi} 
\left(c_3 +2 c_{13}+ 2 
\frac{c_2}{t_\theta}\right)\, , 
\nonumber \\ 
&&\lambda_\gamma=\lambda_Z\equiv\frac{6 \pi\, g\, v^2}{\Lambda^2} c_{WWW}
\,. \nonumber
\end{eqnarray}
Electromagnetic gauge invariance enforces $g_1^\gamma=1$, both in the 
SM and in the presence of the new operators.  
In Eq.~(\ref{eq:classical}), $V \equiv \{\gamma, Z\}$, 
$g_{WW\gamma} = e$, $g_{WWZ} = g
\cos\theta_W$,  and  $W^\pm_{\mu\nu}$ and
$V_{\mu\nu}$ refer exclusively to the kinetic part of the gauge field 
strengths. 

The combination of all the most sensitive searches 
for anomalous TGV deviations in
$WV$ diboson production has been performed in Ref.~\cite{Butter:2016cvz},
presenting the results obtained in the SMEFT framework. These results show that at
present the most stringent constraints on the anomalous TGV are set
by the LHC Run~I searches, whose combined sensitivity has
clearly surpassed that of LEP. 
Even more relevant is the fact that, while
the LHC Higgs data and gauge boson pair production searches are able
to separately set stringent constraints on the HEFT operators,
the combined study of the two sets of data could be used to improve
the understanding of the nature of the Higgs boson state, as
already emphasised in Ref.~\cite{Brivio:2013pma}. 

In brief, three CP-even SMEFT operators with $d=6$ can lead to 
to sizeable corrections to the TGV vertices after
considering all bounds from EWPD~\cite{Corbett:2012dm,Corbett:2012ja,
Corbett:2013pja,Corbett:2015ksa,Butter:2016cvz}:
\beq
\begin{gathered} 
{\cal O}_{W}=\dfrac{ig}{2}(D_{\mu} \Phi)^{\dagger} W^{\mu
\nu} (D_{\nu} \Phi)\,,\qquad\qquad
{\cal O}_{B}=\dfrac{ig'}{2}(D_{\mu} \Phi)^{\dagger}
B^{\mu \nu} (D_{\nu} \Phi)\,,\\
{\cal O}_{WWW}=-\dfrac{ig^3}{8}\mbox{Tr}\left(W_{\mu
\nu}W^{\nu\rho}W_{\rho}^{\mu} \right)\,,
\end{gathered}
\eeq 
where the notation of the original papers has been kept.

As pointed out in Ref.~\cite{Brivio:2013pma}, comparing the 
interactions generated by these three operators with those induced
by the relevant operators in the HEFT basis, one finds two differences: (i) 
for the TGV phenomenology ${\cal O_W}$ and $ {\cal O}_{B}$
give corrections to the vertices equivalent to those induced by 
$\cP_2(h)$ and $\cP_3(h)$, while for the HVV couplings their effects are equivalent
to those of $\cP_4(h)$ and $\cP_5(h)$; (ii) 
the $\mathcal O(p^4)$ chiral operator  
$\cP_{13}(h)$ has no equivalent in the linear expansion at dimension
6.

In other words, (i) implies that, as it is well known from the pre-LHC 
times~\cite{Hagiwara:1993qt}, and recently emphasised in some of the
post--Higgs discovery analyses~\cite{Corbett:2013pja,Ellis:2014jta,Falkowski:2015jaa}, 
the  operators ${\cal O}_{W}$ and ${\cal O}_{B}$ lead at the same time to
anomalous contributions to both Higgs physics and TGV anomalous
measurements. Thus, any deviation generated by them should be 
correlated in data from both sectors, and consequently the combined
analysis of Higgs data and TGV measurements becomes mandatory in order to obtain
constraints as strong as possible on their coefficients~\cite{Butter:2016cvz}.  
Conversely, in the HEFT case, the anomalous TGV
deviations induced by ${\cal O}_{W}$ and ${\cal O}_{B}$ are generated
by $\cP_2(h)$ and $\cP_3(h)$, while their effects on Higgs physics originate from 
$\cP_4(h)$ and $\cP_5(h)$. Therefore, deviations 
in TGV and in Higgs physics could  remain completely
uncorrelated in the HEFT context~\cite{Brivio:2013pma}. This means that 
the nature of the Higgs boson can be directly probed by testing the presence of this
(de)-correlated pattern of interactions in the event of an anomalous
observation in any of the two sectors. 

To illustrate the present status of such comparison, a global analysis of 
the data available both on the Higgs interactions and on the searches 
for anomalous TGV has been performed. The analysis spans 
the 10 coefficients relevant for Higgs physics in the HEFT scenario, see
Eq.~(\ref{eq:higgsparameters}), together with the 3 parameters
relevant for the TGV sector, which have an equivalent in the 
SMEFT Lagrangian, $c_2$, $c_3$ and $c_{WWW}$ (i.e. setting  $c_{13}$ to
zero)\footnote{Notice that the operator belonging to the SMEFT expansion which 
contains the same interactions described by $\cP_{13}(h)$, also called ``linear sibling'',
arises only at $d=8$.}.

In what respects the TGV analysis, the simulation of the relevant distributions
and the statistical fit follow those of Ref.~\cite{Butter:2016cvz}. The best fit values and 
95\% C.L. intervals obtained for $c_2$, $c_3$ and $c_{WWW}$ are quoted for
completeness in Table~\ref{tab:higgstgvresults}.
As can be seen comparing the results in Table~\ref{tab:higgstgvresults} 
with Table 4 of Ref.~\cite{Brivio:2013pma}, derived considering only the LEP 
based TGV bounds on $c_2$ and $c_3$, the new
combination of LHC Run I searches is able to improve substantially the
constraints on $\cP_2(h)$ and $\cP_3(h)$.

\begin{figure}[t]
\centering
\hspace*{-2mm}
\includegraphics[width=0.49\textwidth]{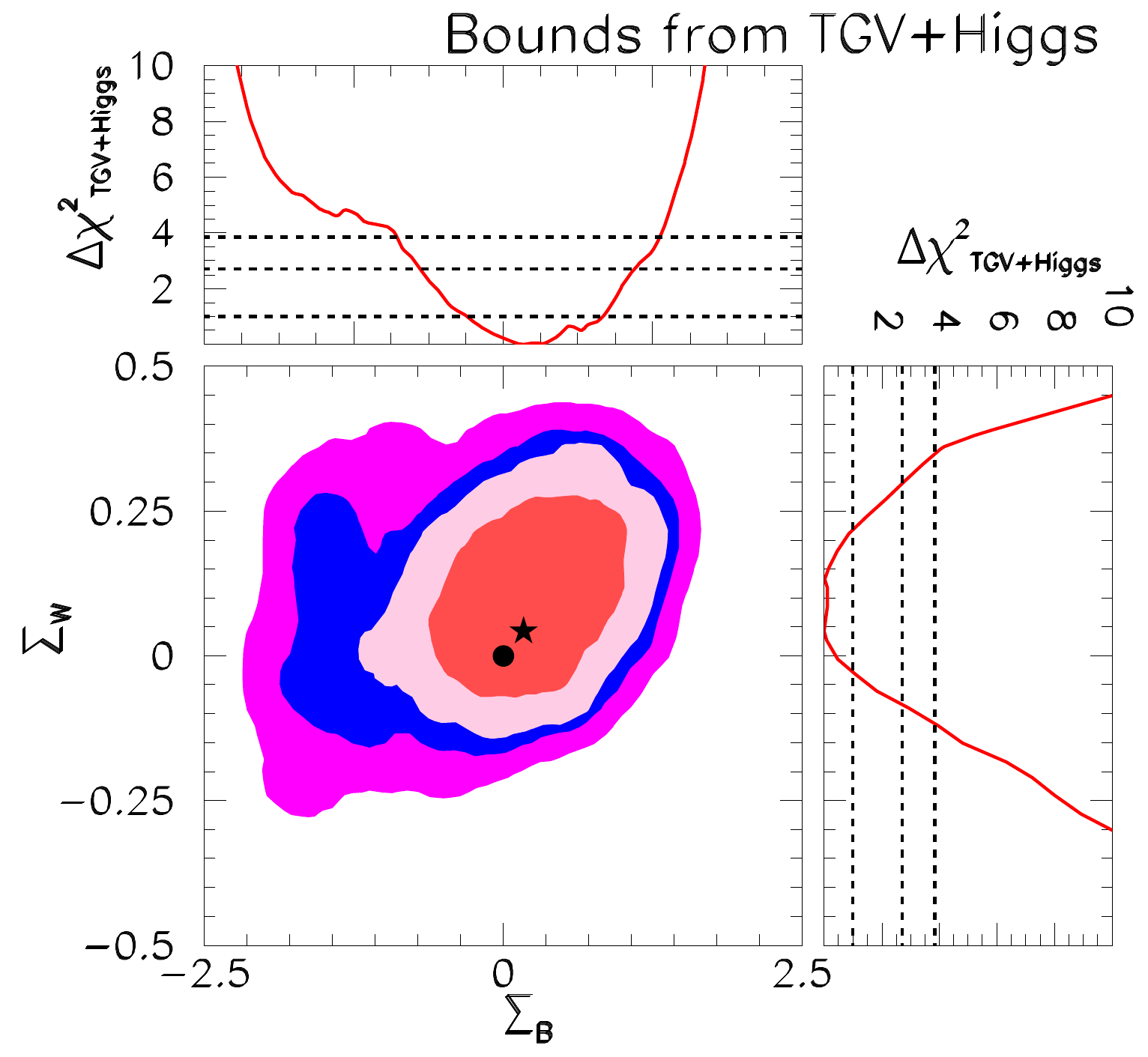}~~~
\includegraphics[width=0.49\textwidth]{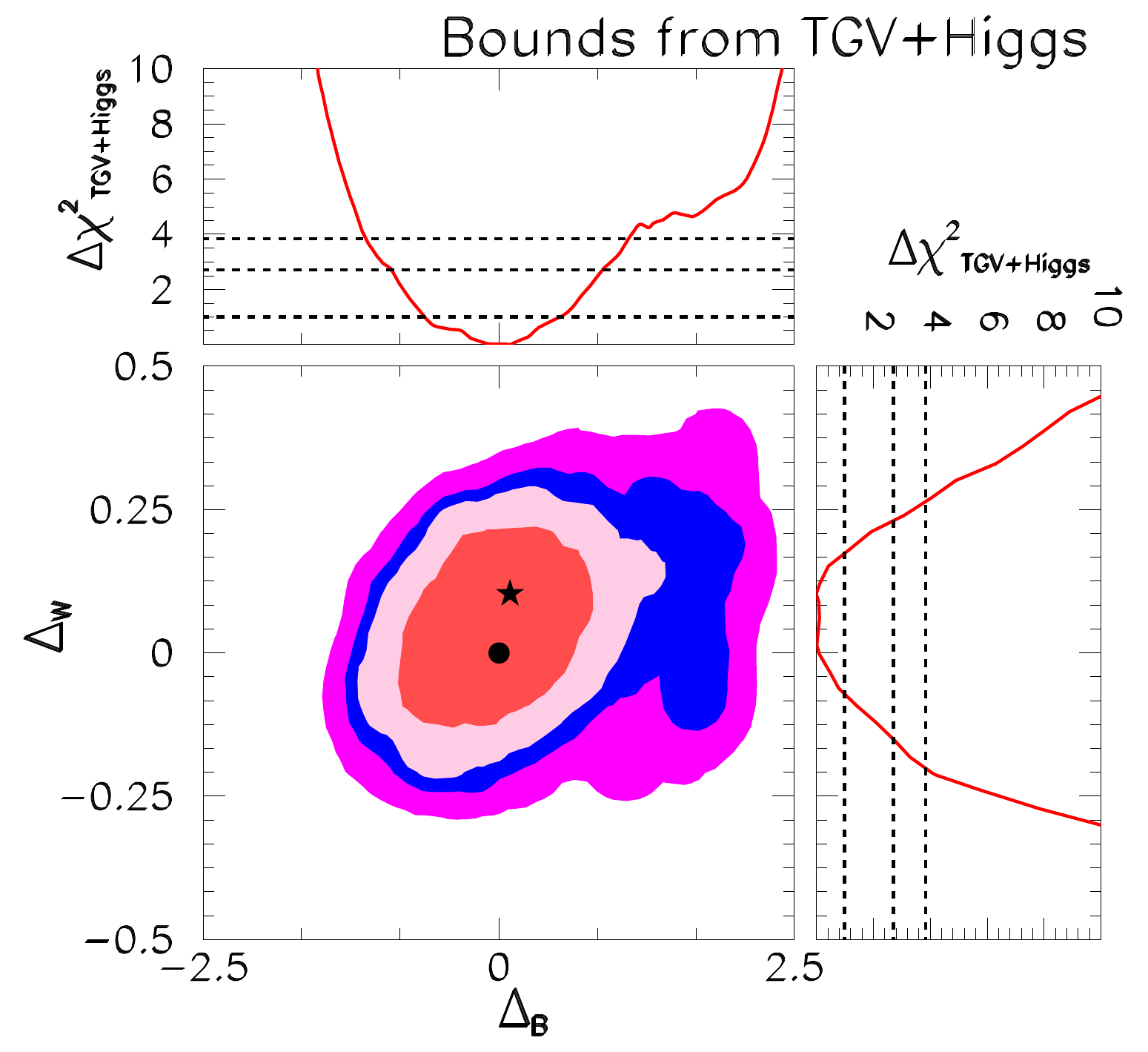}
\caption{\it Present bounds on $\Sigma_B$, $\Sigma_W$, $\Delta_B$ and
$\Delta_W$ (see text for the details on their definition) as obtained
from the most recent combined global analysis of Higgs and TGV
data. The rest of undisplayed parameters spanned in the global analysis
($\Delta a_C,\, a_B,\, a_G,\, a_W,\, , a_{17},\, Y^{(1)}_t,\,
Y^{(1)}_b$, $ Y^{(1)}_\tau$ and $c_{WWW})$ have been profiled. The black dots
signal the $(0,0)$ point, while the stars signal the current best fit point 
obtained in the analysis.}
\label{fig:disentangling}
\end{figure}

It was already shown in Ref.~\cite{Brivio:2013pma} that four specific combinations 
of the coefficients $\cP_2(h)$, $\cP_3(h)$, $\cP_4(h)$ and $\cP_5(h)$  
are meaningful for illustrating the Higgs+TGV results:\footnote{For the sake 
of comparison with Ref.~\cite{Brivio:2013pma}, the four 
combinations have been defined to be quantitatively equivalent to those 
in Ref.~\cite{Brivio:2013pma}, in spite of the different normalisation for
the $c_i$ and $a_i$ coefficients used in here.}
\begin{equation}
\begin{aligned}
\Sigma_B\equiv \frac{1}{\pi g t_\theta}(2c_2+a_4)\,, 
\qquad\qquad 
\Sigma_W\equiv \frac{1}{2\pi g}(2c_3-a_5)\,,\\
\Delta_B\equiv \frac{1}{\pi g t_\theta}(2c_2-a_4)\,, 
\qquad\qquad 
\Delta_W\equiv \frac{1}{2\pi g}(2c_3+a_5)\,.
\end{aligned}
\end{equation}
These four parameters were defined in such a way that, at $d=6$
order in the SMEFT expansion, the two $\Delta$'s
are zero because of gauge invariance and of the doublet nature of the
Higgs, $\Delta_B=\Delta_W=0$. On the other hand, the operators ${\cal
O}_{W}$ and ${\cal O}_{B}$ contribute to the $\Sigma$'s leading to
$\Sigma_B=v^2\frac{f_B}{\Lambda^2}$ and
$\Sigma_W=v^2\frac{f_W}{\Lambda^2}$, being $f_i$ the associated 
Wilson coefficients. In contrast, the HEFT operators could generate independent 
modifications to each of these four variables. Figure~\ref{fig:disentangling} shows the current
status of the bounds on the two relevant planes of coefficients after
taking into consideration all the Higgs measurements included in the
presented Higgs global analysis (based on Ref.~\cite{Corbett:2015ksa}),
together with the most recent combination of TGV searches presented in
the previous subsection (based on Ref.~\cite{Butter:2016cvz}).

As described in Ref.~\cite{Brivio:2013pma}, in the left panel of 
Figure~\ref{fig:disentangling} the $(0,0)$
point corresponds to no deviation from the SM, while in the right one it represents 
the limit in which TGV and HVV couplings show a SMEFT-like correlation.
Therefore, any deviation from $(0,0)$ in the left panel would indicate BSM
physics irrespective of the nature of the EWSB realisation, while
a similar departure in the right panel would disfavour a 
linear  EWSB. As the $\Delta$'s and the  $\Sigma$'s
are orthogonal combinations of parameters, the two panels of Fig.~\ref{fig:disentangling}
are in principle independent of each other. In particular, deviations from $(0,0)$
may occur arbitrarily in only one plane or in both at the same time.

The constraints of $\Sigma_B$, $\Sigma_W$, $\Delta_B$ and
$\Delta_W$ shown in Fig.~\ref{fig:disentangling} present a
significant improvement with respect to the bounds previously 
shown in Fig.~2 of Ref.~\cite{Brivio:2013pma}.
The reason for such a sizeable improvement relies on two key points. 
First, the strength of the derived results is increased by the inclusion of the
 more complete set of run I LHC Higgs event 
rate measurement and by the addition of relevant kinematic distributions, 
that are sensitive to the anomalous SM Lorentz structures generated by $a_3$ and $a_5$~\cite{Corbett:2015ksa}.
Second, the combination of the significant LHC Run I diboson
production analysis as described in Ref.~\cite{Butter:2016cvz} also has a huge
impact in the analysis. The combination of these two ameliorations  
enhances significantly the accuracy of the combined results shown
in Figure~\ref{fig:disentangling}, in spite of the larger
dimensionality of the parameter space considered in the present study
with respect to the global analysis in Ref.~\cite{Brivio:2013pma}.

%
%
\section{Higher order operators and expansion validity}
\label{Sect:EFTValidity}

An important issue for numerical analyses performed
in an EFT approach is that of establishing whether the EFT description is valid
at the typical energies of the processes considered. The task is particularly relevant
when collider data is included in the analysis, as the corresponding measurements are 
typically taken at energies significantly higher than the EW scale.

In general, the validity of the expansion can be discussed studying the impact of operators 
which belong to different expansion orders. In the context of the SMEFT, this is tantamount 
to analysing operators with dimension $d>6$. 
As discussed in Refs.~\cite{Drozd:2015kva,Gorbahn:2015gxa,Biekoetter:2014jwa,Brehmer:2015rna,Biekotter:2016ecg},
this analysis sets different constraints on
the cut-off of the theory, depending on the observables and of the operators considered: the strongest bounds are associated 
to observables that receive contributions from $d=8$ operators 
with a larger number of derivatives, as they induce a strong 
energy-dependence.

Similar general considerations also apply to the HEFT. 
However, in this case the discussion is complicated by the 
simultaneous presence of several characteristic scales and, consequently, 
of multiple expansion parameters. Although the only physical scales of the HEFT are $\Lambda$ 
and $v$, as explained in Sect.~\ref{Sect:Intro}, it is useful to 
keep momentarily the scale $f$ ($\Lambda\leq4\pi f$) as an independent quantity. 
The limit $f\to v$ will be discussed later on. 

In realistic Composite Higgs models, that can be considered as a benchmark for 
understanding the role played by each scale, $v$, $f$ and $\Lambda$ 
enter the low-energy Lagrangian in three different combinations: 
$v/f=\sqrt \xi$, $1/4\pi\leq f/\Lambda\leq1$, and $E/\Lambda$, 
where $E$ is the characteristic energy scale of a given observable. 
As shown in Ref.~\cite{Gavela:2016bzc}, cross 
sections of physical processes only depend on scale suppressions: 
the generic expression, adopting the NDA normalisation of 
Eq.~(\ref{MasterFormula}), is given by
\beq
\sigma\sim \dfrac{\pi(4\pi)^2}{E^2}\left(\dfrac{E^2}{\Lambda^2}\right)^{-N_\Lambda}\,,
\label{GenericCrossSection}
\eeq
where $(-N_\Lambda)$ is the number of powers of $\Lambda$ that suppress
an interaction term. The NDA master formula takes automatically care of all
the $4\pi$ factors appearing in the cross-section (see Ref.~\cite{Gavela:2016bzc} for further details and for
generalisations), so that $(-N_\Lambda)$ actually counts both 
powers of $\Lambda$ and of $f$ indifferently.
As a result, the only quantities that can be considered as proper suppression factors
are $\sqrt \xi$ and $E/\Lambda$. The physical relevance of a given cross-sections is basically determined
by its dependence on these two parameters.

While the dependence on $1/\Lambda$ is explicit in HEFT operators, 
it is less trivial to trace that on $\sqrt{\xi}=v/f$.
To this aim, it is useful to recall (see Sect.~\ref{Sect:Intro}) that $f$ is the scale associated to both the SM 
GBs and the Higgs and, as such, it is always hidden inside the GB matrix 
$\U(x)$ and the generic Higgs functions $\cF(h)$. The dependence on $f$ can be made explicit expanding these structures:
\begin{equation}\label{U_F}
\U = \unity + 2i\frac{\sigma_a \pi^a}{f} + \ldots\,,\qquad\qquad
\cF(h)=1+2a\frac{h}{f}+\ldots\,.
\end{equation}
Within $\V_\mu$ and upon going to unitary gauge, 
the powers on $1/f$ are converted into factors of $\sqrt\xi$.
This is due to the fact that, in the kind of scenarios considered here, 
$\xi$ represents a fine-tuning, that necessarily weights insertions of 
longitudinal components of the gauge bosons~\cite{Gavela:2016bzc}. 
This indeed occurs in composite Higgs models 
(see Refs.~\cite{Alonso:2014wta,Hierro:2015nna}), where analogous 
conclusions are found to hold also for $\partial_\mu\cF(h)$.

It is worth noticing that, while $\U(x)$ and $\cF(h)$, considered globally, are adimensional quantities,
their expansions contain terms with different canonical dimensions that come suppressed
by powers of $f$.
As a result, the leading terms of $\V_\mu$ and $\de_\mu\cF(h)$, obtained applying one derivative to the series of Eq.~\eqref{U_F},
have canonical dimension two: one 
dimension being associated to the derivative and the other
to the first non-vanishing term in the expansion of either $\U$ or $\cF(h)$. 
This observation can be generalised introducing the primary dimension $d_p$, defined in Ref.~\cite{Gavela:2016bzc}
as the canonical dimension of the leading term in the expansion of a given object.
For fundamental elements, such as derivatives, gauge fields and fermions, the primary dimension
coincides with the traditional canonical dimension.
Table~\ref{Tabledp} contains a summary of the primary dimensions for the building blocks used 
in the construction of the HEFT Lagrangian, together with the associated suppression factors.
It follows from the discussion above that a term suppressed by $\xi^{\a/2}(p/\Lambda)^\b$ must have $d_p=\a+\b$.

\begin{table}[b]\centering
\renewcommand{\arraystretch}{1.2}
\begin{tabular}{*4{>{$}c<{$}}}
\text{Building block}& \text{$d_p$}& \text{Factors of $\xi$}& \text{Factors of $p/\Lambda$}\\\hline
\U(x)			&	0&		1&				1\\
\cF(h)		&	0&		1&				1\\
\de_\mu		&	1&		1&		(p/\Lambda)\\
\psi			&   3/2& 		1&	(p/\Lambda)^{3/2}\\
X_{\mu\nu}	&  	2&		1&	      (p/\Lambda)^2\\
\V_\mu		&	2&	\sqrt\xi&		 (p/\Lambda)\\
\de_\mu\cF(h)	&	2&	\sqrt\xi&		 (p/\Lambda)\\
\end{tabular}
\caption{\it Different HEFT building blocks and their primary dimensions. 
The two last columns report the suppression factors associated to each object.}
\label{Tabledp}
\end{table}

\begin{table}[t]
\begin{center}
\scalebox{.9}{
\renewcommand{\arraystretch}{3}
 \begin{tabular}{|>{\columncolor{Gray!10} \boldmath $}c<{$}||*5{p{2.5cm}|}}
 \hline
 \xi ^2 	&\col\parbox[c][2cm]{2.5cm}{$(\de\cF)^2(\V)^2$\\[2mm]  $(\V)^4$\\[2mm] $(\de\cF)^4$}&  &  &  &  \\\hline
 \xi ^{3/2} 	&\col\parbox[c][2cm]{2.5cm}{} 		&  &  &  &  \\\hline
 \xi  		&\col\parbox[c][2cm]{2.5cm}{$(\V^2)(X)^2$\\[2mm] $(\de\cF)(\V)(X)$}  &\col\parbox[c]{2.5cm}{$(\de\cF)(\V)(\bar\psi\psi)$\\[2mm] $(\V)^2(\bar\psi\psi)$\\[2mm] $(\de\cF)^2(\bar\psi\psi)$} &\parbox[c][2cm]{2.5cm}{$(X)^2(\V)^2$\\[2mm] $(\de\cF)(\V)(X)^2$\\[2mm] $(\de\cF)^2(X)^2$ } 	  &  &  \\\hline
 \sqrt{\xi } 	&\col\parbox[c][2cm]{2.5cm}{$(\bar\psi\psi)(\V)$} &\col & \parbox[c][2cm]{2.5cm}{$(\V)(X)(\bar{\psi}\psi)$\\[2mm] $(\de\cF)(X)(\bar\psi\psi)$}		&\parbox[2]{2.5cm}{$(\V)(\bar\psi\psi)^2$\\[2mm] $(\de\cF)(\bar\psi\psi)^2$}  &   \\\hline
 1 		&\col\parbox[c][2cm]{2.5cm}{$(X)^2$}  &\col\parbox[c]{2.5cm}{$(X)(\bar\psi\psi)$}  &\col\parbox[c]{2.5cm}{$(\bar\psi\psi)^2$\\[2mm] $(X)^3$}  & \parbox[c][2cm]{2.5cm}{$(X)^2(\bar\psi\psi)$}  & \parbox[c]{2.5cm}{$(X)(\bar\psi\psi)^2$\\[2mm] $(X)^4$}  \\[3mm]

 \hline\hline 
 \rowcolor{Gray!10}
  &\centering  \bf1 & \centering\boldmath $\left(\frac{p}{\Lambda}\right)$ &\centering\boldmath  $\left(\frac{p}{\Lambda}\right)^2$ &\centering\boldmath  $\left(\frac{p}{\Lambda}\right)^3$ &\centering\boldmath  $\left(\frac{p}{\Lambda}\right)^4$ 
\tabularnewline\hline
  \end{tabular}}
\end{center}
\caption{\it HEFT operators distributed according to their $\xi$ 
and $p/\Lambda$ suppressing factors. A schematic notation has been adopted for
categorising the operators based on the building blocks they contain.
The terms appearing in the cyan boxes correspond to the NLO operators listed
in the previous sections. The other terms refer to operators that usually belong 
to higher Lagrangian orders, but that can have an impact similar to that of the
NLO ones for sufficiently high energies. EOMs have been employed to remove
redundant structures.}
\label{ComparisonAmongOperators}
\end{table}

With the information provided by Table~\ref{Tabledp}, it is easy to infer the dependences 
for all the HEFT operators, that can be thus organised in a two-parameter expansion as indicated, 
schematically, in Table~\ref{ComparisonAmongOperators}.
The colours discriminate between two sets of operators: the structures reported in the cyan boxes
correspond to the NLO Lagrangian considered in this work; the structures in the white cells, instead,
are customarily considered as higher order terms, but their impact may be comparable to that of the 
NLO terms for sufficiently high energies. 
Depending on the observables considered, it may be necessary to include (part of) the second set of operators into the 
phenomenological analysis (see also Ref.~\cite{Eboli:2016kko}), 
even if this would mean working with a ill-defined basis from a renormalisation point of view.
This should not be seen as a concern, as, even considering a complete, non-redundant basis at NNLO, 
only the subcategories listed in Table~\ref{ComparisonAmongOperators} would be physically relevant.
Effects due to operator mixing under the renormalisation 
group running are also expected to be completely negligible at the experimental sensitivities foreseen 
for the near future.

In the limit $f\to v$, the dependence on $\xi$ does not represent
a suppression anymore and
the physical impact of an operator is determined only by the factors of $p/\Lambda$.
In this case, one recovers a pure chiral expansion, which is organised ``horizontally''
in the representation of Table~\ref{ComparisonAmongOperators}. 

On the contrary, in the limit  $p/\Lambda\simeq \sqrt\xi$, all the operators with the same $d_p$ are equally suppressed and therefore one recovers, altogether, the linear expansion organised in canonical (or primary) dimensions. In this case, {\it all} the operators in the white boxes of Table~\ref{ComparisonAmongOperators} should be considered. This condition is for instance fulfilled for $\Lambda=\unit[10]{TeV}$ and $E\simeq\unit[1]{TeV}$, which is within the range of energies that are relevant for processes to be observed at LHC13.

The introduction of the primary dimension, i.e. of a counting on explicit and implicit 
scale suppressions, allows one to link the particular structure of an
operator to the strength of a physical signal in terms of cross sections. Indeed, if 
an observable receives contributions from a single operator, then the corresponding
cross section is uniquely determined by the primary dimension of that operator, according to
Eq.~(\ref{GenericCrossSection}). As a consequence, the $d_p$ is a useful 
phenomenological tool to indicate whether the strength of an observable, that receive contributions
only from operators belonging to higher expansion orders, is expected to be of the same
order or more suppressed with respect to the other processes already considered
in the phenomenological analysis.

An interesting application of the primary dimension
is that if the $d_p$ of an HEFT operator is smaller than the canonical dimension 
of the corresponding linear sibling, then the processes described by these operators 
represent smoking guns to test the linearity of the EWSB realisation. This is the case of 
the operator $\cP_{14}(h)$ discussed in Ref.~\cite{Gavela:2016bzc}: it induces an 
anomalous TGV, commonly called $g_5^Z$, that is expected to be 
strongly suppressed in the SMEFT description, but not in the HEFT one.

%
%
\section{Conclusions}
\label{Sect:Conclusions}

The complete effective Lagrangian for a non-linear realisation 
of the EWSB (shortened into HEFT) has been presented. It provides the most general 
description of the Higgs couplings and it can be used for investigating 
a large spectrum of distinct theories, ranging from the SM to technicolour 
constructions, including Composite Higgs realisations and dilaton-like frameworks.
In contrast with the effective Lagrangian for a linearly realised EWSB (also SMEFT), in which
the Higgs belongs to an exact $SU(2)_L$ doublet, in the HEFT the physical
Higgs is assigned to a singlet representation of the EW group and it is treated as an object independent
of the Goldstone bosons' matrix. 

Assuming invariance under the Lorentz and SM gauge symmetries, 
as well as the conservation of Baryon and Lepton numbers,
the complete chiral basis at the next to leading order
contains a total number of 148 independent, flavour universal terms. When extending
the SM spectrum to include three right-handed neutrinos, 40 more operators
enrich the basis. The generalisation to arbitrary flavour contractions
is straightforward.

Conversely, the SMEFT basis up to $d=6$ consists of only 59 flavour universal terms, in absence
of right-handed neutrinos. 
The different number of operators 
and of building blocks used for the construction of the two bases lead to 
fundamental differences between the SMEFT and the HEFT. 
The possibility
of distinguishing between them has been discussed performing a global fit 
including all the available data from colliders, including EWPD, Higgs and TGV
measurements taken at the LHC Run I. The main 
outcomes are summarised in the following points:
\begin{itemize}
\item The Electroweak precision data analysis together with 
the study of the CKM matrix unitarity allows one to constrain 11
parameters of the HEFT Lagrangian. The corresponding value of the
$\chi^2$ at the minimum is 6. This can be compared with the
corresponding analysis within the SM, whose $\chi^2$ is
18.4.
\item The results for the $S$, $T$ and $U$ parameters are
significantly different from the standard analysis in the
SMEFT with operators up to dimension 6, due to the presence of extra free parameters: 
the allowed range for $S$ is about 4 times broader, while the
bounds on $T$ and $U$ are about 20 times weaker.
\item The analysis of Higgs data depends on a total of 10
parameters, with one bosonic operator more compared to the same analysis in the SMEFT
case at dimension six. Although the final results are quite similar to those obtained for
the SMEFT, the addition of the extra parameter broadens
the allowed range for the remaining 9 coefficients, as expected.
\item The interplay between triple gauge boson vertices and Higgs 
couplings provides an interesting way of investigating the nature of EWSB. 
Although this analysis is not conclusive yet due to the 
limited sensitivity on the observables considered, the introduction of 
kinematic distributions is seen to improve considerably the results.
Would the accuracy of Higgs measurements improve significantly in the future, this kind
of analysis may reveal signatures of non-linearity in the Higgs sector.
\item It has been underlined that with the increase in energy at 
colliders, it may be necessary to consider several operators that, in spite of being usually
considered as higher order effects, may have a non-negligible phenomenological impact. 
The list of the relevant structures has been given in Table~\ref{ComparisonAmongOperators}.
\end{itemize}

In summary, this work extends the chiral basis of Refs.~\cite{Brivio:2013pma,Gavela:2014vra} with the introduction
of fermionic operators. Moreover, the analysis presented here updates and extends that contained in 
Ref.~\cite{Brivio:2013pma} with the inclusion of more recent collider
data and of fermionic observables.
A strategy for disentangling the nature of the EWSB has been discussed, based on the presence of
new anomalous signals and of decorrelations among observables.
It has also been discussed how the phenomenological analysis should be modified
when higher energy data is kept into account, specifying the relevant operator structures
that should be added to the basis in this case.
The analysis presented here represents the first phenomenological study performed 
with the complete HEFT Lagrangian and it could be taken as a reference for dedicated 
experimental analyses aimed at shedding light on the Electroweak symmetry breaking sector and the 
Higgs nature.


\section*{Acknowledgements}
We thank M.B.~Gavela for useful discussions during the development of
the project, as well as Anja Butter, Tilman Plehn and Michael Rauch for
their crucial assistance with both the Higgs and the TGV analyses. We also thank
E.E.~Jenkins, A.V.~Manohar and M.~Trott for interesting comments on the manuscript.
I.B. research was supported by an ESR contract of the EU
network FP7 ITN INVISIBLES (Marie Curie Actions, PITN-GA-2011-289442).  
M.C.G-G is supported by USA-NSF grant
PHY-13-16617, by grants 2014-SGR-104 and by FPA2013-46570 and
consolider-ingenio 2010 program CSD-2008-0037.  L.M. acknowledge
partial support of CiCYT through the project FPA2012-31880 and of the
Spanish MINECO's ``Centro de Excelencia Severo Ochoa'' Programme under
grant SEV-2012-0249.  M.C.G-G and L.M. acknowledge partial support by
FP7 ITN INVISIBLES (PITN-GA-2011-289442), FP10 ITN ELUSIVES 
(H2020-MSCA-ITN-2015-674896)  and INVISIBLES-PLUS
(H2020-MSCA-RISE-2015-690575)\\

\appendix

\section{Additional operators in the presence of RH neutrinos}\label{APP:NR}
Adding right-handed neutrinos to the spectrum amounts to declaring a
non-zero upper component for the $\LR$ doublet, which shall be defined
as $\LR=(N_R,E_R)^T$. Consequently, the lepton Yukawa matrix in the LO
Lagrangian~Eq.~\eqref{Lag0} has to be generalised to account for the masses
and interactions of the neutrinos with the Higgs
\begin{equation}
\cY_{L}(h)\equiv \diag\left(\sum_n Y_{\nu}^{(n)}\dfrac{h^n}{v^n},
\sum_n Y_{\ell}^{(n)}\dfrac{h^n}{v^n}\right)\,. 
\end{equation} 
In addition, the fermionic basis presented in Sec.~\ref{sec.Lfer} must
be enlarged in order to account for the increased number of possible
invariants, as follows:
\begin{align}
\Delta \LLag_{2F}&= \sum_{j=15}^{17} \tflC_j\,\tflName_j +\sum_{j=18}^{28}  
\frac{1}{\Lambda}\left(\tflC_j+i\tilde{n}^\ell_j\right)\,\tflName_j
  +\sum_{j=29}^{31} \frac{4\pi}{\Lambda}\left(\tflC_j+i\tilde{n}^\ell_j\right)\,
\tflName_j\,,\\
  \Delta \LLag_{4F}&= \frac{(4\pi)^2}{\Lambda^2}\,\left[
   \sum_{j=8}^{10} \left(\fflC_j+i\tilde{r}^\ell_j\right)\fflName_j 
+ \sum_{j=1}^{15} \fflC_j\,\fflName_j +
   \sum_{j=24}^{29} \left(\ffqlC_j+i\tilde{r}^{\mathcal{Q}\ell}_j\right)\,
\ffqlName_j + \sum_{j=30}^{38} \ffqlC_j\,\ffqlName_j\right]\,.
\end{align} 
The complete list of additional operators is provided in this Appendix.

\subsection*{Single Leptonic Current Operators}
\noindent
With one derivative
\begin{center}
\renewcommand{\arraystretch}{1.5}
\begin{tabular}{>{\footnotesize}r>{$}l<{$}@{\hspace*{1cm}}>
{\footnotesize}r>{$}l<{$}}
&\tfl(h)\equiv i\LBR \g_\mu \U^\dag\V^\mu\U \LR \cF	\label{Lrr_v}& 
\cancel{CP}&\tfl(h)\equiv\LBR \g_\mu \U^\dag[\V^\mu,\T]\U\LR \cF	
\label{Lrr_cvt}\\
&\tfl(h)\equiv i\LBR \g_\mu \U^\dag\T\V^\mu\T\U\LR \cF 	\label{Lrr_tvt}&
\end{tabular}
\end{center}
With two derivatives
\begin{center}
\renewcommand{\arraystretch}{1.5}
 \begin{tabular}{>{$}l<{$}@{\hspace*{1cm}}>{$}l<{$}}
\tfl(h)\equiv  \LBL\T\U\LR{\de_\mu\cF\de^\mu\cF'}\label{Llr_t_deFdeF} &
\tfl(h)\equiv \LBL \V_\mu \U\LR{\de^\mu\cF}	\label{Llr_v_deF}\\
\tfl(h)\equiv \LBL\T\V_\mu\T\U\LR{\de^\mu\cF}	\label{Llr_tvt_deF}&
\tfl(h)\equiv \LBL \V_\mu\V^\mu\T \U\LR\cF	\label{Llr_vvt}\\
\tfl(h)\equiv \LBL \T\V_\mu\T\V^\mu\T \U\LR\cF  \label{Llr_tvtvt}&
\tfl(h)\equiv \LBL \V_\mu\T\V^\mu \U\LR\cF	\label{Llr_vtv}\\
\tfl(h)\equiv \LBL \V_\mu\T\V^\mu\T \U\LR\cF	\label{Llr_vtvt}&
\tfl(h)\equiv \LBL\ssu\V_\mu\U\LR{\de_\nu\cF}	\label{Llr_s_v_deF}\\
\tfl(h)\equiv \LBL\ssu\T\V_\mu\T\U\LR{\de_\nu\cF}\label{Llr_s_tvt_deF}&
\tfl(h)\equiv \LBL \ssu\V_\mu\T\V_\nu\T\U\LR\cF\label{Llr_s_vtvt_deF}\\
\tfl(h)\equiv \LBL \ssu[\V_\mu,\V_\nu]\T\U\LR\cF \label{Llr_s_vvt_deF}&
\tfl(h)\equiv ig'\,\LBL\ssu\T\U\LR\BBd\cF\label{Llr_s_tb}\\
\tfl(h)\equiv ig\,\LBL\ssu\{\WWd,\T\}\U\LR\cF\label{Llr_s_awt}&
\tfl(h)\equiv ig\,\LBL\ssu\T\WWd\T\U\LR\cF\label{Llr_s_twt}\\
\end{tabular}
\end{center}

\subsection*{Four-fermion Operators}
\noindent
Additional operators with four leptons:
\begin{center}
\renewcommand{\arraystretch}{1.5}
 \begin{tabular}{>{$}l<{$}@{\hspace*{0.5cm}}>{$}l<{$}}
\ffl(h)\equiv(\LBL\s^i\U\LR)(\LBL\s^i\U\LR)\cF \label{Llrlr_ss} &
\ffl(h)\equiv(\LBL\U\LR)(\LBL\T\U\LR)\cF	 \label{Llrlr_1t}\\
\ffl(h)\equiv(\LBL\T\U\LR)(\LBL\T\U\LR)\cF	\label{Llrlr_tt}&
\ffl(h)\equiv(\LBR\g_\mu\LR)(\LBR\g^\mu\U^\dag\T\U\LR)\cF \label{Lrrrr_1t}\\
\ffl(h)\equiv(\LBR\g_\mu\U^\dag\T\U\LR)(\LBR\g^\mu\U^\dag\T\U\LR)\cF \label{Lrrrr_tt}&
\ffl(h)\equiv(\LBL\g_\mu\LL)(\LBR\g^\mu\U^\dag\T\U\LR)\cF \label{Lllrr_1t} \\
\ffl(h)\equiv(\LBL\g_\mu\T\LL)(\LBR\g^\mu\U^\dag\T\U\LR)\cF	\label{Lllrr_tt} &
\ffl(h)\equiv(\LBL\g_\mu\s^i\LL)(\LBR\g^\mu\U^\dag\s^i\U\LR)\cF \label{Lllrr_ss} 
\end{tabular}
\end{center}
\noindent
Additional mixed operators with two quarks and two leptons
\begin{center}
\renewcommand{\arraystretch}{1.5}
\hspace*{-7mm}
 \begin{tabular}{>{$}l<{$}@{\hspace*{5mm}}>{$}l<{$}}
\ffql(h)\equiv{(\LBL\U\QR)(\QBL\T\U\LR)\cF	} \label{QLlrlrmix_t1}&
{\ffql(h)\equiv(\LBL\T\U\LR)(\QBL\U\QR)\cF}	\label{QLlrlr_1t}	\\
{\ffql(h)\equiv(\LBL\T\U\LR)(\QBL\T\U\QR)\cF }	\label{QLlrlr_tt}& 
\ffql(h)\equiv{(\LBL\T\U\QR)(\QBL\T\U\LR)\cF  }	\label{QLlrlrmix_tt}\\
\ffql(h)\equiv{(\LBL\s^i\T\U\LR)(\QBL\s^i\U\QR)\cF}       &
{\ffql(h)\equiv(\LBL\s^i\T\U\QR)(\QBL\s^i\U\LR)\cF}\\
{\ffql(h)\equiv(\LBR\g_\mu\U^\dag\T\U\LR)(\QBR\g^\mu\QR)\cF }\label{QLr_t1}&
{\ffql(h)\equiv(\LBR\g_\mu\U^\dag\T\U\LR)(\QBR\g^\mu\U^\dag\T\U\QR)\cF }\label{QLr_tt}\\
{\ffql(h)\equiv(\LBR\g_\mu\U^\dag\s^j\U\LR)(\QBR\g^\mu\U^\dag\s^j\U\QR)\cF }\label{QLr_ss}&
{\ffql(h)\equiv(\QBL\g_\mu\QL)(\LBR\g^\mu\U^\dag\T\U\LR)\cF	}\\
{\ffql(h)\equiv(\QBL\g_\mu\T\QL)(\LBR\g^\mu\U^\dag\T\U\LR)\cF	}&
{\ffql(h)\equiv(\QBL\g_\mu\s^j\QL)(\LBR\g^\mu\U^\dag\s^j\U\LR)\cF	}\\
{\ffql(h)\equiv(\QBL\g^\mu\LL)(\LBR\g_\mu\U^\dag\T\U\QR)\cF }	\label{QLrrllmix_1t}&
{\ffql(h)\equiv(\QBL\g_\mu\T\LL)(\LBR\g^\mu\U^\dag\T\U\QR)\cF	}\\
{\ffql(h)\equiv(\QBL\g^\mu\s^j\T\LL)(\LBR\g_\mu\U^\dag\s^j\U\QR)\cF}
\end{tabular}
\end{center}

%
\section{Removal of \texorpdfstring{$\mathbf{\cF(h)}$}{F(h)}  from the Higgs and fermions kinetic terms}
\label{APP:LOLag}
All the kinetic terms in the LO Lagrangian, Eq.~\eqref{Lag0}, are canonically normalised, despite the fact that the singlet nature of the $h$ field in principle allows one to couple them to a function $\cF(h)$. In the case of the gauge-boson kinetic term, the absence of a Higgs-dependence is justified in the assumption that the transverse components of the gauge fields do not interact with the Higgs sector at LO. On the other hand, in the cases of the Higgs and of the fermions' kinetic terms, the dependence $\cF(h)$ is completely redundant, as it can be removed via a field redefinition (analogously to what was done in Ref.~\cite{Giudice:2007fh}. See also Ref.~\cite{Buchalla:2013rka}).
This appendix provides more details about this redefinition.

The coupling of the fermionic kinetic term to a generic Higgs function would be of the form
\begin{equation}
 \frac{i}{2}\left(\bar{\psi}\slashed{D}\psi 
-\bar{\psi}\overleftarrow{\slashed{D}}\psi\right)\left(1+ \cF_\psi(h)\right)\,,
\end{equation}
where $\psi=\{Q,L\}$ and 
\begin{equation}
 \cF_\psi(h) = c_\psi +2a_\psi \frac{h}{v}+b_\psi\frac{h^2}{v^2}+\dots
\end{equation} 
The dependence of $\cF_\psi(h)$ can therefore be removed via the redefinition
\begin{equation}\label{psi.transf}
 \psi\to \psi' \left[1+\cF_\psi(h)\right]^{-1/2}\,.
\end{equation}
As this substitution is applied to the whole
Lagrangian, it induces a modification of all the couplings between fermionic and Higgs fields,
which can be reabsorbed in redefinitions of the functions $\cF_i(h)$ that accompany fermionic operators.  
In particular, this is also true for the LO Yukawa couplings, as they are accompanied by arbitrary polynomials $\cY_{Q,L}(h)$.
In conclusion, the insertion of a function $\cF_\psi(h)$ in the fermionic kinetic term is redundant in the LO Lagrangian of Eq.~\eqref{Lag0}.

The Higgs kinetic term may also be written as
\beq
\dfrac{1}{2}\de_\mu h\de^\mu h\left(1+\cF_H(h)\right)\,,
\eeq
with 
\begin{equation}
\cF_H(h)= c_H+2a_H \frac{h}{v}+b_H\frac{h^2}{v^2}+\dots
\end{equation} 
In this case, the $\cF_H(h)$ function can be removed by the field redefinition 
\begin{equation}\label{h.transf}
 h' \to \int_0^{h}{\sqrt{1+\cF_H(s)}\,ds}
\end{equation} 
in fact
\begin{align}
  \frac{1}{2}\de_\mu h' \de^\mu h'&=
   \frac{1}{2}\left[\de_\mu h\sqrt{1+\cF_H(h)}\right]^2=\cP_H(h)\,.
\end{align} 
Although this redefinition looks quite involved, it clearly induces modifications of all the Higgs couplings in the Lagrangian.
As these are always described by arbitrary coefficients, the redefinition~\eqref{h.transf} can be entirely reabsorbed into redefinitions of the functions $\cF_i(H)$ that appear in the Lagrangian.
As seen for the case of $\cF_\psi(h)$ above, the presence of $\cF_H(h)$ in the Higgs kinetic term is redundant within the LO Lagrangian chosen in Eq.~\eqref{Lag0}.

\subsubsection*{A practical example}
In order to give a practical illustration, one can consider a specific function 
\begin{equation}\label{FH_example}
 \cF_H(h) = 2a_H\frac{h}{v}\,.
\end{equation} 
Then the following equation
\begin{equation}
 h'=\int_0^{h}{\sqrt{1+2 a_H s/v} \,ds} = \frac{v}{3 a_H}\left[\left(\frac{2a_H h}{v}+1\right)^{3/2}-1\right]
\end{equation} 
can be solved analytically, obtaining
 \begin{equation}
  h=\frac{v}{2 a_H}\left[\left( \frac{3 a_H}{v}h'+1\right)^{2/3}-1\right]\,.
 \end{equation} 
Plugging this result into $\cF_C(h)$ and re-expanding in $h'/v$, it
gives\footnote{In this computation $a_H>0$ is assumed. For negative
values the third roots give some complications.}:
\begin{equation}
 \cF_C(h) = 1+2a_C\frac{h'}{v}+\left(b_C-a_C a_H\right)\left(\frac{h'}{v}\right)^2\,,
\end{equation}
so that the impact of the redefinition can be entirely reabsorbed
defining primed coefficients
\begin{equation}\label{Cprimed}
 a'_C = a_C\,,\qquad\qquad
  b'_C = b_C-a_C a_H\,.
\end{equation} 
An analogous redefinition allows one to reabsorb  inside the function $\cY_{\psi}^{(n)}$ the effects on the Yukawa interactions.

\section{Construction of the fermionic basis}
\label{APP:ReductionFermionBasis}
This appendix provides additional information about the construction
of the fermionic basis specifying, in particular, the relation between
the structures present in the operators presented in
Sec.~\ref{sec.Lfer} and those that have been removed.  In the
following, generic fermion fields are denoted by $\psi=\{Q,L\}$ while
$\Gamma$ stands for an arbitrary $SU(2)$ structure, combination of the
blocks $\{\T,\,\V_\mu,\,D_\mu,\,\s^j\}$.  The Lorentz contractions are
always explicited and, whenever they are not specified, chiralities
are arbitrary. The correspondence between classes of operators is
indicated schematically by an arrow ($\to$); signs and numerical 
coefficients are not specified in these relations.

\subsection{Useful identities}
\label{APP:Formulae}
A list of useful identities is provided below.
Since the building blocks $A=\{\T,\, \V_\mu,\,\D_\mu\V^\mu\}$ are traceless, 
they can be generically rewritten as
$A=\frac{1}{2}\tr[A\s^a]\s^a$. This yields the relations:
\begin{align}\label{identities_TV}
 [\T,\V_\mu] &=\frac{i}{2}\e^{ijk} \tr(\T\s^i)\tr(\V_\mu\s^j)\s^k \\
 \{\T,\V_\mu\} &= \tr(\T\V_\mu)\unity\\
 \T\V_\mu\T &=\frac{1}{2}\left[\tr(\T\V_\mu)
\tr(\T\s^i)-\tr(\V_\mu\s^i)\right]\s^i= \T \tr(\T\V_\mu)-\V_\mu
\end{align}
The properties of the $SU(2)$ generators additionally lead to the 
following identities:
\begin{equation}
\label{other_structures}
  \begin{aligned}
 \T\V_\mu\V^\mu &=\V_\mu\V^\mu\T\\
  \T\V_{[\mu}\T\V_{\nu]} &=\V_{[\mu}\T\V_{\nu]}\T\\
  \T\V_{[\mu}\T\V_{\nu]}\T &=\V_{[\mu}\T\V_{\nu]}\\
  \T[\V_\mu,\V_\nu] &=-[\V_\mu,\V_\nu]\T-2\V_{[\mu}\T\V_{\nu]}\,.
 \end{aligned}
\end{equation} 
The transformation properties of $\T$ and $\V_\mu$ ensure:
\begin{align}
 \D_\mu\T&=[\V_\mu,\T] \label{DmuT}\\
 \V_{\mu\nu} = \D_\mu\V_\nu-\D_\nu\V_\mu &= ig \WWd - ig'\BBd/2 + [\V_\mu,\V_\nu] \label{Vmunu}
\end{align}

\noindent
Fierz identities for chiral (anticommuting) fields have been employed 
for the reduction of the four fermion basis
\begin{align}
 (\bar{A}_L B_R)(\bar{C}_L D_R)=&   -\frac{1}{2} (\bar{A}_L D_R)(\bar{C}_L B_R)
 -\frac{1}{8}(\bar{A}_L\ssu D_R)(\bar{C}_L\s_{\mu\nu}B_R) 
\label{fierz.tensor}\\
 (\bar{A}_L B_R)(\bar{C}_R D_L)=&   -\frac{1}{2}  (\bar{A}_L\g_\mu D_L)
(\bar{C}_R\g^\mu B_R)   
\label{fierz.scalar}\\
 (\bar{A}_L\g_\mu B_L)(\bar{C}_L\g^\mu D_L)=& (\bar{A}_L\g_\mu D_L)
(\bar{C}_L\g^\mu B_L)  
\label{fierz.vectorL}\\
 (\bar{A}_R\g_\mu B_R)(\bar{C}_R\g^\mu D_R)=& (\bar{A}_R\g_\mu D_R)
(\bar{C}_R\g^\mu B_R)  
\label{fierz.vectorR}\,.
\end{align}
Whenever they are applied to $SU(2)$ doublets (and $SU(3)$ triplets),
these identities must be applied together with the completeness
relations for the generators of $SU(2)$ (and of $SU(3)$)
\begin{align}
\s^a_{ij} \s^a_{mn} &= 2\d_{in}\d_{mj}-\d_{ij}\d_{mn}\label{dec.su2}\\
\lambda^A_{ij} \lambda^A_{mn} &= 2\d_{in}\d_{mj}-\frac{2}{3}\d_{ij}\d_{mn} 
\label{dec.su3}
\end{align}
in order to recover the correct gauge contractions.  For example,
combining Eq.~\eqref{fierz.scalar} with Eqs.~\eqref{dec.su2}
and~\eqref{dec.su3}, the scalar identity for quark doublets reads
\begin{equation}
 \begin{aligned}
 (\bar{Q}_{1L} Q_{2R})(\bar{Q}_{3R}Q_{4L})   
 &=-\frac{1}{12}(\bar{Q}_{1L}\g_\mu Q_{4L})
(\bar{Q}_{3R}\g^\mu Q_{2R})-\frac{1}{12}
(\bar{Q}_{1L}\g_\mu \s^kQ_{4L})(\bar{Q}_{3R}\g^\mu \s^kQ_{2R})+\\
 &-\frac{1}{8}(\bar{Q}_{1L}\g_\mu \lambda^AQ_{4L})
(\bar{Q}_{3R}\g^\mu\lambda^A Q_{2R})-\frac{1}{8}(\bar{Q}_{1L}
\g_\mu\lambda^A \s^kQ_{4L})(\bar{Q}_{3R}\g^\mu\lambda^A \s^kQ_{2R})\,.
 \end{aligned}
\end{equation}

\boldmath
\subsection{Construction of  \texorpdfstring{$\Delta\LLag_{2F}$}{L2F}}
\unboldmath

\label{APP_2F}
\begin{itemize}
\item Since for traceless matrices $\tr(AB)\unity=\{A,B\}$, the
operators of the type $\bar{\psi}\g_\mu \psi \tr(\Gamma_1\Gamma_2)\cF$
with $\Gamma_i=\{\T,\,\V_\mu,\,\D_\mu\V_\nu\}$ are always equivalent
to the bilinears $\bar{\psi}\g_\mu\{\Gamma_1,\Gamma_2\}\psi\cF$.

\item Bilinears with a derivative on the fermion field and a vector
current (e.g. $(D_\mu\bar{\psi}) \g^\mu X \psi$) can been removed via
integration by parts and application of the EOMs (see
Appendix~\ref{APP:EOM})

\item Operators with a derivative on the fermion field but with no
gamma matrices (of the type $\bar{\psi} \Gamma^\mu D_\mu \psi$) are
removed using the relation $g^{\mu\nu}=\{\g^\mu,\g^\nu\}/2$,
integration by parts and the EOMs:
 \begin{equation}
  \begin{aligned}
 \bar{\psi}\Gamma_\mu D^\mu \psi &=\bar{\psi}\, g_{\mu\nu} \Gamma^\mu
D^\nu \psi=
\bar{\psi}\slashed{\Gamma}(\slashed{D}\psi)+\bar{\psi}\g^\nu
\slashed{\Gamma}(D_\nu \psi)=\\
 &= \bar{\psi}\slashed{\Gamma}(\slashed{D}\psi) -
(\bar{\psi}\overleftarrow{\slashed{D}})\slashed{\Gamma}\psi-\bar{\psi}
(\slashed{D}\slashed{\Gamma})\psi   \\
 &\to\, \bar{\psi}\slashed{\Gamma}\psi + \bar{\psi}D_\mu \Gamma^\mu \psi 
+i \bar{\psi}\ssu D_\mu \Gamma_\nu \psi.
  \end{aligned}
 \end{equation}

\item bilinears with the structure $\g^\mu \g^\nu\D_\mu\V_\nu$ can be
reduced to a combination of dipole operators (containing field
strengths), terms with the structure $\ssu\V_\mu\V_\nu$ and bilinears
with the direct contraction $\D_\mu\V^\mu$. In fact:
 \begin{equation}
\g^\mu \g^\nu\D_\mu\V_\nu = \left(g^{\mu\nu}-i\ssu\right)\D_\mu\V_\nu
= \D_\mu\V^\mu -\frac{i}{2}\ssu \V_{\mu\nu}
 \end{equation} 
where Eq.~\eqref{Vmunu} shall be applied on the latter
term. The former can also be removed using the EOMs.

 \item the commutator $[D_\mu,D_\nu]$ is always vanishing when applied
to $SU(2)$ invariants (right-handed fermions, $B$ and $G$ fields,
$\cF(h)$ functions), while it is traded for a field strength when it
acts on a quantity $X$ with non-trivial isospin transformations:
${[D_\mu,D_\nu]X= ig [\WWd, X]}$.

 \item further combinations of $\T$ and $\V_\mu$ that do not appear in
the basis reported in Sec.~\ref{sec.Lfer} have been traded for others
using the identities~\eqref{identities_TV} and~\eqref{other_structures}.

\end{itemize}


\boldmath
\subsection{Construction of   \texorpdfstring{$\Delta\LLag_{4F}$}{L4F}}
\unboldmath
\label{APP_4F}

Details as regards the construction and reduction of the four-fermion
operators basis are provided in this section. None of the terms of
$\Delta\LLag_{4F}$ have been removed via the EOMs, while the Fierz
identities~\eqref{fierz.tensor}-\eqref{fierz.vectorR} have been
extensively employed for removing redundant structures.
In particular, operators with tensor currents
($(\bar{\psi}\ssu\psi)^2$) were not included in the final basis, as
they are always equivalent to combinations of scalar contractions via
the Fierz identity~\eqref{fierz.tensor}. Similarly, operators with the
scalar contraction $(\bar{\psi}_L\psi_R)(\bar{\psi}_R\psi_L)$ have
been traded for terms with the vector structure $(\bar{\psi}_L\g_\mu
\psi_L)(\bar{\psi}_R\g^\mu \psi_R)$ employing the identity~\eqref{fierz.scalar}. 

\subsubsection*{Four-quark (lepton) operators} 
\begin{itemize}
 \item There are four independent $SU(2)$ contractions of four quarks that can be constructed with the scalar structure $(\bar{\psi}_L\psi_R)(\bar{\psi}_L\psi_R)$. They are easily identified in unitary gauge by the $U(1)_\text{em}$ invariants
$$
(uu)(uu),\qquad (dd)(dd),\qquad (uu)(dd),\qquad(ud)(du)\,.
$$
Keeping colour contractions into account, the total number of independent operators in this category is 8.

With four leptons there is only one invariant with this Lorentz structure, due to the absence of right-handed neutrinos: $(ee)(ee)$.
 
We do not provide the expressions of all the possible $SU(2)$ structures in terms of the invariants selected for the basis of Sect.~\ref{sec.Lfer}. However, it is worth commenting on two contractions that can be constructed without the explicit insertion of Goldstone bosons: in the four-quark case they are
\begin{equation}\label{oper_eps}
\begin{aligned}
 \ffqName_{\e1}&=\e_{ij}\e_{ab}(\QBL_i\QR_a)(\QBL_j\QR_b)\cF\\
 \ffqName_{\e2}&=\e_{ij}\e_{ab}(\QBL_i\lambda^A\QR_a)(\QBL_j\lambda^A\QR_b)\cF\,.
\end{aligned}
\end{equation} 
In the four-leptons case, it is possible to introduce a structure analogous to the first one, but only in presence of right-handed neutrinos. this would read:
\begin{equation}\label{oper_eps_l}
 \fflName_{\e N} = \e_{ij}\e_{ab}(\LBL_i\LR_a)(\LBL_j\LR_b)\cF\,.
\end{equation} 
The operators of Eqs.~\eqref{oper_eps} and~\eqref{oper_eps_l} are redundant in the basis of Sect.~\ref{sec.Lfer}: in fact, exploiting the properties of the Pauli matrices and the 
completeness relation~\eqref{dec.su2} one has
\begin{equation}\label{Ueps_relation}
 \U_{ia}\U_{jb}-\left(\U\s^k\right)_{ia}\left(\U\s^k\right)_{jb} = 
2\e_{ij}\e_{ab}\left(c_\pi^2+\frac{|\vec{\pi}|^2}{v^2}s_{\pi}^2\right)\,,
\end{equation}
where we have used the compact notation $s_\pi\equiv\sin(|\vec{\pi}|/v)$, 
$c_\pi\equiv\cos(|\vec{\pi}|/v)$. From Eq.~\eqref{Ueps_relation} it follows 
immediately that 
\begin{equation}
\begin{aligned}
 \frac{\Offq{Qlrlr_11}-\Offq{Qlrlr_ss}}{2} &=\ffqName_{\e1}
\left(c_\pi^2+\frac{|\vec{\pi}|^2}{v^2}s_\pi^2\right)
 =\ffqName_{\e1}\left(1-\frac{|\vec{\pi}|^2}{v^2}+\frac{4}{3}
\frac{|\vec{\pi}|^4}{v^4}+\dots\right)\\
 \frac{\Offq{Qlrlr_ll}-\Offq{Qlrlr_lsls}}{2} &= \ffqName_{\e2}
\left(c_\pi^2+\frac{|\vec{\pi}|^2}{v^2}s_\pi^2\right)
 =\ffqName_{\e2}\left(1-\frac{|\vec{\pi}|^2}{v^2}+
\frac{4}{3}\frac{|\vec{\pi}|^4}{v^4}+\dots\right)\\
\frac{\Offl{Llrlr_11}-\Offl{Llrlr_ss}}{2} &= \fflName_{\e N}
\left(c_\pi^2+\frac{|\vec{\pi}|^2}{v^2}s_\pi^2\right)
 =\fflName_{\e N}\left(1-\frac{|\vec{\pi}|^2}{v^2}+\frac{4}{3}
\frac{|\vec{\pi}|^4}{v^4}+\dots\right)
 \end{aligned}
\end{equation} 
Therefore, the interactions contained in $\ffqName_{\e1}$,
$\ffqName_{\e2}$ and $\fflName_{\e N}$ are already described by linear combinations of operators in the
basis.

 \item The class of four-fermion operators with two left-handed currents contains four independent operators in both the four-quarks and four-leptons cases:
 $$
\begin{aligned}
  (\QBL \g_\mu \QL)^2: \quad&	(uu)(uu),\quad (dd)(dd),\quad  (uu)(dd),\quad (ud)(du)\\
  (\LBL\g_\mu \LL)^2: \quad&	(\nu\nu)(\nu\nu),\quad (ee)(ee),\quad  (\nu\nu)(ee),\quad (\nu e)(e\nu)
\end{aligned}
 $$
 Notice that in this case the octet colour contraction $(\QBL \g_\mu \lambda^A \QL)^2$ is not independent. In fact it is equivalent to a combination of invariants with singlet colour contractions. Using Eqs.~\eqref{dec.su3} and~\eqref{fierz.vectorL}:
 \begin{align}
 (\QBL\g_\mu  \lambda^a \QL)(\QBL\g_\mu  \lambda^a \QL)&=
   \frac{1}{3}(\QBL\g_\mu  \QL)(\QBL\g_\mu  \QL)+(\QBL\g_\mu \s^j \QL)(\QBL\g_\mu  \s^j\QL)
\end{align}
  An analogous relation holds for the structures with right-handed currents.
  
\item The class of four-fermion operators with two right-handed currents contains four independent operators in the four-quark case but only one in the four-lepton sector:
 $$
\begin{aligned}
  (\QBR\g_\mu \QR)^2: \quad&	(uu)(uu),\quad (dd)(dd),\quad  (uu)(dd),\quad (ud)(du)\\
  (\LBR\g_\mu \LR)^2: \quad&	(ee)(ee)\\
\end{aligned}
 $$
 
\item Finally, there are five independent $SU(2)$ contractions for quark vector currents of opposite chirality $(\bar{\psi}_L\g_\mu \psi_L)(\bar{\psi}_R\g^\mu \psi_R)$, to be doubled when including octet colour contractions:
 $$
 (uu)(uu),\quad (dd)(dd),\quad (uu)(dd),\quad(dd)(uu),\quad(ud)(du)+(du)(ud)
 $$
 The four-lepton counterpart, instead, contains two invariants corresponding to the interactions
  $$
 (ee)(ee),\quad (\nu\nu)(ee)\,.
 $$
\end{itemize}

\subsubsection*{Mixed quark-lepton operators} 
\begin{itemize}
\item Operators with the scalar contraction $(\bar{\psi}_L\psi_R)(\bar{\psi}_L\psi_R)$ can have either the structure $(\bar{Q}Q)(\bar{L}L)$ or $(\bar{Q}L)(\bar{L}Q)$. Each of these yield three independent invariants, that are most easily identified in unitary gauge by the interactions:
\begin{equation}
\begin{aligned}
  (\bar{Q}Q)(\bar{L}L): \quad&(uu)(ee),\quad (dd)(ee),\quad  (du)(\nu e)\\
  (\bar{Q}L)(\bar{L}Q): \quad&(ue)(eu),\quad (de)(ed),\quad (de)(\nu u)\\
\end{aligned}
\end{equation} 

\item The two combinations $(\bar{Q}_L\g_\mu Q_L)(\bar{L}_L\g^\mu
L_L)$, $(\bar{Q}_L\g_\mu L_L)(\bar{L}_L\g^\mu Q_L)$ are related by the
Fierz identity~\eqref{fierz.vectorL}, and therefore only the former
structure has been retained. The same holds for the analogous terms
constructed with right-handed currents, that are connected by
Eq.~\eqref{fierz.vectorR}.

This class includes five independent left-handed invariants,
identified by the hermitian combinations
$$(uu)(ee),\quad (dd)(ee),\quad  (uu)(\nu\nu),\quad  (dd)(\nu\nu),\quad  (du)(\nu e)+(ud)(e\nu)\,.$$
and two right-handed ones:
$$(uu)(ee),\quad (dd)(ee).$$

\item Operators with one left-handed and one right-handed current can be
constructed in either of the combinations
$(\QBL\g_\mu\QL)(\LBR\g^\mu\LR)$, $(\LBL\g_\mu\LL)(\QBR\g^\mu\QR)$ and
\\ $(\QBL\g_\mu\LL)(\LBR\g^\mu\QR)$. 
These provide, respectively, 2 + 5 + 3 independent interactions:
\begin{equation}
\begin{aligned}
  (\bar{Q}Q)(\bar{L}L): \quad&(uu)(ee),\quad (dd)(ee)\\
  (\bar{L}L)(\bar{Q}Q): \quad&(ee)(uu),\quad (ee)(dd),
\quad  (\nu\nu)(uu),\quad  (\nu\nu)(dd),\quad  (\nu e)(du)+(e\nu)(ud)\\
  (\bar{L}Q)(\bar{Q}L): \quad&(eu)(ue),\quad (ed)(de),\quad  (\nu u)(de)
\end{aligned}
\end{equation} 
\end{itemize}

\section{Application of the EOMs}
\label{APP:EOM}
Given the LO Lagrangian in Eq.~(\ref{Lag0}), the fields satisfy the following 
EOMs:
\begin{align}
&\begin{aligned}
 i\slashed{D}\psi_L = \frac{v}{\sqrt{2}} \U \cY_\psi(h)  \psi_R\\[2mm]
 i\slashed{D}\psi_R= \frac{v}{\sqrt{2}} \cY^\dag_Q(h) \U^\dag  \psi_L 
\label{EOM_psi}
 \end{aligned}\\
 \nn\\
& (D^\mu\WWd)^a = \sum_{\psi=Q,L}\frac{g}{2}\bar{\psi}_L\s^a \g_\nu \psi_L 
+\dfrac{igv^2}{4}\tr[\V_\nu \s^a] \cF_C(h)\label{EOM_W}\\[2mm]
& \de^\mu \BBd =g\ct\sum_{\parbox{12mm}{\scriptsize$i=L,R$\\
$\psi=Q,L$}} \bar{\psi}_i\mathbf{h}_{\psi_i}\g_\nu \psi_i 
-\frac{ig\ct v^2}{4}\tr[\T\V_\mu]\cF_C(h)\label{EOM_B}\\
 \nn\\
&  \square h = - V'(h)-\frac{v^2}{4}\tr[\V_\mu\V^\mu]  
\cF'_C(h)-\sum_{\psi=Q,L}\frac{v}{\sqrt2}\left(\bar\psi_L\U\cY'_\psi(h)
\psi_R+\text{h.c.}\right)\label{EOM_h}
\end{align}
where $\mathbf{h}_{\psi_i}$ are the hypercharges in the $2\times2$ matrix 
notation:
\begin{equation}
 \begin{aligned}
  h_{Q_L}&=\diag\left(1/6,1/6\right)\,,\qquad&	h_{Q_R}&
=\diag\left(2/3,-1/3\right)\,,\\
  h_{L_L}&= \diag\left(-1/2,-1/2\right)\,,&	h_{L_R}&
=\diag\left(0,-1\right)\,,
 \end{aligned}
\end{equation} 
and the prime denotes the first derivative with respect to $h$.
A consequence of Eqs.~\eqref{EOM_W} and~\eqref{EOM_psi} is 
\begin{equation}
 \D_\mu \left(\V^\mu \cF_C\right) = \frac{i}{v^2}D_\mu
\left(\sum_{\psi=Q,L}\bar{\psi}_L\s^j \g^\mu \psi_L\right)\s^j=
 \frac{1}{\sqrt2 v}\sum_{\psi=Q,L}\left(\bar\psi_L\s^j\U\cY_\psi(h) \psi_R
-\bar\psi_R\cY^\dag_\psi(h)\U^\dag\s^j \psi_L\right)\s^j
\end{equation} 
which can be recast in the form
\begin{equation}\label{DmuVmu}
\tr(\s^j\D_\mu\V^\mu)\cF(h) =
\frac{\sqrt2}{v}\sum_{\psi=Q,L}\left(\bar\psi_L\s^j\U\cY_\psi(h) \psi_R
-\bar\psi_R\cY^\dag_\psi(h)\U^\dag\s^j \psi_L\right)-\tr(\s^j\V_\mu)
\de^\mu\cF(h)\,,
\end{equation}
which is valid order by order in the $h$ expansion.
\newpage
\subsection*{Operators that have been removed via EOM}
The EOMs relate the purely bosonic and the fermionic
sectors, and they have been used to eliminate operators that are
redundant when both sectors are considered at the same time. In this
section we list the categories of operators that have been removed.

\subsubsection*{Bosonic sector}
\begin{itemize}
 \item Operators containing $\square\cF(h)$.\hfill
 
 Applying the EOM for the Higgs, Eq.~\eqref{EOM_h}, these terms can
 be traded for a combination of other bosonic operators plus fermionic 
bilinears and four-fermion operators. 
 The following CP even terms have been removed, compared to the basis 
of Ref.~\cite{Brivio:2013pma}:
\begin{equation}
\begin{aligned}
 \cP_{\square H}(h) &= 
 \frac{\square h \square h}{v^2}\cF \\
 \cP_{7}(h) &= \tr(\V_\mu\V^\mu)\square \cF \\
 \cP_{25}(h)&=\tr(\T\V_\mu)\tr(\T\V^\mu)\square\cF\\
 \end{aligned}
\end{equation}
and the CP odd operator
\begin{equation}
 \cS_{13} = \tr(\T\V_\mu)\de^\mu\square\cF
\end{equation}

\item Operators containing $\D_\mu\V^\mu$.\hfill

Rewriting the traceless matrix $\D_\mu\V^\mu$ as
 \begin{equation}
  \D_\mu\V^\mu = \frac{\s^a}{2}\tr(\s^a\D_\mu\V^\mu)
 \end{equation}
  and applying the identity~\eqref{DmuVmu}, these bosonic operators
can be traded by combinations of fermion bilinears, four-fermion
operators and other bosonic terms that already belong to the
basis. The following CP even terms have been eliminated, in the
notation of Ref.~\cite{Brivio:2013pma}:
\begin{equation}
\begin{aligned}
 \cP_{9}(h) &= \tr((\D_\mu\V^\mu)^2)\cF \\
 \cP_{10}(h) &= \tr(\V_\nu\D_\mu\V^\mu)\de^\nu \cF\\
 \cP_{15}(h)&=\tr(\T\D_\mu\V^\mu)\tr(\T\D_\nu\V^\nu)\cF\\
 \cP_{16}(h)&=\tr([\T,\V_\nu]\D_\mu\V^\mu)\tr(\T\V^\nu)\cF\\
 \cP_{19}(h) &= \tr(\T\D_\mu\V^\mu)\tr(\T\V_\nu)\de^\nu\cF\,.
 \end{aligned}
\end{equation}
Analogously, five CP odd operators have been traded for others: 
in the notation of Ref.~\cite{Gavela:2014vra} they are
\begin{equation}
 \begin{aligned}
 \mathcal{S}_{10} &= i\tr(\V_\nu\D_\mu\V^\mu)\tr(\T\V^\nu)\cF \\
 \mathcal{S}_{11} &= i\tr(\T\D_\mu\V^\mu)\tr(\V_\nu\V^\nu)\cF \\
 \mathcal{S}_{12} &= i\tr([\V^\mu,\T]\D_\nu\V^\nu)\de_\mu\cF \\
 \mathcal{S}_{14} &= i\tr(\T\D_\mu\V^\mu)\de_\nu\cF(h)\de^\nu\cF'  \\
 \mathcal{S}_{16} &= i\tr(\T\D_\mu\V^\mu)\tr(\T\V_\nu)\tr(\T\V^\nu)\cF\,.
\end{aligned}
\end{equation}

\end{itemize}

\subsubsection*{Fermionic sector}
\begin{itemize}
\item Bilinears of the type $\bar{\psi}\Gamma\g_\mu\psi\,\de^\mu\cF$.\hfill

 Applying the EOMs for fermions (Eq.~\eqref{EOM_psi}), these operators can 
be schematically rewritten as
 \beq
 \begin{aligned}
  \bar{\psi}\Gamma\g_\mu\psi\de^\mu\cF &= 
-\bar{\psi}\overleftarrow{\slashed{D}}\Gamma\psi\cF-\bar{\psi}\g_\mu 
(D^\mu\Gamma)\psi\cF-\bar{\psi}\Gamma\slashed{D}\psi\cF\\
  &\to\bar{\psi}\g_\mu (D^\mu\Gamma)\psi\cF \bar{\psi}\Gamma\psi\cF
 \end{aligned}
 \eeq

\item  Bilinears containing $\square\cF$.\hfill

 Operators in this category are removed applying the EOM for the Higgs
field, Eq.~\eqref{EOM_h} and traded for other bilinears plus
four-fermion operators.

\item {\boldmath invariants containing $\D_\mu\V^\mu$}\hfill

 As in the bosonic sector, these operators are removed applying the
identity~\eqref{DmuVmu}. and traded for other bilinears plus
four-fermion operators.

\item Finally, the EOMs for the gauge
(Eqs.~\eqref{EOM_W},~\eqref{EOM_B}) and Higgs (Eq.~\eqref{EOM_h})
fields imply the following additional relations (signs and numeric
coefficients not specified):
\begin{align}
 \cP_B+\cP_1+\cP_2+\cP_4+\cP_T\to\,&i\bar L_{L_i} \g_\mu \{\V^\mu,\T\} L_{L_i} \cF+
\Otfq{Qll_avt}+\Otfq{Qrr_avt}\nn\\
 \cP_W+\cP_1+\cP_3+\cP_5+\tr(\V_\mu\V^\mu)\cF\to\,& i\bar L_{L_i} \g_\mu \V^\mu L_{L_i} \cF
+\Otfq{Qll_v}
\label{EOMForLeptonicDominantOperators}\\
 \cP_T+\cP_1+\cP_3+\cP_{12}+\cP_{13}+\cP_{17}\to\,& 
i\bar L_{L_i} \g_\mu \T\V^\mu\T L_{L_i} \cF+i\bar L_{L_i} \g_\mu \V^\mu L_{L_i} \cF 
+\Otfq{Qll_tvt}+\Otfq{Qll_v}\,.\nn
\end{align}
These have been employed to remove the three (flavour-diagonal contractions of the) leptonic operators
specified on the right-hand side. This choice simplifies the renormalisation procedure.
\end{itemize}

\section{Feynman rules}\label{APP:FR}
This appendix provides a complete list of all the Feynman rules resulting from both fermionic and bosonic operators considered in the present work and listed in Sections~\ref{sec.L_bos} and~\ref{sec.Lfer}. For compactness we omit CP violating terms, that are not relevant for the phenomenological study presented. 
The rules are derived in unitary gauge and only vertices with up to four legs are shown.
The SM contribution and the renormalization effects are also included, up to first order in the effective coefficients. The latter are sometimes encoded in the quantities $\Delta g_1$, $\Delta g_2$, $\Delta g_W$ and $\Delta M_W$ defined in Eqs.~\eqref{delta_g_s},~\eqref{d_mW} in the text. 

A few comments about the notation and conventions used:
\begin{itemize}
 \item All momenta are flowing inwards and the convention $\de_\mu\to -ip_\mu$ has been used in the derivation.
 \item We use a shorthand notation for the products $c_ia_i$: for the bosonic operators, we replace $a_i c_i\to a_i$ and $b_i c_i\to a_i$.
 For the fermionic operators, we write $a^f_i n^f_i\to (na)^f_i$. The structure $\bar{\psi}\psi\de\cF\de\cF^\prime$  gives couplings $hhff$ with the coefficients $n^f_i a^f_i a^{\prime f}_i$. This notation has been shortened in $ (naa')^f_i$.
 For the coefficients of the function $\cF_C(h)$, defined in Eq.~\eqref{FC}, the notation $ a_C = 1+\Delta a_C$, $b_C= 1+ \Delta b_C$ is adopted.
\item We have fixed $V_\text{CKM} =\unity$ for compactness. At the same level, all the effective coefficients are implicitly taken to be flavor-diagonal.
\item In the vertices with a single fermion current the spin contractions are obvious. 
For those with four fermions we use a notation with square brackets and lowercase indices: for example $[P_R]_{ab}[P_L]_{cd}$ means that the right chirality projector contracts the spins of the $a$ and $b$ particle, and the left chirality one shall be inserted between the $c$ and $d$ fields.
Note that, in four-fermion vertices, negative signs for the Wick contractions that require commuting the fermionic creation and annihilation operators are not indicated.
\item Uppercase indices indicate color and are assumed to be summed over when repeated. Whenever they are not specified, the color (and flavor) contractions go with those of the spin.
\end{itemize}

\clearpage

\begin{landscape}
\scriptsize

\addtolength{\hoffset}{-8mm}
\addtolength{\voffset}{1cm}

\newcommand{\nr}{\stepcounter{diagram}(FR.\arabic{diagram})}
\newcommand{\Anr}{\stepcounter{diagramDV}(A.\arabic{diagramDV})}
\addtolength{\linewidth}{3cm}
\newcounter{diagram}
\renewcommand{\arraystretch}{6}
\renewcommand{\headrulewidth}{0pt}

\subsection*{FR: propagators}
\begin{center}
\begin{tabular}{c@{\hspace*{2mm}}p{4cm}>{$}l<{$}}

\nr& \parbox{4cm}{\centering
\begin{fmffile}{Fdiagrams/Aprop}
\begin{fmfgraph*}(35,25)

\fmfleft{i1}
\fmfright{o1}
  \fmf{boson,label=$\gamma$}{i1,o1}
  
\fmfv{lab=$\nu$}{o1}
\fmfv{lab=$\mu$}{i1}
\end{fmfgraph*}
\end{fmffile}
}   	& \dfrac{-i}{ p^2}\left[g^{\mu\nu}-(1-\eta)\dfrac{p^\mu p^\nu}{p^2}\right]\quad(\eta \text{ is the gauge fixing parameter})\\

\nr&  \parbox{4cm}{\centering
\begin{fmffile}{Fdiagrams/Zprop}
\begin{fmfgraph*}(35,25)

\fmfleft{i1}
\fmfright{o1}
  \fmf{boson,label=$Z$}{i1,o1}
  
\fmfv{lab=$\nu$}{o1}
\fmfv{lab=$\mu$}{i1}
\end{fmfgraph*}
\end{fmffile}
}	& \dfrac{-i}{p^2-m_Z^2}\left[g^{\mu\nu}-\dfrac{p^\mu p^\nu}{m_Z^2}\right]	\\

\nr&  \parbox{4cm}{\centering
\begin{fmffile}{Fdiagrams/Wprop}
\begin{fmfgraph*}(35,25)

\fmfleft{i1}
\fmfright{o1}
  \fmf{boson,label=$W$}{i1,o1}
  
\fmfv{lab=$\nu$}{o1}
\fmfv{lab=$\mu$}{i1}
\end{fmfgraph*}
\end{fmffile}
}	& \dfrac{-i}{p^2-M_W^2}\left[g^{\mu\nu}-\dfrac{p^\mu p^\nu}{M_W^2}\right];\quad 
  M_W^2 =  m_Z^2\ct^2\left(1+\dfrac{\Delta M_W^2}{M_W^2}\right)= m_Z^2\ct^2 \left(1+2t_{2\theta} c_1+\dfrac{2c_T \ct^2 }{\cdt}-4 c_{12}+\dfrac{64\sqrt2\pi^2\st^2}{\cdt}\dfrac{v^2}{\Lambda^2}(\Cffl{Lllll_11}-\Cffl{Lllll_tt})\right)
  \\

\nr& \parbox{4cm}{\centering
\begin{fmffile}{Fdiagrams/Gprop}
\begin{fmfgraph*}(35,25)
\fmfset{curly_len}{5}
\fmfleft{i1}
\fmfright{o1}
  \fmf{gluon,label=$G$,l.side=left}{i1,o1}
  
\fmfv{lab=$\nu$}{o1}
\fmfv{lab=$\mu$}{i1}
\end{fmfgraph*}
\end{fmffile}
}	& \dfrac{-ig^{\mu\nu}}{ p^2}\left[g^{\mu\nu}-(1-\eta)\dfrac{p^\mu p^\nu}{p^2}\right]\\
 
\nr& \parbox{4cm}{\centering
\begin{fmffile}{Fdiagrams/hprop}
\begin{fmfgraph*}(35,25)

\fmfleft{i1}
\fmfright{o1}
  \fmf{dashes,label=$h$}{i1,o1}
  
\end{fmfgraph*}
\end{fmffile}
}	& \dfrac{-i}{ p^2-m_h^2} \\
\end{tabular}
\end{center}

\newpage
\subsection*{FR: Bosonic}
\begin{center}
\begin{longtable}{m{1cm}m{3.7cm}>{$}l<{$}}
\nr&  \parbox{4cm}{\centering
\begin{fmffile}{Fdiagrams/wwacoupl}
\begin{fmfgraph*}(35,25)

\fmfleft{i1,i2,i3}
\fmfright{o3,o2,o1}
  \fmf{boson}{i2,v1}
  \fmf{boson}{v1,o1}
  \fmf{boson}{v1,o3}
  
\fmfv{lab=$W^+_\mu$}{o1}
\fmfv{lab=$W^-_\nu$}{o3}
\fmfv{lab=$A_\rho$}{i2}
\end{fmfgraph*}
\end{fmffile}
} &   \begin{array}{l}ie\Big[g_{\mu\nu}(p_+-p_-)_\rho-g_{\nu\rho}p_{-\mu}+g_{\mu\rho}p_{+\nu}-(g_{\nu\rho} p_{A\mu}-g_{\mu\rho}p_{A\nu})\left(1- 2\left(\dfrac{c_1}{t_\theta}+2c_{12}\right)  +\dfrac{g}{4\pi}\left(\dfrac{2c_2}{t_\theta}+c_3+2c_{13}\right)\right)\Big] +\\[-1cm]
-\dfrac{24i\pi\st}{\Lambda^2} c_{WWW}\Big[g_{\rho\mu}\left((p_+\cdot p_-)p_{A\nu}-(p_A\cdot p_-) p_{+\nu}\right) +g_{\mu\nu}\left((p_A\cdot p_- )p_{+\rho}-(p_A\cdot p_+)p_{-\rho}\right)+ g_{\rho\nu}\left((p_A\cdot p_+) p_{-\mu}-(p_+\cdot p_-)p_{A\mu}\right)+\\[-1.2cm]
+p_{A\mu}p_{+\nu}p_{-\rho}-p_{A\nu}p_{+\rho}p_{-\mu}\Big]
\end{array}\\
\end{longtable}

\end{center}

\clearpage
\vspace*{-7mm}
\subsection*{FR: Fermionic}
\subsubsection*{Single quark current}
\begin{center}
 \vspace*{-5mm}
\begin{longtable}{m{1cm}m{3.7cm}>{$}l<{$}}
\nr&  \parbox{4cm}{\centering\begin{fmffile}{Fdiagrams/hddcoupl}
\begin{fmfgraph*}(35,25)
\fmfset{arrow_len}{2mm}
\fmfleft{i1,i2,i3}
\fmfright{o3,o2,o1}
  \fmf{dashes}{i2,v1}
  \fmf{fermion}{o1,v1}
  \fmf{fermion}{v1,o3}
  
\fmfv{lab=$h$}{i2}
\fmfv{lab=$d$}{o1,o3}
\end{fmfgraph*}
\end{fmffile}
} & -\dfrac{i}{\sqrt2}Y^{(1)}_D \\

\nr&  \parbox{4cm}{\centering\begin{fmffile}{Fdiagrams/huucoupl}
\begin{fmfgraph*}(35,25)
\fmfset{arrow_len}{2mm}
\fmfleft{i1,i2,i3}
\fmfright{o3,o2,o1}
  \fmf{dashes}{i2,v1}
  \fmf{fermion}{o1,v1}
  \fmf{fermion}{v1,o3}
  
\fmfv{lab=$h$}{i2}
\fmfv{lab=$u$}{o1,o3}
\end{fmfgraph*}
\end{fmffile}
} & -\dfrac{i}{\sqrt2}Y^{(1)}_U\\

\nr&  \parbox{4cm}{\centering\begin{fmffile}{Fdiagrams/zddcoupl}
\begin{fmfgraph*}(35,25)
\fmfset{arrow_len}{2mm}
\fmfleft{i1,i2,i3}
\fmfright{o3,o2,o1}
  \fmf{boson}{i2,v1}
  \fmf{fermion}{o1,v1}
  \fmf{fermion}{v1,o3}
  
\fmfv{lab=$Z^\mu$}{i2}
\fmfv{lab=$d$}{o1,o3}
\end{fmfgraph*}
\end{fmffile}
}	&{\Sas \begin{array}{ll}    
					       -\dfrac{ig}{\ct}\Bigg[\left(-\dfrac12+\dfrac{\st^2}{3}\right)\left(1+\Delta g_1+\dfrac{2}{3-2\st^2}\Delta g_2
					       +\dfrac{6}{3-2\st^2}(\Ctfq{Qll_v}-2\Ctfq{Qll_avt}+\Ctfq{Qll_tvt})\right)\g_\mu P_L
					       +\dfrac{\st^2}{3}\left(1+\Delta g_1-\dfrac{1}{\st^2}\Delta g_2-\dfrac{3}{\st^2}(\Ctfq{Qrr_v}-2\Ctfq{Qrr_avt}+\Ctfq{Qrr_tvt})\right)\g^\mu P_R+\\
					       -4\pi \ct^2 \left[(\Ctfq{Qlr_s_w}-2\Ctfq{Qlr_s_atw}+\Ctfq{Qlr_s_twt})+2(\Ctfq{Qlr_s_b}-\Ctfq{Qlr_s_tb})t^2_\theta\right]\, \dfrac{p_{Z\nu}}{\Lambda}\; \ssu\g_5  \Bigg]
                                                   	       \end{array}}\\
\end{longtable}
\end{center}
\clearpage
\subsubsection*{Single lepton current}
 \begin{center}
\vspace*{-5mm}
 \begin{longtable}{m{1cm}m{3.7cm}>{$}l<{$}}
 
\nr&  \parbox{4cm}{\centering\begin{fmffile}{Fdiagrams/hllcoupl}
\begin{fmfgraph*}(35,25)
\fmfset{arrow_len}{2mm}
\fmfleft{i1,i2,i3}
\fmfright{o3,o2,o1}
  \fmf{dashes}{i2,v1}
  \fmf{fermion}{o1,v1}
  \fmf{fermion}{v1,o3}
  
\fmfv{lab=$h$}{i2}
\fmfv{lab=$l^-$}{o3}
\fmfv{lab=$l^+$}{o1}
\end{fmfgraph*}
\end{fmffile}
} & -\dfrac{i}{\sqrt2}Y_E^{(1)} \\

\nr&  \parbox{4cm}{\centering\begin{fmffile}{Fdiagrams/zllcoupl}
\begin{fmfgraph*}(35,25)
\fmfset{arrow_len}{2mm}
\fmfleft{i1,i2,i3}
\fmfright{o3,o2,o1}
  \fmf{boson}{i2,v1}
  \fmf{fermion}{o1,v1}
  \fmf{fermion}{v1,o3}
  
\fmfv{lab=$Z^\mu$}{i2}
\fmfv{lab=$l^-$}{o3}
\fmfv{lab=$l^+$}{o1}
\end{fmfgraph*}
\end{fmffile}
}	& {\Sas \begin{array}{ll}   
						-\dfrac{ig}{\ct}\Bigg[\left(-\dfrac12+\st^2\right)\left(1+\Delta g_1+\dfrac{2}{1-2\st^2}\Delta g_2\right)\g_\mu P_L
						+\st^2\left(1+\Delta g_1-\dfrac{1}{\st^2}\Delta g_2 +\dfrac{2}{\st^2}\Ctfl{Lrr_avt}\right) \g_\mu P_R\\
					        -4\pi\ct^2\left[\Ctfl{Llr_s_w}+2\Ctfl{Llr_s_b}t^2_\theta\right]\, \dfrac{p_{Z\nu}}{\Lambda}\; \ssu\g_5  \Bigg]
                                                   	       \end{array}}\\

\nr&  \parbox{4cm}{\centering\begin{fmffile}{Fdiagrams/znncoupl}
\begin{fmfgraph*}(35,25)
\fmfset{arrow_len}{2mm}
\fmfleft{i1,i2,i3}
\fmfright{o3,o2,o1}
  \fmf{boson}{i2,v1}
  \fmf{fermion}{o1,v1}
  \fmf{fermion}{v1,o3}
  
\fmfv{lab=$Z^\nu$}{i2}
\fmfv{lab=$\nu$}{o1,o3}
\end{fmfgraph*}
\end{fmffile}
}	& -\dfrac{ig}{2}\g_\mu P_L\left(1+\Delta g_1\right)\\
\end{longtable}
\end{center} 

\clearpage
\subsubsection*{Four quarks}
\renewcommand{\arraystretch}{7}
\begin{center}
\begin{tabular}{m{1cm}m{3.7cm}>{$}l<{$}}

\nr&  \parbox{4cm}{\centering\begin{fmffile}{Fdiagrams/uuuucoupl}
\begin{fmfgraph*}(35,25)
\fmfset{arrow_len}{2mm}
\fmfleft{i1,i2,i3,i4}
\fmfright{o4,o3,o2,o1}
  \fmf{fermion}{i1,v1,i4}
  \fmf{fermion}{o4,v1,o1}
  
\fmfv{lab=$u_{a}^A$}{i4}
\fmfv{lab=$u_{b}^B$}{i1}
\fmfv{lab=$u_{d}^D$}{o1}
\fmfv{lab=$u_{e}^E$}{o4}
\end{fmfgraph*}
\end{fmffile}
} &  {\Sasp\begin{array}{l}
							\dfrac{16\pi^2}{\Lambda^2}\Bigg[
                                                        2i (\Cffq{Qlrlr_11}+\Cffq{Qlrlr_ss}+\Cffq{Qlrlr_t}+\Cffq{Qlrlr_tt})([P_L]_{ab}[P_L]_{de}+[P_R]_{ab}[P_R]_{de}+[P_L]_{ae}[P_L]_{db}+[P_R]_{ae}[P_R]_{db})+\\
                                                        +4i(\Cffq{Qrrrr_11}+\Cffq{Qrrrr_1t}+\Cffq{Qrrrr_tt}+\Cffq{Qrrrr_ss}) ( [\g^\mu P_R]_{ab}\,[\g_\mu P_R]_{de}+[\g^\mu P_R]_{ae}\,[\g_\mu P_R]_{db} ) 
                                                        +4i(\Cffq{Qllll_11}+\Cffq{Qllll_1t}+\Cffq{Qllll_tt}+r^Q_{9}) ( [\g^\mu P_L]_{ab}\,[\g_\mu P_L]_{de}+[\g^\mu P_L]_{ae}\,[\g_\mu P_L]_{db} ) +\\
                                                        +2i(\Cffq{Qllrr_11}+\Cffq{Qllrr_1t}+\Cffq{Qllrr_t1}+\Cffq{Qllrr_tt}+\Cffq{Qllrr_ss}) ( [\g^\mu P_L]_{ab}\,[\g_\mu P_R]_{de}+[\g^\mu P_R]_{ab}\,[\g_\mu P_L]_{de}+
                                                         [\g^\mu P_L]_{ae}\,[\g_\mu P_R]_{db}+[\g^\mu P_R]_{ae}\,[\g_\mu P_L]_{db} ) +\\
                                                        +8i(\Cffq{Qlrlr_ll}+\Cffq{Qlrlr_lsls}+\Cffq{Qlrlr_llt}+\Cffq{Qlrlr_ltlt})\Big[
                                                         \lambda^a_{AB}\lambda^a_{DE} \,([P_L]_{ab}\,[P_L]_{de}+[P_R]_{ab}\,[P_R]_{de})+
                                                        \lambda^a_{AE}\lambda^a_{DB} \,([P_L]_{ae}\,[P_L]_{db}+[P_R]_{ae}\,[P_R]_{db})\Big]+\\
                                                        +8i(\Cffq{Qllrr_ll}+\Cffq{Qllrr_llt}+\Cffq{Qllrr_ltl}+\Cffq{Qllrr_ltlt}+\Cffq{Qllrr_lsls})\Big[
                                                         \lambda^a_{AB}\lambda^a_{DE} \,([\g^\mu P_L]_{ab}\,[\g^\mu P_R]_{de}+[\g^\mu P_R]_{ab}\,[\g^\mu P_L]_{de})
                                                         +\lambda^a_{AE}\lambda^a_{DB} \,([\g^\mu P_L]_{ae}\,[\g^\mu P_R]_{db}+[\g^\mu P_R]_{ae}\,[\g^\mu P_L]_{db})\Big]\Bigg]
                                                           \end{array}}\\
\end{tabular}
\end{center}

\clearpage
\subsubsection*{Four leptons}
\begin{center} 
\begin{tabular}{m{1cm}m{3.7cm}>{$}l<{$}}
 
\nr&  \parbox{4cm}{\centering\begin{fmffile}{Fdiagrams/llllcoupl}
\begin{fmfgraph*}(35,25)
\fmfset{arrow_len}{2mm}
\fmfleft{i1,i2,i3,i4}
\fmfright{o4,o3,o2,o1}
  \fmf{fermion}{i1,v1,i4}
  \fmf{fermion}{o4,v1,o1}
  
\fmfv{lab=$l_{a}^+$}{i4}
\fmfv{lab=$l_{b}^-$}{i1}
\fmfv{lab=$l_{d}^-$}{o1}
\fmfv{lab=$l_{e}^+$}{o4}
\end{fmfgraph*}
\end{fmffile}
} &{\Sas\begin{array}{l}
							 \dfrac{16\pi^2}{\Lambda^2}\Bigg[
                                                         2i \Cffl{Llrlr_11}( [P_L]_{ab}[P_L]_{de}+[P_L]_{ae}[P_L]_{db} +[P_R]_{ab}[P_R]_{de}+[P_R]_{ae}[P_R]_{db} ) +\\
                                                         +4i\Cffl{Lrrrr_11} ( [\g^\mu P_R]_{ab}\,[\g_\mu P_R]_{de}+[\g^\mu P_R]_{ae}\,[\g_\mu P_R]_{db} ) 
                                                         +4i(\Cffl{Lllll_11}-\Cffl{Lllll_1t}+\Cffl{Lllll_tt}) ( [\g^\mu P_L]_{ab}\,[\g_\mu P_L]_{de}+[\g^\mu P_L]_{ae}\,[\g_\mu P_L]_{db} ) +\\
                                                         +2i(\Cffl{Lllrr_11}-\Cffl{Lllrr_t1}) ( [\g^\mu P_L]_{ab}\,[\g_\mu P_R]_{de}+[\g^\mu P_R]_{ab}\,[\g_\mu P_L]_{de}+
                                                         [\g^\mu P_L]_{ae}\,[\g_\mu P_R]_{db}+[\g^\mu P_R]_{ae}\,[\g_\mu P_L]_{db} ) 
                                                         \dfrac{1}{\Lambda^2}\Bigg]
                                                            \end{array}}\\

\nr&  \parbox{4cm}{\centering\begin{fmffile}{Fdiagrams/nnnncoupl}
\begin{fmfgraph*}(35,25)
\fmfset{arrow_len}{2mm}
\fmfleft{i1,i2,i3,i4}
\fmfright{o4,o3,o2,o1}
  \fmf{fermion}{i1,v1,i4}
  \fmf{fermion}{o4,v1,o1}
  
\fmfv{lab=$\nu_{a}$}{i4}
\fmfv{lab=$\nu_{b}$}{i1}
\fmfv{lab=$\nu_{d}$}{o1}
\fmfv{lab=$\nu_{e}$}{o4}
\end{fmfgraph*}
\end{fmffile}
} & \dfrac{16\pi^2}{\Lambda^2}\Bigg[4i(\Cffl{Lllll_11}+\Cffl{Lllll_1t}+\Cffl{Lllll_tt}) ( [\g^\mu P_L]_{ab}\,[\g_\mu P_L]_{de}+[\g^\mu P_L]_{ae}\,[\g_\mu P_L]_{db} ) \Bigg] \\
\end{tabular}
\end{center}

\subsubsection*{Two quark-two leptons}

\begin{center} 
\begin{tabular}{m{1cm}m{3.7cm}>{$}l<{$}}
 
\nr&  \parbox{4cm}{\centering\begin{fmffile}{Fdiagrams/uullcoupl}
\begin{fmfgraph*}(35,25)
\fmfset{arrow_len}{2mm}
\fmfleft{i1,i2,i3,i4}
\fmfright{o4,o3,o2,o1}
  \fmf{fermion}{i1,v1,i4}
  \fmf{fermion}{o4,v1,o1}
  
\fmfv{lab=$u_{a}$}{i4}
\fmfv{lab=$u_{b}$}{i1}
\fmfv{lab=$l_{d}^-$}{o1}
\fmfv{lab=$l_{e}^+$}{o4}
\end{fmfgraph*}
\end{fmffile}
} &  {\Sas\begin{array}{l}
							\dfrac{16\pi^2}{\Lambda^2}\Bigg[
                                                        i(\Cffql{QLlrlr_11}+\Cffql{QLlrlr_t1}-\Cffql{QLlrlr_ss}) ( [P_L]_{ab}\,[P_L]_{de}+[P_R]_{ab}\,[P_R]_{de})+\\
                                                        +2i(\Cffql{QLl_11}+\Cffql{QLl_1t}-\Cffql{QLl_t1}-\Cffql{QLl_tt}-\Cffql{QLl_ss})[\g^\mu P_L]_{ab}\,[\g_\mu P_L]_{de}
						        +2i(\Cffql{QLr_11}+\Cffql{QLr_1t})[\g^\mu P_R]_{ab}\,[\g_\mu P_R]_{de}+\\
						        +2i(\Cffql{QLrrll_11}+\Cffql{QLrrll_t1})[\g^\mu P_L]_{ab}\,[\g_\mu P_R]_{de}
						        +2i(\Cffql{QLllrr_11}-\Cffql{QLllrr_t1}+\Cffql{QLllrr_1t}-\Cffql{QLllrr_tt}-\Cffql{QLllrr_ss})[\g^\mu P_R]_{ab}\,[\g_\mu P_L]_{de}+\\
						        +2i\Cffql{QLlrlrmix_ss}([P_L]_{ae}[P_L]_{db} +[P_R]_{ae}[P_R]_{db})  
                                                        +i \Cffql{QLrrllmix_ss}([\g^\mu P_L]_{ae}\,[\g_\mu P_R]_{db}+[\g^\mu P_R]_{ae}\,[\g_\mu P_L]_{db})  
                                                        \Bigg]
                                                        \end{array}}\\
\end{tabular}
\end{center}
\end{landscape}


\newpage
\providecommand{\href}[2]{#2}\begingroup\raggedright\endgroup


\begin{thebibliography}{10}

\bibitem{Aad:2012tfa}
{\bf ATLAS} Collaboration, G.~Aad {\em et.~al.}, {\it {Observation of a New
  Particle in the Search for the Standard Model Higgs Boson with the ATLAS
  Detector at the LHC}},  Phys. Lett. {\bf B716} (2012) 1--29,
  [\href{http://arxiv.org/abs/1207.7214}{{\tt arXiv:1207.7214}}].

\bibitem{Chatrchyan:2012xdj}
{\bf CMS} Collaboration, S.~Chatrchyan {\em et.~al.}, {\it {Observation of a
  New Boson at a Mass of 125 GeV with the CMS Experiment at the LHC}},  Phys.
  Lett. {\bf B716} (2012) 30--61, [\href{http://arxiv.org/abs/1207.7235}{{\tt
  arXiv:1207.7235}}].

\bibitem{Englert:1964et}
F.~Englert and R.~Brout, {\it {Broken Symmetry and the Mass of Gauge Vector
  Mesons}},  Phys. Rev. Lett. {\bf 13} (1964) 321--323.

\bibitem{Higgs:1964ia}
P.~W. Higgs, {\it {Broken Symmetries, Massless Particles and Gauge Fields}},
  Phys. Lett. {\bf 12} (1964) 132--133.

\bibitem{Higgs:1964pj}
P.~W. Higgs, {\it {Broken Symmetries and the Masses of Gauge Bosons}},  Phys.
  Rev. Lett. {\bf 13} (1964) 508--509.

\bibitem{Buchmuller:1985jz}
W.~Buchmuller and D.~Wyler, {\it {Effective Lagrangian Analysis of New
  Interactions and Flavor Conservation}},  Nucl. Phys. {\bf B268} (1986)
  621--653.

\bibitem{Grzadkowski:2010es}
B.~Grzadkowski, M.~Iskrzynski, M.~Misiak, and J.~Rosiek, {\it {Dimension-Six
  Terms in the Standard Model Lagrangian}},  JHEP {\bf 10} (2010) 085,
  [\href{http://arxiv.org/abs/1008.4884}{{\tt arXiv:1008.4884}}].

\bibitem{Kaplan:1983fs}
D.~B. Kaplan and H.~Georgi, {\it {$SU(2)\times U(1)$ Breaking by Vacuum
  Misalignment}},  Phys. Lett. {\bf B136} (1984) 183.

\bibitem{Kaplan:1983sm}
D.~B. Kaplan, H.~Georgi, and S.~Dimopoulos, {\it {Composite Higgs Scalars}},
  Phys. Lett. {\bf B136} (1984) 187.

\bibitem{Banks:1984gj}
T.~Banks, {\it {Constraints on $SU(2)\times U(1)$ Breaking by Vacuum
  Misalignment}},  Nucl. Phys. {\bf B243} (1984) 125.

\bibitem{Agashe:2004rs}
K.~Agashe, R.~Contino, and A.~Pomarol, {\it {The Minimal Composite Higgs
  Model}},  Nucl. Phys. {\bf B719} (2005) 165--187,
  [\href{http://arxiv.org/abs/hep-ph/0412089}{{\tt hep-ph/0412089}}].

\bibitem{Gripaios:2009pe}
B.~Gripaios, A.~Pomarol, F.~Riva, and J.~Serra, {\it {Beyond the Minimal
  Composite Higgs Model}},  JHEP {\bf 04} (2009) 070,
  [\href{http://arxiv.org/abs/0902.1483}{{\tt arXiv:0902.1483}}].

\bibitem{Halyo:1991pc}
E.~Halyo, {\it {Technidilaton Or Higgs?}},  Mod. Phys. Lett. {\bf A8} (1993)
  275--284.

\bibitem{Goldberger:2008zz}
W.~D. Goldberger, B.~Grinstein, and W.~Skiba, {\it {Distinguishing the Higgs
  Boson from the Dilaton at the Large Hadron Collider}},  Phys. Rev. Lett. {\bf
  100} (2008) 111802, [\href{http://arxiv.org/abs/0708.1463}{{\tt
  arXiv:0708.1463}}].

\bibitem{Appelquist:1980vg}
T.~Appelquist and C.~W. Bernard, {\it {Strongly Interacting Higgs Bosons}},
  Phys. Rev. {\bf D22} (1980) 200.

\bibitem{Longhitano:1980iz}
A.~C. Longhitano, {\it {Heavy Higgs Bosons in the Weinberg-Salam Model}},
  Phys. Rev. {\bf D22} (1980) 1166.

\bibitem{Longhitano:1980tm}
A.~C. Longhitano, {\it {Low-Energy Impact of a Heavy Higgs Boson Sector}},
  Nucl. Phys. {\bf B188} (1981) 118.

\bibitem{Feruglio:1992wf}
F.~Feruglio, {\it {The Chiral Approach to the Electroweak Interactions}},  Int.
  J. Mod. Phys. {\bf A8} (1993) 4937--4972,
  [\href{http://arxiv.org/abs/hep-ph/9301281}{{\tt hep-ph/9301281}}].

\bibitem{Weinberg:1978kz}
S.~Weinberg, {\it {Phenomenological Lagrangians}},  Physica {\bf A96} (1979)
  327.

\bibitem{Grinstein:2007iv}
B.~Grinstein and M.~Trott, {\it {A Higgs-Higgs Bound State Due to New Physics
  at a TeV}},  Phys. Rev. {\bf D76} (2007) 073002,
  [\href{http://arxiv.org/abs/0704.1505}{{\tt arXiv:0704.1505}}].

\bibitem{Contino:2010mh}
R.~Contino, C.~Grojean, M.~Moretti, F.~Piccinini, and R.~Rattazzi, {\it {Strong
  Double Higgs Production at the Lhc}},  JHEP {\bf 05} (2010) 089,
  [\href{http://arxiv.org/abs/1002.1011}{{\tt arXiv:1002.1011}}].

\bibitem{Alonso:2012px}
R.~Alonso, M.~B. Gavela, L.~Merlo, S.~Rigolin, and J.~Yepes, {\it {The
  Effective Chiral Lagrangian for a Light Dynamical "Higgs Particle"}},  Phys.
  Lett. {\bf B722} (2013) 330--335, [\href{http://arxiv.org/abs/1212.3305}{{\tt
  arXiv:1212.3305}}]. [Erratum: Phys. Lett.B726,926(2013)].

\bibitem{Alonso:2012pz}
R.~Alonso, M.~B. Gavela, L.~Merlo, S.~Rigolin, and J.~Yepes, {\it {Flavor with
  a Light Dynamical "Higgs Particle"}},  Phys. Rev. {\bf D87} (2013), no.~5
  055019, [\href{http://arxiv.org/abs/1212.3307}{{\tt arXiv:1212.3307}}].

\bibitem{Brivio:2013pma}
I.~Brivio, T.~Corbett, O.~\'Eboli, M.~Gavela, J.~Gonz\'alez-Fraile, {\em
  et.~al.}, {\it {Disentangling a Dynamical Higgs}},  JHEP {\bf 1403} (2014)
  024, [\href{http://arxiv.org/abs/1311.1823}{{\tt arXiv:1311.1823}}].

\bibitem{Gavela:2014vra}
M.~B. Gavela, J.~Gonz\'alez-Fraile, M.~C. Gonz\'alez-Garc\'ia, L.~Merlo,
  S.~Rigolin, and J.~Yepes, {\it {CP violation with a dynamical Higgs}},  JHEP
  {\bf 10} (2014) 44, [\href{http://arxiv.org/abs/1406.6367}{{\tt
  arXiv:1406.6367}}].

\bibitem{Buchalla:2013rka}
G.~Buchalla, O.~Cat\`a, and C.~Krause, {\it {Complete Electroweak Chiral
  Lagrangian with a Light Higgs at NLO}}, Nucl. Phys. {\bf B880} (2014) 552,
  [\href{http://arxiv.org/abs/1307.5017}{{\tt arXiv:1307.5017}}].

\bibitem{Yepes:2015zoa}
J.~Yepes, {\it {Spin-1 Resonances in a Non-Linear Left-Right Dynamical Higgs
  Context}},  \href{http://arxiv.org/abs/1507.03974}{{\tt arXiv:1507.03974}}.

\bibitem{Yepes:2015qwa}
J.~Yepes, R.~Kunming, and J.~Shu, {\it {CP Violation from Spin-1 Resonances in
  a Left-Right Dynamical Higgs Context}}, Commun. Theor. Phys. {\bf 66} (2016) 93,
  [\href{http://arxiv.org/abs/1507.04745}{{\tt arXiv:1507.04745}}].

\bibitem{Feruglio:2016zvt}
F.~Feruglio, B.~Gavela, K.~Kanshin, P.~A.~N. Machado, S.~Rigolin, and S.~Saa,
  {\it {The minimal linear sigma model for the Goldstone Higgs}}, JHEP {\bf 06} (2016) 038,
  [\href{http://arxiv.org/abs/1603.05668}{{\tt arXiv:1603.05668}}].

\bibitem{Alonso:2014wta}
R.~Alonso, I.~Brivio, B.~Gavela, L.~Merlo, and S.~Rigolin, {\it {Sigma
  Decomposition}},  JHEP {\bf 12} (2014) 034,
  [\href{http://arxiv.org/abs/1409.1589}{{\tt arXiv:1409.1589}}].

\bibitem{Hierro:2015nna}
I.~M. Hierro, L.~Merlo, and S.~Rigolin, {\it {Sigma Decomposition: the CP-Odd
  Lagrangian}}, JHEP {\bf 04} (2016) 016 [\href{http://arxiv.org/abs/1510.07899}{{\tt
  arXiv:1510.07899}}].

\bibitem{Brivio:2014pfa}
I.~Brivio, O.~J.~P. \'Eboli, M.~B. Gavela, M.~C. Gonz\'alez-Garcia, L.~Merlo,
  and S.~Rigolin, {\it {Higgs ultraviolet softening}},  JHEP {\bf 12} (2014)
  004, [\href{http://arxiv.org/abs/1405.5412}{{\tt arXiv:1405.5412}}].

\bibitem{Brivio:2015kia}
I.~Brivio, M.~B. Gavela, L.~Merlo, K.~Mimasu, J.~M. No, R.~del Rey, and
  V.~Sanz, {\it {Non-linear Higgs portal to Dark Matter}},  JHEP {\bf 04}
  (2016) 141, [\href{http://arxiv.org/abs/1511.01099}{{\tt arXiv:1511.01099}}].

\bibitem{Murayama:2014yja}
H.~Murayama, V.~Rentala, and J.~Shu, {\it {Probing strong electroweak symmetry
  breaking dynamics through quantum interferometry at the LHC}},  Phys. Rev.
  {\bf D92} (2015), no.~11 116002, [\href{http://arxiv.org/abs/1401.3761}{{\tt
  arXiv:1401.3761}}].

\bibitem{Delgado:2013hxa}
R.~L. Delgado, A.~Dobado, and F.~J. Llanes-Estrada, {\it {One-loop $W_LW_L$ and
  $Z_LZ_L$ scattering from the electroweak Chiral Lagrangian with a light
  Higgs-like scalar}},  JHEP {\bf 02} (2014) 121,
  [\href{http://arxiv.org/abs/1311.5993}{{\tt arXiv:1311.5993}}].

\bibitem{Delgado:2014jda}
R.~L. Delgado, A.~Dobado, M.~J. Herrero, and J.~J. Sanz-Cillero, {\it {One-loop
  $\gamma\gamma \to$ W$_{L}^{+}$ W$_{L}^{-}$ and $\gamma\gamma \to$ Z$_{L}$
  Z$_{L}$ from the Electroweak Chiral Lagrangian with a light Higgs-like
  scalar}},  JHEP {\bf 07} (2014) 149,
  [\href{http://arxiv.org/abs/1404.2866}{{\tt arXiv:1404.2866}}].

\bibitem{Alonso:2012jc}
R.~Alonso, M.~B. Gavela, L.~Merlo, S.~Rigolin, and J.~Yepes, {\it {Minimal
  Flavour Violation with Strong Higgs Dynamics}},  JHEP {\bf 06} (2012) 076,
  [\href{http://arxiv.org/abs/1201.1511}{{\tt arXiv:1201.1511}}].

\bibitem{Gavela:2016bzc}
B.~M. Gavela, E.~E. Jenkins, A.~V. Manohar, and L.~Merlo, {\it {Analysis of
  General Power Counting Rules in Effective Field Theory}},
  \href{http://arxiv.org/abs/1601.07551}{{\tt arXiv:1601.07551}}.

\bibitem{Manohar:1983md}
A.~Manohar and H.~Georgi, {\it {Chiral Quarks and the Nonrelativistic Quark
  Model}},  Nucl. Phys. {\bf B234} (1984) 189.

\bibitem{Cohen:1997rt}
A.~G. Cohen, D.~B. Kaplan, and A.~E. Nelson, {\it {Counting 4 Pis in Strongly
  Coupled Supersymmetry}},  Phys. Lett. {\bf B412} (1997) 301--308,
  [\href{http://arxiv.org/abs/hep-ph/9706275}{{\tt hep-ph/9706275}}].

\bibitem{Biekoetter:2014jwa}
A.~Biekötter, A.~Knochel, M.~Krämer, D.~Liu, and F.~Riva, {\it {Vices and
  virtues of Higgs effective field theories at large energy}},  Phys. Rev. {\bf
  D91} (2015) 055029, [\href{http://arxiv.org/abs/1406.7320}{{\tt
  arXiv:1406.7320}}].

\bibitem{Contino:2016jqw}
R.~Contino, A.~Falkowski, F.~Goertz, C.~Grojean, and F.~Riva, {\it {On the
  Validity of the Effective Field Theory Approach to SM Precision Tests}},
  \href{http://arxiv.org/abs/1604.06444}{{\tt arXiv:1604.06444}}.

\bibitem{Georgi:1986df}
H.~Georgi, D.~B. Kaplan, and L.~Randall, {\it {Manifesting the Invisible Axion
  at Low-Energies}},  Phys. Lett. {\bf B169} (1986) 73.

\bibitem{MSStoappear}
L.~Merlo, S.~Saa, and M.~Sacristan, {\it {B and L non-Conserving Effective
  Lagrangian for a Dynamical Higgs}},  to appear.

\bibitem{Hagiwara:1993ck}
K.~Hagiwara, S.~Ishihara, R.~Szalapski, and D.~Zeppenfeld, {\it {Low-Energy
  Effects of New Interactions in the Electroweak Boson Sector}},  Phys. Rev.
  {\bf D48} (1993) 2182--2203.

\bibitem{Hagiwara:1996kf}
K.~Hagiwara, T.~Hatsukano, S.~Ishihara, and R.~Szalapski, {\it {Probing
  Nonstandard Bosonic Interactions via W Boson Pair Production at Lepton
  Colliders}},  Nucl. Phys. {\bf B496} (1997) 66--102,
  [\href{http://arxiv.org/abs/hep-ph/9612268}{{\tt hep-ph/9612268}}].

\bibitem{Corbett:2012dm}
T.~Corbett, O.~J.~P. \'Eboli, J.~Gonz\'alez-Fraile, and M.~C.
  Gonz\'alez-Garcia, {\it {Constraining anomalous Higgs interactions}},  Phys.
  Rev. {\bf D86} (2012) 075013, [\href{http://arxiv.org/abs/1207.1344}{{\tt
  arXiv:1207.1344}}].

\bibitem{Corbett:2012ja}
T.~Corbett, O.~J.~P. \'Eboli, J.~Gonz\'alez-Fraile, and M.~C.
  Gonz\'alez-Garcia, {\it {Robust Determination of the Higgs Couplings: Power
  to the Data}},  Phys. Rev. {\bf D87} (2013) 015022,
  [\href{http://arxiv.org/abs/1211.4580}{{\tt arXiv:1211.4580}}].

\bibitem{Corbett:2013pja}
T.~Corbett, O.~J.~P. \'Eboli, J.~Gonz\'alez-Fraile, and M.~C.
  Gonz\'alez-Garcia, {\it {Determining Triple Gauge Boson Couplings from Higgs
  Data}},  Phys. Rev. Lett. {\bf 111} (2013) 011801,
  [\href{http://arxiv.org/abs/1304.1151}{{\tt arXiv:1304.1151}}].

\bibitem{Agashe:2014kda}
{\bf Particle Data Group} Collaboration, K.~A. Olive {\em et.~al.}, {\it
  {Review of Particle Physics}},  Chin. Phys. {\bf C38} (2014) 090001.

\bibitem{Aad:2015zhl}
{\bf ATLAS, CMS} Collaboration, G.~Aad {\em et.~al.}, {\it {Combined
  Measurement of the Higgs Boson Mass in $pp$ Collisions at $\sqrt{s}=7$ and 8
  TeV with the ATLAS and CMS Experiments}},  Phys. Rev. Lett. {\bf 114} (2015)
  191803, [\href{http://arxiv.org/abs/1503.07589}{{\tt arXiv:1503.07589}}].

\bibitem{Peskin:1990zt}
M.~E. Peskin and T.~Takeuchi, {\it {A New constraint on a strongly interacting
  Higgs sector}},  Phys. Rev. Lett. {\bf 65} (1990) 964--967.

\bibitem{Peskin:1991sw}
M.~E. Peskin and T.~Takeuchi, {\it {Estimation of oblique electroweak
  corrections}},  Phys. Rev. {\bf D46} (1992) 381--409.

\bibitem{Pomarol:2013zra}
A.~Pomarol and F.~Riva, {\it {Towards the Ultimate SM Fit to Close in on Higgs
  Physics}},  JHEP {\bf 01} (2014) 151,
  [\href{http://arxiv.org/abs/1308.2803}{{\tt arXiv:1308.2803}}].

\bibitem{Ciuchini:2014dea}
M.~Ciuchini, E.~Franco, S.~Mishima, M.~Pierini, L.~Reina, and L.~Silvestrini,
  {\it {Update of the electroweak precision fit, interplay with Higgs-boson
  signal strengths and model-independent constraints on new physics}},  in {\em
  {International Conference on High Energy Physics 2014 (ICHEP 2014) Valencia,
  Spain, July 2-9, 2014}}, 2014.
\newblock \href{http://arxiv.org/abs/1410.6940}{{\tt arXiv:1410.6940}}.

\bibitem{ALEPH:2005ab}
{\bf SLD Electroweak Group, DELPHI, ALEPH, SLD, SLD Heavy Flavour Group, OPAL,
  LEP Electroweak Working Group, L3} Collaboration, S.~Schael {\em et.~al.},
  {\it {Precision electroweak measurements on the $Z$ resonance}},  Phys. Rept.
  {\bf 427} (2006) 257--454, [\href{http://arxiv.org/abs/hep-ex/0509008}{{\tt
  hep-ex/0509008}}].

\bibitem{Group:2012gb}
{\bf CDF, D0} Collaboration, T.~E.~W. Group, {\it {2012 Update of the
  Combination of CDF and D0 Results for the Mass of the W Boson}},
  \href{http://arxiv.org/abs/1204.0042}{{\tt arXiv:1204.0042}}.

\bibitem{ALEPH:2010aa}
{\bf Tevatron Electroweak Working Group, CDF, DELPHI, SLD Electroweak and Heavy
  Flavour Groups, ALEPH, LEP Electroweak Working Group, SLD, OPAL, D0, L3}
  Collaboration, L.~E.~W. Group, {\it {Precision Electroweak Measurements and
  Constraints on the Standard Model}},
  \href{http://arxiv.org/abs/1012.2367}{{\tt arXiv:1012.2367}}.

\bibitem{Isidori:2013cla}
G.~Isidori, A.~V. Manohar, and M.~Trott, {\it {Probing the nature of the
  Higgs-like Boson via $h \to V \mathcal{F}$ decays}},  Phys. Lett. {\bf B728}
  (2014) 131--135, [\href{http://arxiv.org/abs/1305.0663}{{\tt
  arXiv:1305.0663}}].

\bibitem{Isidori:2013cga}
G.~Isidori and M.~Trott, {\it {Higgs form factors in Associated Production}},
  JHEP {\bf 02} (2014) 082, [\href{http://arxiv.org/abs/1307.4051}{{\tt
  arXiv:1307.4051}}].

\bibitem{Gonzalez-Alonso:2015bha}
M.~Gonzalez-Alonso, A.~Greljo, G.~Isidori, and D.~Marzocca, {\it {Electroweak
  bounds on Higgs pseudo-observables and $h \to 4 \ell$ decays}},  Eur. Phys.
  J. {\bf C75} (2015) 341, [\href{http://arxiv.org/abs/1504.04018}{{\tt
  arXiv:1504.04018}}].

\bibitem{Corbett:2015ksa}
T.~Corbett, O.~J.~P. \'Eboli, D.~Goncalves, J.~Gonz\'alez-Fraile, T.~Plehn, and
  M.~Rauch, {\it {The Higgs Legacy of the LHC Run I}},  JHEP {\bf 08} (2015)
  156, [\href{http://arxiv.org/abs/1505.05516}{{\tt arXiv:1505.05516}}].

\bibitem{Masso:2012eq}
E.~Massó and V.~Sanz, {\it {Limits on anomalous couplings of the Higgs boson
  to electroweak gauge bosons from LEP and the LHC}},  Phys. Rev. {\bf D87}
  (2013), no.~3 033001, [\href{http://arxiv.org/abs/1211.1320}{{\tt
  arXiv:1211.1320}}].

\bibitem{Banerjee:2013apa}
S.~Banerjee, S.~Mukhopadhyay, and B.~Mukhopadhyaya, {\it {Higher dimensional
  operators and the LHC Higgs data: The role of modified kinematics}},  Phys.
  Rev. {\bf D89} (2014), no.~5 053010,
  [\href{http://arxiv.org/abs/1308.4860}{{\tt arXiv:1308.4860}}].

\bibitem{Ellis:2014dva}
J.~Ellis, V.~Sanz, and T.~You, {\it {Complete Higgs Sector Constraints on
  Dimension-6 Operators}},  JHEP {\bf 07} (2014) 036,
  [\href{http://arxiv.org/abs/1404.3667}{{\tt arXiv:1404.3667}}].

\bibitem{Ellis:2014jta}
J.~Ellis, V.~Sanz, and T.~You, {\it {The Effective Standard Model after LHC Run
  I}},  JHEP {\bf 03} (2015) 157, [\href{http://arxiv.org/abs/1410.7703}{{\tt
  arXiv:1410.7703}}].

\bibitem{Edezhath:2015lga}
R.~Edezhath, {\it {Dimension-6 Operator Constraints from Boosted VBF Higgs}},
  \href{http://arxiv.org/abs/1501.00992}{{\tt arXiv:1501.00992}}.

\bibitem{Lafaye:2009vr}
R.~Lafaye, T.~Plehn, M.~Rauch, D.~Zerwas, and M.~Duhrssen, {\it {Measuring the
  Higgs Sector}},  JHEP {\bf 08} (2009) 009,
  [\href{http://arxiv.org/abs/0904.3866}{{\tt arXiv:0904.3866}}].

\bibitem{Klute:2012pu}
M.~Klute, R.~Lafaye, T.~Plehn, M.~Rauch, and D.~Zerwas, {\it {Measuring Higgs
  Couplings from LHC Data}},  Phys. Rev. Lett. {\bf 109} (2012) 101801,
  [\href{http://arxiv.org/abs/1205.2699}{{\tt arXiv:1205.2699}}].

\bibitem{Plehn:2012iz}
T.~Plehn and M.~Rauch, {\it {Higgs Couplings after the Discovery}},  Europhys.
  Lett. {\bf 100} (2012) 11002, [\href{http://arxiv.org/abs/1207.6108}{{\tt
  arXiv:1207.6108}}].

\bibitem{Klute:2013cx}
M.~Klute, R.~Lafaye, T.~Plehn, M.~Rauch, and D.~Zerwas, {\it {Measuring Higgs
  Couplings at a Linear Collider}},  Europhys. Lett. {\bf 101} (2013) 51001,
  [\href{http://arxiv.org/abs/1301.1322}{{\tt arXiv:1301.1322}}].

\bibitem{Lopez-Val:2013yba}
D.~Lopez-Val, T.~Plehn, and M.~Rauch, {\it {Measuring extended Higgs sectors as
  a consistent free couplings model}},  JHEP {\bf 10} (2013) 134,
  [\href{http://arxiv.org/abs/1308.1979}{{\tt arXiv:1308.1979}}].

\bibitem{Corbett:2015mqf}
T.~Corbett, O.~J.~P. Eboli, D.~Goncalves, J.~Gonzalez-Fraile, T.~Plehn, and
  M.~Rauch, {\it {The Non-Linear Higgs Legacy of the LHC Run I}},
  \href{http://arxiv.org/abs/1511.08188}{{\tt arXiv:1511.08188}}.

\bibitem{Eboli:2010qd}
O.~J.~P. Eboli, J.~Gonzalez-Fraile, and M.~C. Gonzalez-Garcia, {\it
  {Scrutinizing the ZW+W- vertex at the Large Hadron Collider at 7 TeV}},
  Phys. Lett. {\bf B692} (2010) 20--25,
  [\href{http://arxiv.org/abs/1006.3562}{{\tt arXiv:1006.3562}}].

\bibitem{Hagiwara:1986vm}
K.~Hagiwara, R.~D. Peccei, D.~Zeppenfeld, and K.~Hikasa, {\it {Probing the Weak
  Boson Sector in $e^+ e^- \to W^+ W^-$}},  Nucl. Phys. {\bf B282} (1987) 253.

\bibitem{Butter:2016cvz}
A.~Butter, O.~J.~P. \'Eboli, J.~Gonz\'alez-Fraile, M.~C. Gonz\'alez-Garcia, and
  T.~Plehn, {\it {The Gauge-Higgs Legacy of the LHC Run I}},
  \href{http://arxiv.org/abs/1604.03105}{{\tt arXiv:1604.03105}}.

\bibitem{Hagiwara:1993qt}
K.~Hagiwara, R.~Szalapski, and D.~Zeppenfeld, {\it {Anomalous Higgs boson
  production and decay}},  Phys. Lett. {\bf B318} (1993) 155--162,
  [\href{http://arxiv.org/abs/hep-ph/9308347}{{\tt hep-ph/9308347}}].

\bibitem{Falkowski:2015jaa}
A.~Falkowski, M.~Gonzalez-Alonso, A.~Greljo, and D.~Marzocca, {\it {Global
  constraints on anomalous triple gauge couplings in effective field theory
  approach}},  Phys. Rev. Lett. {\bf 116} (2016), no.~1 011801,
  [\href{http://arxiv.org/abs/1508.00581}{{\tt arXiv:1508.00581}}].

\bibitem{Drozd:2015kva}
A.~Drozd, J.~Ellis, J.~Quevillon, and T.~You, {\it {Comparing EFT and Exact
  One-Loop Analyses of Non-Degenerate Stops}},  JHEP {\bf 06} (2015) 028,
  [\href{http://arxiv.org/abs/1504.02409}{{\tt arXiv:1504.02409}}].

\bibitem{Gorbahn:2015gxa}
M.~Gorbahn, J.~M. No, and V.~Sanz, {\it {Benchmarks for Higgs Effective Theory:
  Extended Higgs Sectors}},  JHEP {\bf 10} (2015) 036,
  [\href{http://arxiv.org/abs/1502.07352}{{\tt arXiv:1502.07352}}].

\bibitem{Brehmer:2015rna}
J.~Brehmer, A.~Freitas, D.~Lopez-Val, and T.~Plehn, {\it {Pushing Higgs
  Effective Theory to its Limits}},  Phys. Rev. {\bf D93} (2016) 075014,
  [\href{http://arxiv.org/abs/1510.03443}{{\tt arXiv:1510.03443}}].

\bibitem{Biekotter:2016ecg}
A.~Biekötter, J.~Brehmer, and T.~Plehn, {\it {Pushing Higgs Effective Theory
  over the Edge}},  \href{http://arxiv.org/abs/1602.05202}{{\tt
  arXiv:1602.05202}}.

\bibitem{Eboli:2016kko}
O.~J.~P. Eboli and M.~C. Gonzalez-Garcia, {\it {Clarifying the bosonic
  quartic couplings}}, Phys. Rev. {\bf D93} (2016) 093013,  [\href{http://arxiv.org/abs/1604.03555}{{\tt
  arXiv:1604.03555}}].

\bibitem{Giudice:2007fh}
G.~F. Giudice, C.~Grojean, A.~Pomarol, and R.~Rattazzi, {\it {The
  Strongly-Interacting Light Higgs}},  JHEP {\bf 06} (2007) 045,
  [\href{http://arxiv.org/abs/hep-ph/0703164}{{\tt hep-ph/0703164}}].

\bibitem{Berthier:2016tkq}
L.~Berthier, M.~Bj\o rn, and M.~Trott, 
{\it {Incorporating doubly resonant $W^\pm$ data in a global 
fit of SMEFT parameters to lift flat directions}},  JHEP {\bf 06} (2007) 045,
[\href{http://arxiv.org/abs/1606.06693}{{\tt arXiv: 1606.06693}}].  



\end{thebibliography}
\end{document}